\documentclass[editor]{aa}
\usepackage{natbib,amssymb,graphicx,graphics}
\usepackage{multirow}
\newcommand{\excs}{\extracolsep{\fill}}

\usepackage{graphicx}
\usepackage{txfonts}
\begin{document} 

\title{The 2008 outburst in the Young Stellar System Z CMa}
\subtitle{III - Multi-epoch high-angular resolution images and spectra of the components in near-infrared}
\titlerunning{Multi-epoch images and spectra of the components of Z CMa in the near-infrared during the 2008 outburst}
\author{M. Bonnefoy\inst{1, 2}
 \and
 G. Chauvin\inst{1,2}
 \and
 C. Dougados\inst{1,2}
 \and
\'A. K\'osp\'al\inst{3}
 \and
 M. Benisty \inst{1,2}
 \and
 G. Duch\^{e}ne \inst{1,2,4}
 \and
 J. Bouvier \inst{1,2}
 \and
 P. J. V. Garcia \inst{5}
 \and
 E. Whelan \inst{6}
 \and
 S. Antoniucci \inst{7}
 \and
 L. Podio \inst{7}}

\offprints{Micka\"{e}l Bonnefoy: mickael.bonnefoy@univ-grenoble-alpes.fr}

\institute{Univ. Grenoble Alpes, IPAG, F-38000 Grenoble, France 
\and
 CNRS, IPAG, F-38000 Grenoble, France \\
\email{mickael.bonnefoy@obs.ujf-grenoble.fr}
\and
Konkoly Observatory, Research Centre for Astronomy and Earth Sciences, Hungarian Academy of Sciences, PO Box 67, 1525 Budapest, Hungary
%\and
%Centro de Astrobiolog\'{i}a, INTA-CSIC, PO Box 78, 28691 Villanueva de la Ca\~{n}ada, Madrid, Spain 
\and
Astronomy Department, UC Berkeley, 501 Campbell Hall, Berkeley CA 94720-3411, USA
\and
Universidade do Porto, Faculdade de Engenharia, Rua Dr. Roberto Frias, s/n, P-4200-465 Porto, Portugal
\and
Department of Experimental Physics, National University of Ireland Maynooth, Maynooth, Co. Kildare, Ireland
\and
INAF - Osservatorio Astrofisico di Arcetri - L.go E. Fermi 5, I-50125 Firenze, Italy}

\date{Received April 12, 2016 ; accepted June 22, 2016}

%------------------------------------------------------------------------
%------------------------------------------------------------------------
  \abstract 
% context heading (optional)
    {Z CMa is a complex pre-main sequence binary with a current
      separation of 110~mas, known to consist of an FU Orionis star (SE
      component) and an embedded Herbig Be star (NW component). Although it represents a well-studied and characterized  system, the origin of photometric variabilities, the component properties, and the physical configuration
      of the system remain mostly unknown.}
% aims heading (mandatory)
  {Immediately when the late-2008 outburst of Z CMa was announced to the
    community, we initiated a high angular resolution imaging campaign
    aimed at characterizing the outburst state of both components of
    the system in the near-infrared.}
% methods heading (mandatory) 
  {We used the VLT/NACO and the Keck/NIRC2 near-infrared adaptive
    optics instrument to monitor the astrometric position and the
    near-infrared photometry of the Z CMa components during the
    outburst phase and one year after. The VLT/SINFONI and Keck/OSIRIS integral field
    spectroscrographs were in addition used to characterize for the
    first time the resolved spectral properties of the FU Orionis and the
    Herbig Be component during and after the outburst.}
% results heading  (mandatory) 
  {We confirm that the  NW star dominates the system flux in the 1.1-3.8 $\mu$m range and is responsible for the photometric outburst.   We extract the first medium-resolution (R$\sim$2000-4000) near-infrared (1.1-2.4 $\mu$m) spectra of the individual components.  The SE component has a spectrum  typical of FU Orionis objects. The NW component spectrum is characteristic of embedded outbursting protostars and EX Or objects.  It displays numerous emission lines whose intensity correlates with the system activity. In particular, we find a correlation between the Br$\gamma$ equivalent width and the system brightness. The bluing of the continuum of the NW component along with the absolute flux and color-variation of the system during the outburst suggests that the outburst was caused by a complex interplay between a variation of the extinction in the line of sight of the NW component on one hand, and the emission of shocked regions close to the NW component on the other. We confirm the recently reported wiggling of the SE component jet from $[Fe II]$ line emission. We find a point-like structure associated with a peak emission at 2.098$\mu$m coincidental with the clump or arm seen in broadband polarization differential imaging as well as additional diffuse emission along a PA=$214^{\circ}$. The origin of these two structures is unclear and deserves further investigation.}
 {}
 
 
\maketitle

%
%________________________________________________________________

\section{Introduction}
\object{Z Canis Majoris} (\object{Z CMa}) is a complex pre-main sequence binary member of the CMa OB1 association (age $<$ 1 Myr; Herbst et al. 1978), with an estimated distance ranging from 950 pc to 1150 pc \citep{1974A&A....37..229C, 1978ApJS...38..309H, 2000MNRAS.312..753K}. The system is known for its recurring  EX Orionis (EX Or)-like outbursts  with recent events recorded in 1987, 2000, 2004, 2008, 2011, 2015, and 2016. 

 Z CMa was classified by \cite{1989ApJ...338.1001H} as
a FU Orionis variable based on its broad, doubled optical line absorptions,
optical spectral type of F-G, and CO first overtone. The authors were able to
reproduce the UV-optical spectrum by a model of an optically thick
accretion disk surrounding an 1-3~$M_{\odot}$ star with an accretion
rate of $10^{-3}$~$M_{\odot}.\rm{yr}^{-1}$. The model failed
to explain the near-infrared (NIR) part of the Z CMa spectrum, however, as well as a strong increase
in brightness in 1987. During the 1987 outburst, the optical spectrum
was characterized by a featureless and bluer continuum, a rise of
Balmer lines, and emission lines of Fe II, Cr II,
and Ti II \citep{1991ApJ...370..384H}, which is unconsistent with typical FU Orionis
activity.

The problem was partly solved by \cite{1991AJ....102.2073K} who 
discovered that Z CMa consists of a 110 mas binary. The SW component of the pair is the FU Orionis star that dominates the optical flux. The NW component dominates the NIR ($>$ 2 $\mu$m) to sub-mm spectrum, and the total luminosity. Follow-up observations \citep[][hereafter VDA04]{1993ApJ...417..687W, 1999A&A...346..892G, VDA04} permitted the conclusion that this component is an Herbig Be star (hereafter HBe). VDA04 obtained a NIR (1.9--4.1~$\mu$m) spectrum while the system was returning to its quiescent state. This spectrum shows Br$\alpha$, Br$\gamma$, and Pf$\gamma$ in emission and the CO band-head in absorption around 2.3~$\mu$m. The authors suggested that the emission lines are formed into an accreting circumstellar disk surrounding the HBe, and
also in an extended envelope above and below the disk plane. The emission line analysis led VDA04 to infer a B0IIIe spectral type. Luminosity and effective temperature estimates enables the authors to derive a mass of $16-38$~M$_{\odot}$ for the HBe using various evolutionary model predictions for ages of $3.10^5$~yr and $4.10^4$~yr respectively.

At large scales, this system reveals a rich outflow activity. A
 collimated optical outflow (3.9~pc; P.A. = 240$^{\circ}$) and a bow-shock-shaped feature at 60~$\!''$ have been reported by \cite{1989A&A...224L..13P}. Multiple outflow component profiles traced by optical
forbidden lines are present close to the source. \cite{2010ApJ...720L.119W} confirmed the existence of
a jet driven by the HBe star (paper 2), which is unambiguously the driving
source of the Z CMa parsec-scale outflow. The HBe star jet is seen oscillating around a given position angle (jet wiggling), probably due to a closer companion
to the HBe star. A twin jet driven by the FUOR component was also
clearly identified. \cite{2001ApJ...546..358M} also reported
a cavity-like structure in adaptive optics $J$-band imaging, extending
to the SW of the system, which they interpreted as light scattered off the wall of
a jet-blown cavity aligned with the Z CMa large-scale outflows. 

\cite{2009A&A...497..117A} modeled the spectral energy distribution of the system by contributions from free-free emission, a disk of size  $180^{+250}_{-140}$ au tilted by  $30^{+40}_{-20}$ degrees, and an  infalling envelope (spherical or toroidal) surrounding each star that possibly extends from 2000 to 5000 au and might be carved by the outflows. The HBe itself is embedded in a dust cocoon \citep{1993ApJ...417..687W}. A sketch of the system can be found in \cite{2012A&A...543A..70C}.  The dust distribution at the 500 au scale around Z CMa was  investigated using NIR imaging polarimetry  \citep{2015A&A...578L...1C, 2016SciA....200875L}. It reveals an extended filamentary structure (up to 2") observed by \cite{2002ApJ...580L.167M}, whose origin is unclear, as well as a polarized clump  closer  to the stars (0.3-0.5'').  

At the Astronomical Unit scale,  the FUOR is strongly resolved by the Keck-I interferometer to a
level that is difficult to explain with  thermal emission of
the accretion disk alone \citep{2006ApJ...641..547M}. Hence its close environment appears to be more
complex than expected. The HBe component has also been  observed with
Keck-I \citep{2005ApJ...624..832M} and with AMBER on the VLTI \citep{2008A&A...479..589L, 2010A&A...517L...3B}. Based on the
Keck observations, the authors have modelled the NIR emission
as coming from a uniform ring. Nonetheless, they have finally concluded that
the ring model was doubtful for this star, considering
the large uncertainty of its spectral type. Morover,
the amount of infrared emission coming from this source alone at a given epoch is also
poorly known, which adds to the difficulty of interpreting the
interferometric measurements.

From January 2008 to October 2009, the system began to experience an  optical outburst, the largest reported in the past 90 years of available observations \citep{2009IBVS.5905....1G}. This outburst triggered additional observations of the system.   \cite{2013ApJ...763L...9H} presented adaptive optics JHKL band photometry of the individual components. They concluded that the embedded HBe component is solely responsible for the outburst.   \cite{2010A&A...517L...3B} obtained spectrally resolved interferometric observations of the HBe  to study the hot gas emitting across the Br$\gamma$ emission line (paper 1). They found that the line profile, the astrometric signal, and the visibilities are consistent with the emission of a bipolar wind that may be partly seen through a disk hole inside the dust sublimation
radius at the au scale. Their multi-epoch observations led them to suggest that the outburst is related to a period of strong mass-loss and not to a change in  the extinction along the line of sight. The spectrophotometric, spectropolarimetric, and polarimetric imaging  observations of the system \citep{2010A&A...509L...7S, 2012A&A...543A..70C} suggest in contrast that the outburst was caused by changes in extinction of the dust cocoon surrounding the HBe star.

\begin{figure*}[t]
\centering
\vspace{-0.1cm}
\includegraphics[width=\linewidth]{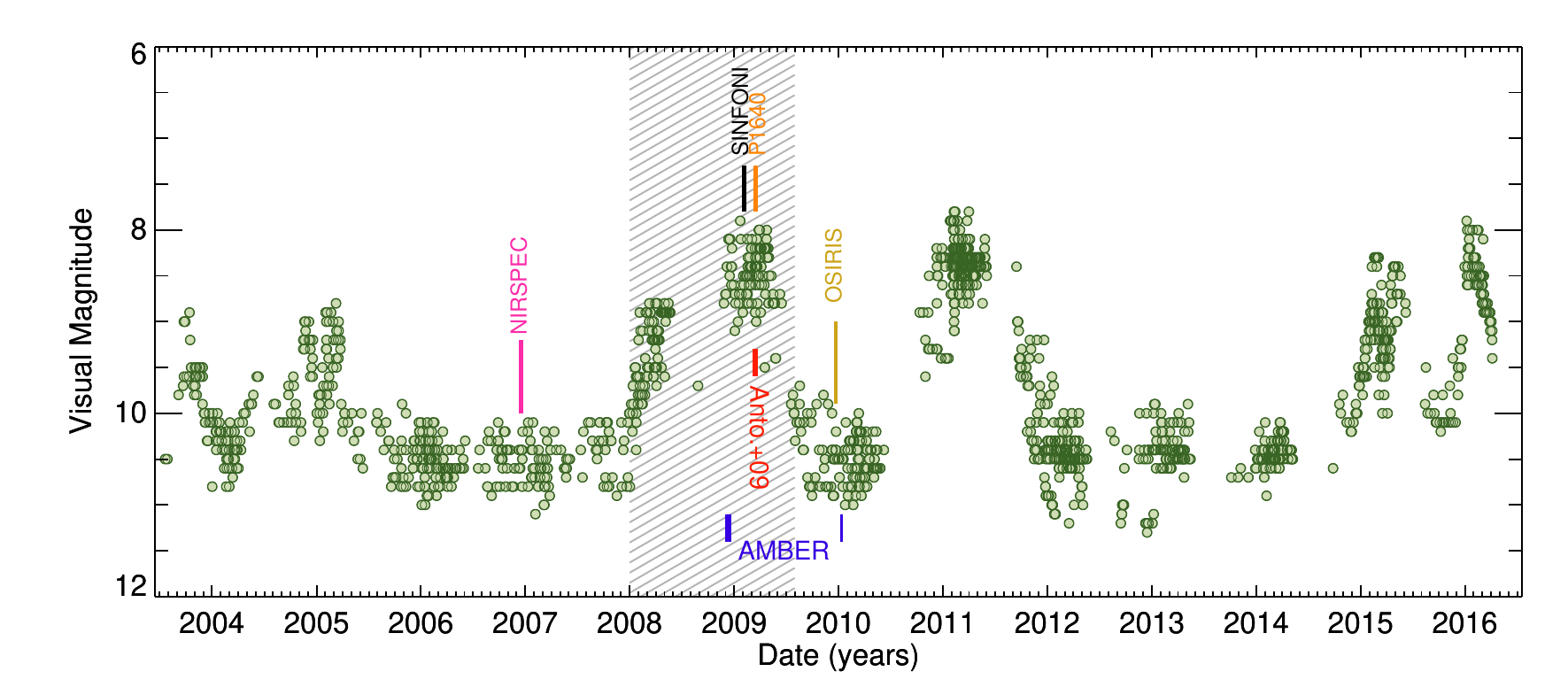}
\caption{Light curve of Z CMa inferred from AAVSO observations. The spectroscopic observations of the HBe and FUOR are reported in the figure. The dashed zone corresponds to the 2008 outbust of the system.}
\label{fig:lightcurve}
\end{figure*}

We present the results of a complementary high angular resolution imaging campaign to resolve and characterize the properties of each component of Z CMa in the NIR during and after its 2008 outburst phase and to study its close environnement, in an attempt to better understand the origins of the bursts (extinction, or accretion and ejection scenarios). Our new observations were made from January to March 2009, that is, about one year after the start of the two-year outburst. We obtained an additional observation when the system returned to quiescence (see Fig. \ref{fig:lightcurve}).  We  extracted the first medium-resolution (R$\sim$2000-4000) spectra  from 1.1 to 2.5 $\mu$m of each components of the system. We present  additional JHKL band photometric and astrometric data of the binary. We report in Sect. \ref{section: observations} our observations and detail the associated data reduction and analysis in Sect. \ref{section:reduction}. We analyze  the evolution of the NIR photometric and spectroscopic properties of the components in Sects. \ref{section:AOimaging} and Sect. \ref{section:IFU}, respectively. We discuss these results in the context of other variable objects, and of past outbursts of the system in Sect. \ref{section:discussion}. We summarize our results in Sect. \ref{section:summary}.

\begin{table*}[t]
\caption{Observing log. Sr-2.2$\mu$m corresponds to the Strehl ratio measured at 2.2$\mu$m.}             % title of Table
\label{tab:obslog}
\centering
\renewcommand{\footnoterule}{}  % to avoid a line before footnotes
\begin{tabular*}{\textwidth}{@{\excs}lllllllll}     % 7 columns
\hline\hline\noalign{\smallskip}
UT Date     & Name      & Instr. & Filter (Grism) &   Camera  & Exp. Time    & Sr-2.2$\mu$m      & Airmass     & Comment\\ 
\noalign{\smallskip}\hline\noalign{\smallskip}
2009-01-31  & Z CMa      & NACO & $J, H, K_s, L'$  & S13, L27  & 90s          & (32\%)            & 1.19        &       \\
            & HD54335   & NACO & $J, H, K_s, L'$  & S13, L27  & 90s          & 22\%              & 1.25        & psf-ref   \\
\noalign{\smallskip}
2009-02-06  & Z CMa      & SINFONI & $J$     & 0.025       & 240s         & n.a.               & 1.05         &    \\
				  & Z CMa      & SINFONI & $H$     & 0.025       & 240s         & n.a.               & 1.09         &    \\
				  &Z CMa		&	SINFONI	&	$K$	& 0.025		&	240s	       & n.a.               & 1.22         &    \\
2009-01-28	&	HIP072241	&	SINFONI &	$J$	&	0.25	&	10s &  n.a. & 1.06 & telluric \\
2009-02-22	&	HIP054970	&	SINFONI & $H$	&	0.25	&	12s & n.a.	&	1.09	& telluric \\
2009-02-07	&	HIP071218	&	SINFONI & $K$	&	0.25	&	10s & n.a.	&	1.22	& telluric \\
\noalign{\smallskip}
2009-02-26  & Z CMa      & NACO & $J, H, K_s, L'$  & S13, L27  & 90s          & (42\%)            & 1.06        &   \\
            & HD54335   & NACO & $J, H, K_s, L'$  & S13, L27  & 90s          & 41\%              & 1.11        & psf-ref   \\
\noalign{\smallskip}
2009-03-11  & Z CMa      & NACO & $J, H, K_s, L'$  & S13, L27  & 90s          & (41\%)            & 1.02        &  \\
            & HD54335   & NACO & $J, H, K_s, L'$  & S13, L27  & 90s          & 34\%              & 1.04        &  psf-ref \\
\noalign{\smallskip}
2009-12-07  & Z CMa      & NIRC2 & $J_{\rm{cont}}, H_{\rm{cont}}, K_{\rm{cont}}$   & 9.96mas/pix       & 8.6s       & -              & 1.29     &         \\
            & HIP33998  & NIRC2 & $J_{\rm{cont}}, H_{\rm{cont}}, K_{\rm{cont}}$   & 9.96mas/pix       & 8.6s       & 45\%           & 1.51     & psf-ref         \\ 
\noalign{\smallskip}\hline
\end{tabular*}
\footnotetext[1]{Seeing evaluated at 0.55 $\mu$m.}
\end{table*}

%------------------------------------------------------------------------
%------------------------------------------------------------------------
\section{Observations}
\label{section: observations}
%% ----------------------------
%% 2. Observations
%% 2.1 VLT/NACO
%% 2.2 Keck/NIRSPEC
%% 2.3 VLT/SINFONI

%% 3. Data reduction and analysis
%% 3.1 High-contrast imaging
%% 3.2 Integral field spectroscopy

%% 4. Results
%% 4.1 Astrometry and orbit constraints
%% 4.2 NIR photometry
%% 4.2 Resolved Jet-like structure
%% 4.3 Spectroscopic analysis
%% 4.4 Spectro-astrometry

%% 5. Discussion
%% ----------------------------

%----------------------------------
\subsection{VLT/NACO}

\begin{figure}[t]
\centering
\vspace{-0.1cm}
\includegraphics[width =\columnwidth]{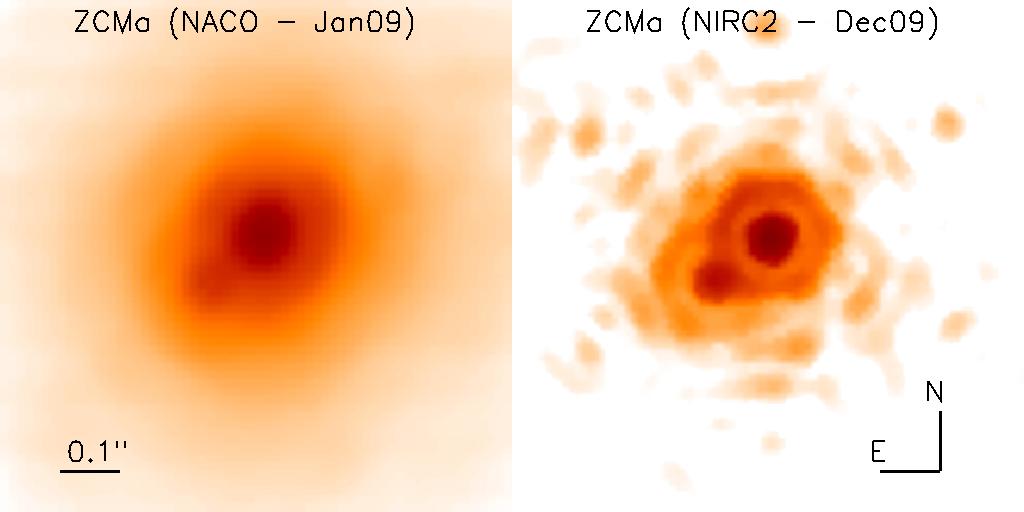}
\caption{Left: VLT/NACO $K_s$-band image of Z CMa on Januray 31,
  2009. Right: Keck/NIRC2 $K_{\rm{cont}}$ image on December 7,
  2009. }
\label{fig:image}
\end{figure}

The Z CMa 2008 outburst was monitored with the NACO high-contrast
adaptive optics (AO) imager of the VLT-UT4. The NAOS AO system
\citep{2003SPIE.4839..140R} is equipped with a tip-tilt mirror, a 185 piezo
actuator deformable mirror and two wavefront sensors (visible and
IR). Attached to NAOS, CONICA \citep{2003SPIE.4841..944L} is the NIR ($1-5\,\mu$m domain) imaging camera equipped with a
$1024\times1024$ pixel Aladdin InSb array. Observations were obtained
at three different epochs between January 2009 and March 2009
(outburst phase). During the different observing campaigns, the
atmospheric conditions were sufficiently stable to close the AO loop
and resolve both components (see Fig.~\ref{fig:image},
\textit{Left panel}). The fainter FU Orionis component in $K$-band is clearly visible on both Keck and VLT images. Both images were
normalized in flux to show the flux variation that is discernable
between January and December 2009 in K band. The relative position of
the FUOR  with respect to that of the HBe component could be monitored well.  The
typical observing sequence included a set of five jittered images
obtained using the J, H, and K$_{s}$ bands with the S13 camera CONICA (mean
plate scale of 13.25~mas/pixel) and using the L' filter with L27 (mean
plate scale of 27.10~mas/pixel), leading to a total exposure time of
$\sim5$~min on source. The corresponding setups are reported in
Table~\ref{tab:obslog}.

%----------------------------------
\subsection{Keck/NIRC2}

On December 17, 2009 (transient phase),  imaging observations were
also obtained using the AO system on the 10 m Keck
II telescope \citep{2004ApOpt..43.5458V}. We obtained direct images
from 1.0 to 2.5 $\mu$m that clearly resolved the two components of
the system with the facility AO-dedicated NIR camera NIRC2, a $1024\times1024$ detector for
which we used the $9.963\pm0.005$ mas pixel$^{-1}$ scale 
and whose absolute orientation on the sky is $0.13\pm0.02^{\circ}$ \citep{2008ApJ...689.1044G}.  We
used the narrowband  filters $J_{cont}$, $H_{cont}$, and $K_{cont}$ centered on 1.2132, 1.5804, and 2.2706~$\mu$m  to avoid the saturation of the star. The corresponding setups are also reported in Table~\ref{tab:obslog}. Both components
were easily resolved to determine the relative flux and position (Fig.~\ref{fig:image}, \textit{Right panel}).

%----------------------------------
\subsection{VLT/SINFONI}

We used the SINFONI instrument \citep[Spectrograph for INfrared Field
Observations, see ][]{2003SPIE.4841.1548E, 2004Msngr.117...17B}, located
at the Cassegrain focus of the VLT UT4 Yepun to observe Z CMa during
the outburst phase on February 6, 2009. The instrument provides
AO-assisted integral field spectroscopy. It uses a modified version of
the Multi-Applications Curvature Adaptive Optics system \citep[MACAO,][]{2003SPIE.4839..329B} designed to feed the SPectrograph for Infrared
Faint Field Imaging \citep[SPIFFI, ][]{2003Msngr.113...17E}. SPIFFI pre-slit
optics where chosen to provide a spatial pixel scale of  12.5$\times$25~mas per
pixel. Three different
gratings were used to cover the J (1.1--1.4~$\mu$m), H
(1.45--1.85~$\mu$m), and K band (1.95--2.45~$\mu$m) at medium
resolving powers (2000, 3000, and 4000 respectively).

The instrument was rotated to orient the binary horizontally in the
field of view (FoV) (see Fig.~2). Sky exposures were recorded
following a ABBA pattern. Additional offsets on the object were
chosen to increase the FOV to $850\times860$ mas in the J
band, $850\times900$ mas in the H band, and $950\times900$ mas in the
K band. This also permitted us to artificially double the vertical spatial
sampling. The source was bright enough at optical wavelengths to be
used as guide probe for the wavefront sensing. Hipparcos standard
stars were also acquired soon after Z CMa to correct the spectra for
telluric features (see Table~\ref{tab:obslog}).

\begin{figure}[t]
\centering
\vspace{-0.1cm}
\includegraphics[width = \columnwidth]{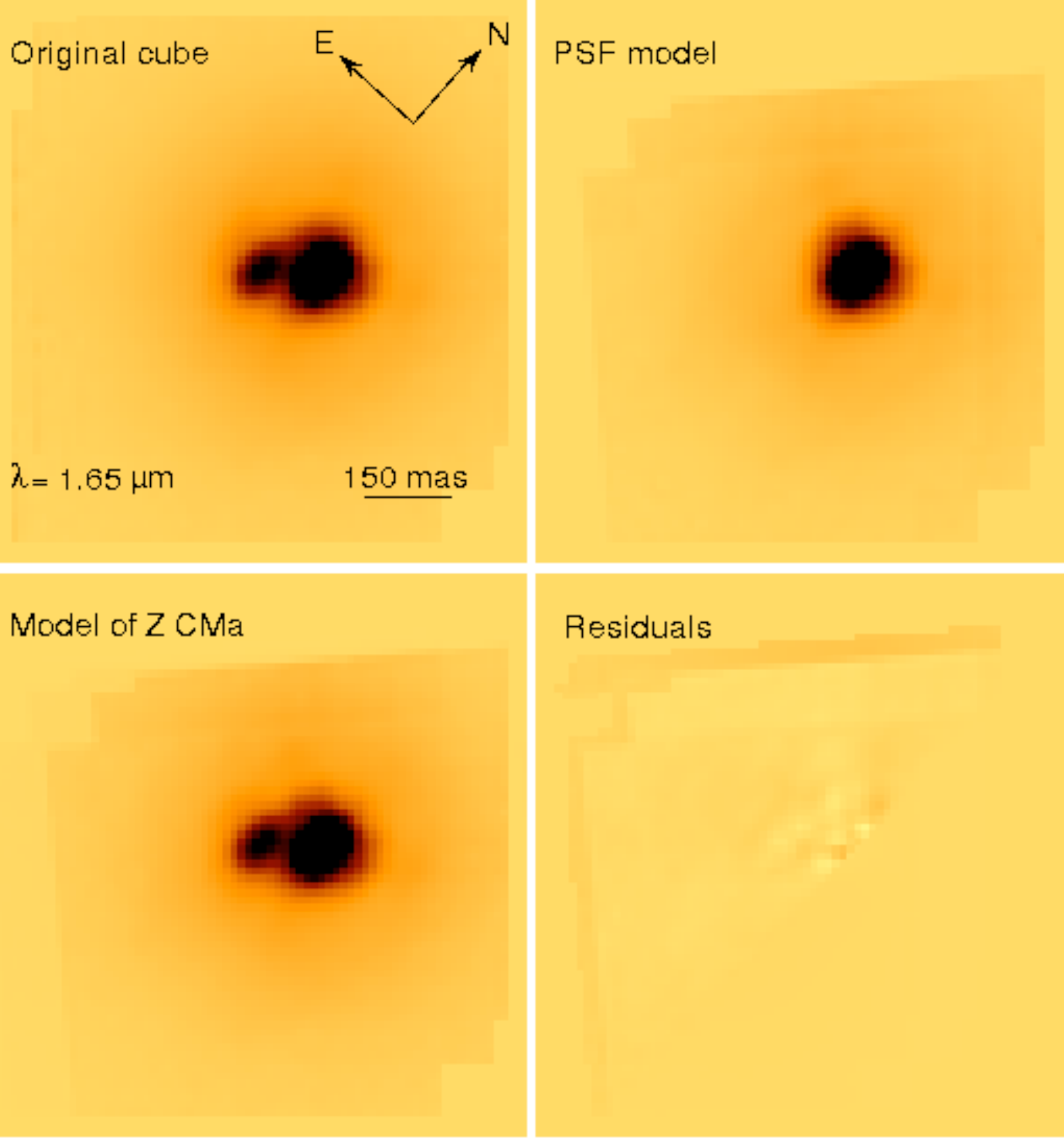}

\caption{Illustration of the spectral deblending process at 1.65
  $\mu$m. Upper left: The initial Z CMa datacube where the two
  components are resolved. Upper right. During the observation, the
  adaptive-optics corrected PSF showed a strong
  astigmatism. To achieve a proper extraction of the spectra of the Z
  CMa components, we duplicated the HBe star profile to create a PSF
  model. Lower left: The position and the flux of the individual
  sources were then retrieved using a modified version of the CLEAN
  algorithm. Lower right: The extraction error is estimated from the
  residuals.}

\label{fig:extraction}
\end{figure}

%------------------------------------------------------------------------
%------------------------------------------------------------------------
\section{Data analysis}
\label{section:reduction}

\begin{figure*}[t]
\centering
\vspace{-0.1cm}
\includegraphics[height = 8cm]{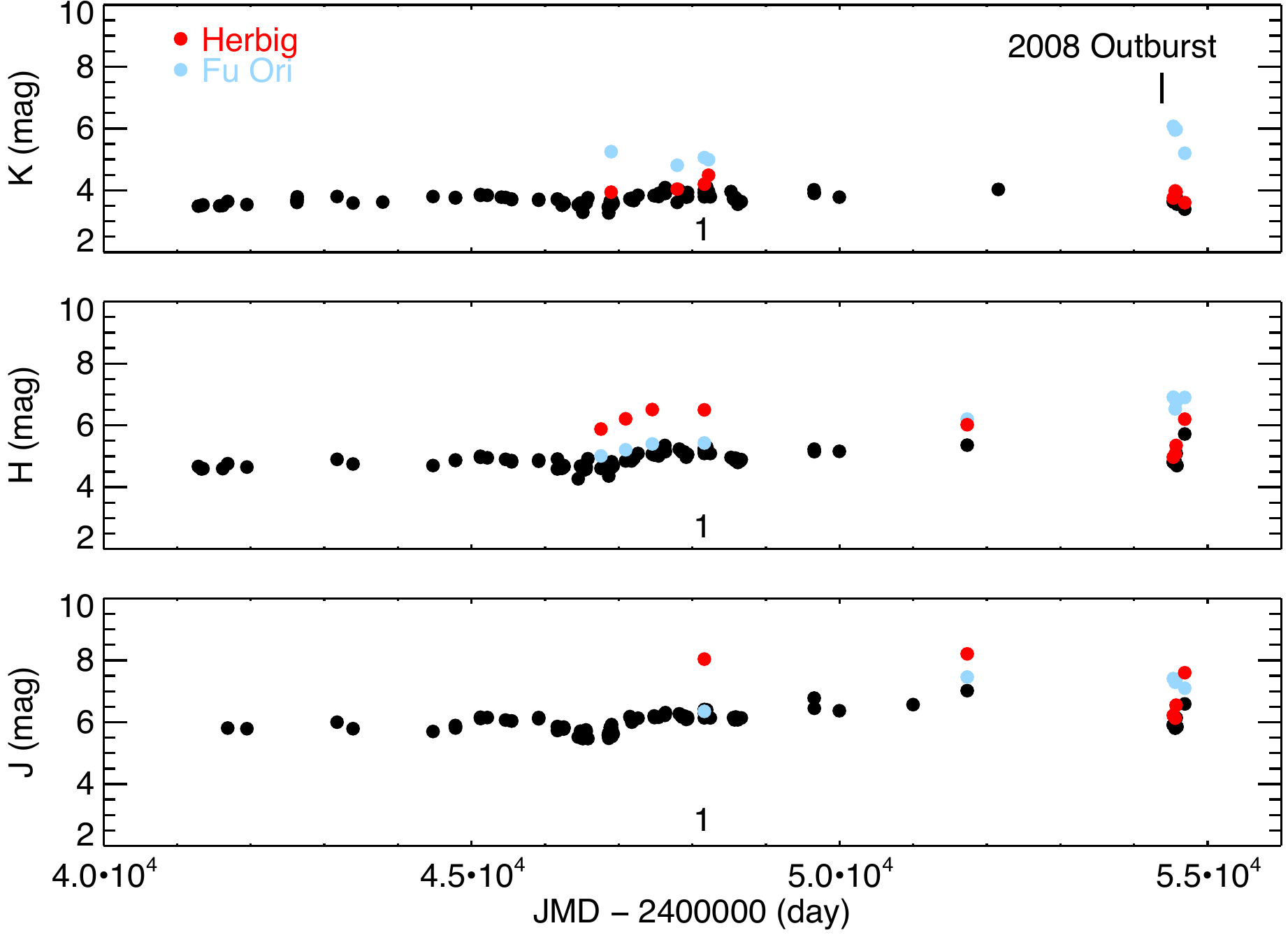}
\includegraphics[height = 8cm]{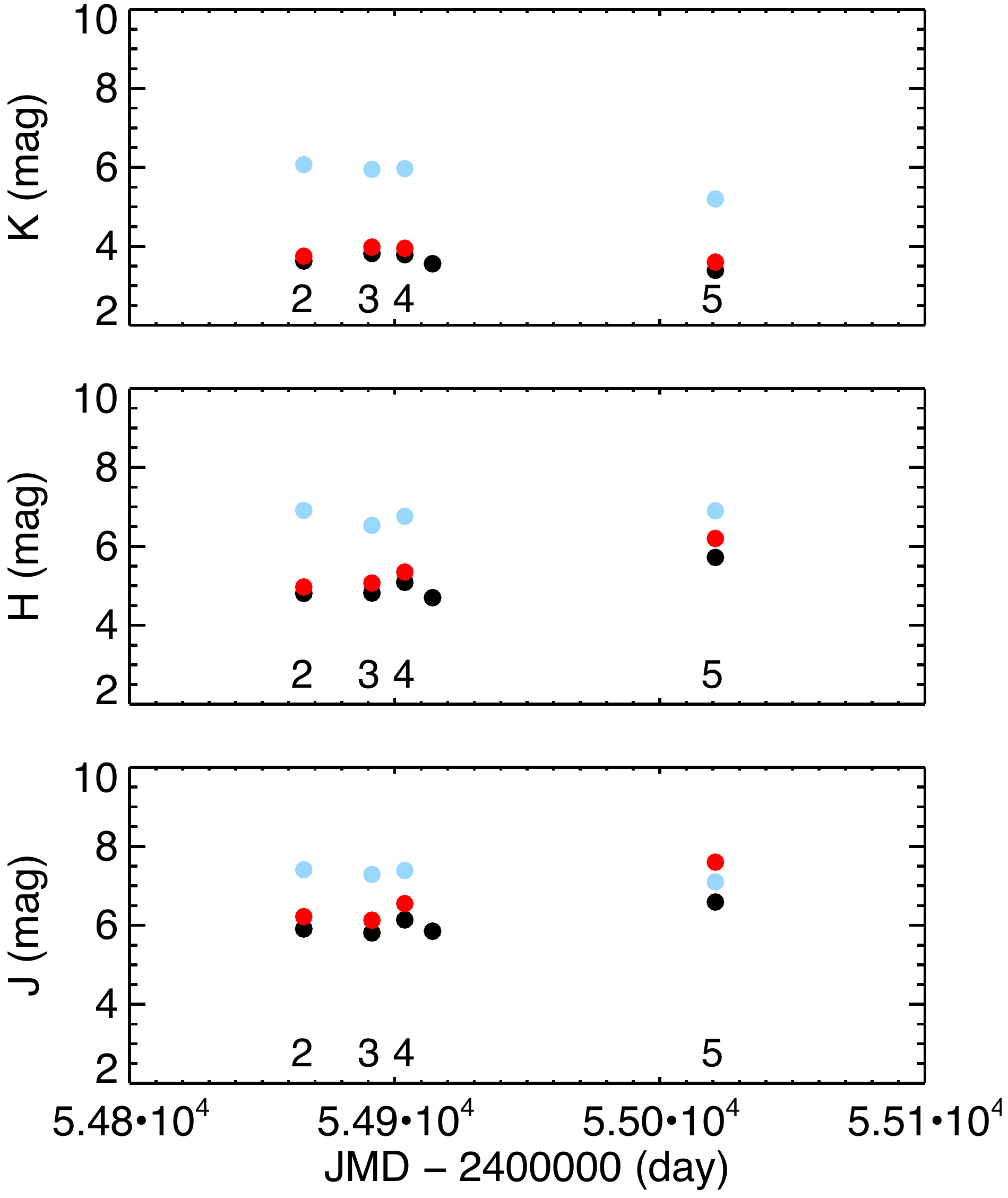}
\caption{Resolved Naco and Keck JHK photometry of FUOR component relative to the HBe compared to previous measurements (unresolved components) from the literature \citep[VDA4,][and ref.  therein]{1991AJ....102.2073K, 1993A&A...269..282H}. The epochs with simultaneous JHK-band photometry are: [1] October 1990 (quiescence);  [2-3-4], January 31, February 26, and March 11, 2009 respectively (outburst); [5] December 2009 (returning to quiescence).}
\label{fig:zcmaphot}
\end{figure*}

\begin{table*}[t]
\caption{NaCo and Keck relative astrometry and photometry of FUOR relative to the HBe star in $JHK_s$}             % title of Table
\label{tab:relastrophot}
\centering
\renewcommand{\footnoterule}{}  % to avoid a line before footnotes
\begin{tabular*}{\textwidth}{@{\excs}lllllllllll}     % 7 columns
\hline\hline\noalign{\smallskip}
UT Date     & $\Delta$          & PA            & platescale     & true North    & $\Delta J$          & $\Delta H$          & $\Delta K_s$          & $\Delta L'$          \\
           & (mas)             & (deg)         & (mas)          & (deg)         & (mag)               & (mag)               & (mag)               & (mag)                  \\   
\noalign{\smallskip}\hline\noalign{\smallskip}
09-01-31    & $111\pm3$         & $132.7\pm1.0$ &                &               & $1.19\pm0.13$       & $1.94\pm0.10$       & $2.42\pm0.11$       & $2.9\pm0.3$           \\
09-02-26    & $111\pm3$         & $132.4\pm1.0$ & $13.25\pm0.04$ & $0.10\pm0.15$ & $1.16\pm0.04$       & $1.46\pm0.04$       & $1.97\pm0.05$       &                       \\
09-03-11    & $112\pm3$         & $132.1\pm1.0$ &                &               & $0.82\pm05$         & $1.41\pm0.05$       & $2.01\pm0.04$       &                      \\
09-12-07    & $111\pm3$         & $133.3\pm1.0$ & 9.963$\pm$0.005& 0.13$\pm$0.02 & $-0.47\pm0.20$      & $0.64\pm0.05$       & $1.51\pm0.05$       &                      \\
\noalign{\smallskip}\hline
\end{tabular*}
\footnotetext[1]{Seeing evaluated at 0.55 $\mu$m.}
\end{table*}

\begin{table*}[t]
\caption{NaCo and Keck photometry of HBe, FUOR, and of the system in $JHK_s$}             % title of Table
\label{tab:phot}
\centering
\renewcommand{\footnoterule}{}  % to avoid a line before footnotes
\begin{tabular*}{\textwidth}{@{\excs}lllllllllll}     % 7 columns
\hline\hline\noalign{\smallskip}
UT Date     & $\mathrm{J_{HBe}}$          & $\mathrm{H_{HBe}}$  & $\mathrm{Ks_{HBe}}$  & $\mathrm{J_{FUOR}}$          & $\mathrm{H_{FUOR}}$  & $\mathrm{Ks_{FUOR}}$ & $\mathrm{J_{syst}}$          & $\mathrm{H_{syst}}$  & $\mathrm{Ks_{syst}}$ \\
\noalign{\smallskip}\hline\noalign{\smallskip}
09-01-31    & 6.22$\pm$0.08	& 4.97$\pm$0.08	& 3.75 $\pm$ 0.09 &	7.41$\pm$0.08 &	6.91$\pm$0.08 &	6.07$\pm$0.09 &	5.91$\pm$0.07 &  4.81$\pm$0.07 & 3.63$\pm$0.07 \\
09-02-26    & 6.13$\pm$0.08 	& 5.07$\pm$0.08 &	3.98 $\pm$ 0.09 &	7.29$\pm$0.08 &	6.53$\pm$0.08 & 5.95$\pm$0.09 & 5.81$\pm$0.07 &	4.82$\pm$0.07 &	3.82$\pm$0.07   \\
09-03-11    & 6.55$\pm$0.09 	& 5.35$\pm$0.10 &	3.95 $\pm$ 0.07 &	7.39$\pm$0.09 &	6.76$\pm$0.10 & 5.97$\pm$0.07 & 6.14$\pm$0.07 & 5.09$\pm$0.07 &	3.79$\pm$0.07   \\
09-12-07    & 7.6$\pm$0.2     &  6.2$\pm$0.2   &  3.6 $\pm$0.2  &   7.1$\pm$0.2 &  6.9$\pm$0.2 & 5.2$\pm$0.2  & 6.59$\pm$0.2   & 5.72$\pm$0.2    &  3.39$\pm$0.2    \\
\noalign{\smallskip}\hline
\end{tabular*}
\end{table*}

%----------------------------------
\subsection{High-contrast imaging}

After cosmetic reductions (bad pixels, dark subtraction, flat fielding) using
\textit{eclipse} \citep{1997Msngr..87...19D}, we applied the deconvolution
algorithm of \cite{1998SPIE.3353..426V} to obtain the position of the
HBe component relative to the FUOR at each epoch for both NaCo and
Keck data. The star HD54335 ($V=7.4$, $K=3.6$) and HIP33998 ($V=7.99$,
$K=7.895$) of similar brightness were observed consecutively to serve
as point spread function (PSF) estimate for the NaCo and Keck data,
respectively. The results are reported in
Table~\ref{tab:relastrophot}.  The total system brightness was calibrated using  the zero points provided by ESO for each observing night, or using the PSF standard HIP33998 observed with NIRC2.

%----------------------------------
\subsection{Integral field spectroscopy}
\label{subsec:IFS}
SINFONI data were reduced with the ESO data reduction pipeline version
1.9.8 \citep{2006NewAR..50..398A}. We used in addition custom routines to
correct raw images from several electronic effects that affect the
detector, as previously reported in \cite{2014A&A...562A.127B}. The pipeline
carried out cube reconstruction from corrected detector images. Hot
and nonlinear pixels were tagged using dark and flat-field exposures.
Arc lamp frames acquired during the days following the observations
enabled us to calibrate the distortion, the wavelength scale, and the
slitlet positions. Slitlet distances were measured with north-south
scanning of the detector illuminated with an optical fiber. Object-sky
frame pairs were subtracted, flat-fielded and corrected for bad pixels
and distortions. In each band, four datacubes were finally reconstructed
from clean science images and merged into a master cube. The quality
of the reduction was checked using the ESO trending parameters.  We
obtained standard star datacubes in a similar way.  However, their  low signal to noise ratio (S/N) would have
degraded the final Z CMa spectra, therefore we decided not to use it. We instead reduced datacubes of the
telluric standards HIP072241 (B3V) in the J band, HIP054970 (B5III) in
the H band, and of HIP HIP071218 (G2V) in the K band observed at the
airmass of Z CMa on different nights and with the instrument pre-optics offering a 0.25 arcsecond/spaxel sampling. The standard star spectra and the Z CMA spectra were  smoothed to the same resolution. This change of resolution was estimated from the arc lamp calibration files taken the same days as the observations. Those telluric standards were found to provide the best  possible removal of the telluric lines. 

Although Z CMa components are resolved in our final datacubes, the
 two sources contaminate each other. We therefore
developed an improved version of the spectral extraction algorithm
 described in \cite{2009A&A...506..799B} to retrieve deblended spectra of
the HBe and FUOR stars in each band (see Fig. 2). Our algorithm, \texttt{CLEAN 3D}  has now been applied successfully on different datasets \citep{2014A&A...562A.127B, 2016MNRAS.456.2576B}. We provide a reference concise description in this paper.  The positions of the two sources were initially fit with a Moffat function in
each monochromatic cube slice. Their variation with wavelength caused
by the atmospheric refraction was fit with a low-order
polynomial. A modified version of the CLEAN algorithm \citep{2001AJ....121.1163D} was then applied on each slice to create a model of Z CMa
while keeping the position of the binary fixed. The standard star
datacubes were not reprentative of the PSF of Z CMa, which was affected
by a strong astigmatism and could therefore not be used as input
of the algorithm. We therefore chose to create a PSF template duplicating  the Herbig profile following the observed direction of
the PSF lengthening.  The position of the sources were then refined,
an improved PSF model was created and a second CLEAN step was
performed on the original cube. The algorithm produced two final
datacubes containing the model of the HBe and the FUOR
components. The typical error introduced by the spectral extraction
and computed from residual maps ranges from 0.5 to 5\%. Z CMa spectra were finally integrated inside circular appertures of
390 mas in the J band and 1250 mas in the H and K band. Spectra were
divided by the standard star spectra corrected for intrinsic features
and from their black body. The standard star spectra are found to
provide a good correction of telluric features in the J band and a
moderate correction in the H and K band. Spectra were finally scaled
in flux by interpolating the NACO photometry of January 31, 2009 and February 26, 2009 on the date of the SINFONI observations. We estimate that the S/N measured on line-free regions on the Herbig spectra is 205 in the J band, 122 in the H band, and 120 in the K band. \\

We repeated this procedure on the  $Jbb$ (1.18--1.416 $\mu$m) and $Hbb$ (1.473--1.803 $\mu$m) Keck/OSIRIS datacubes obtained on December 22, 2009  \citep{2010ApJ...720L.119W} when the system had returned to its quiescent state. We also reduced and analyzed an additional datacube obtained on the same night with the  $Kbb$ (1.965--2.381 $\mu$m) setting. The cubes have $20\times20$ mas spaxels that sample the system separation well.  The limited FoV of the cubes along the binary PA and residual artifacts limited the accuracy of the extraction to $\sim$10 \%. We obtained the spectra of each component by integrating the flux of the sources over circular apertures (R=5 pixels) in the extracted datacubes.  The spectra were corrected for the atmospheric transmission using the spectra of A0V stars observed on the same night (HD 67213 for the $Jbb$ and $Kbb$ datacubes, HD 109615 for the $Kbb$ datacubes). The  standard star spectra were previously multiplied by a  high-resolution  (R=500000) spectrum of  Vega smoothed at the resolution of OSIRIS (R$\sim$3800) and shifted to the radial velocity of the stars\footnote{\textit{http://kurucz.harvard.edu/stars/VEGA/}}. The standard star was located at the edge of the FoV of the H-band cube, which introduced strong differential flux losses that translate into an additional slope in the final spectra. We estimated and corrected the spectra for this slope considering that the FUOR spectral slope  should be identical to the one of the SINFONI spectrum. The OSIRIS spectra have a lower S/N than those of the SINFONI spectra (S/N=31 in the J band, 82 in the H band, and 70 in the K band).  Finally, we note that the K-band spectra are also slightly affected by an oscillating pattern  (with a 0.08 $\mu$m period)  that might be related to numerical artifacts found on the spaxels close to the edges of the FoV.

%------------------------------------------------------------------------
%------------------------------------------------------------------------
\section{AO imaging of Z CMa}
\label{section:AOimaging}
%----------------------------------
\subsection{Resolved photometry}
\label{subsection:resphot}
The photometry of the FUOR component relative the HBe at
each epoch for both NaCo and Keck data is reported in Table \ref{tab:phot} and compared in
Fig.~\ref{fig:zcmaphot} with previous total and individual photometry. From the literature, the FUOR is brighter
in the quiescent phase in J and H bands, whereas the HBe dominates in
K band. The system brightness also seems  to decay significantly
since 1986, and presumably since the last FUOR burst. During the 2008
outburst, our NaCo data show that the HBe component became much
brighter in all  NIR  wavelengths than the FUOR, as shown by a
flux ratio inversion in J and H bands (see
Table~\ref{tab:relastrophot}). This confirms  that
the HBe component is at the origin of the 2008 outburst, in agreement with \cite{2013ApJ...763L...9H}. Our Keck
observations, taken one year later, indicate that the system did not
completely return to its quiescent phase in December 2009 as the HBe
component remains brighter in H-band. 

%----------------------------------
\subsection{Extended structures}
\label{section:extstruc}

\begin{figure}[b]
\centering
\vspace{-0.1cm}
\includegraphics[width = 7cm]{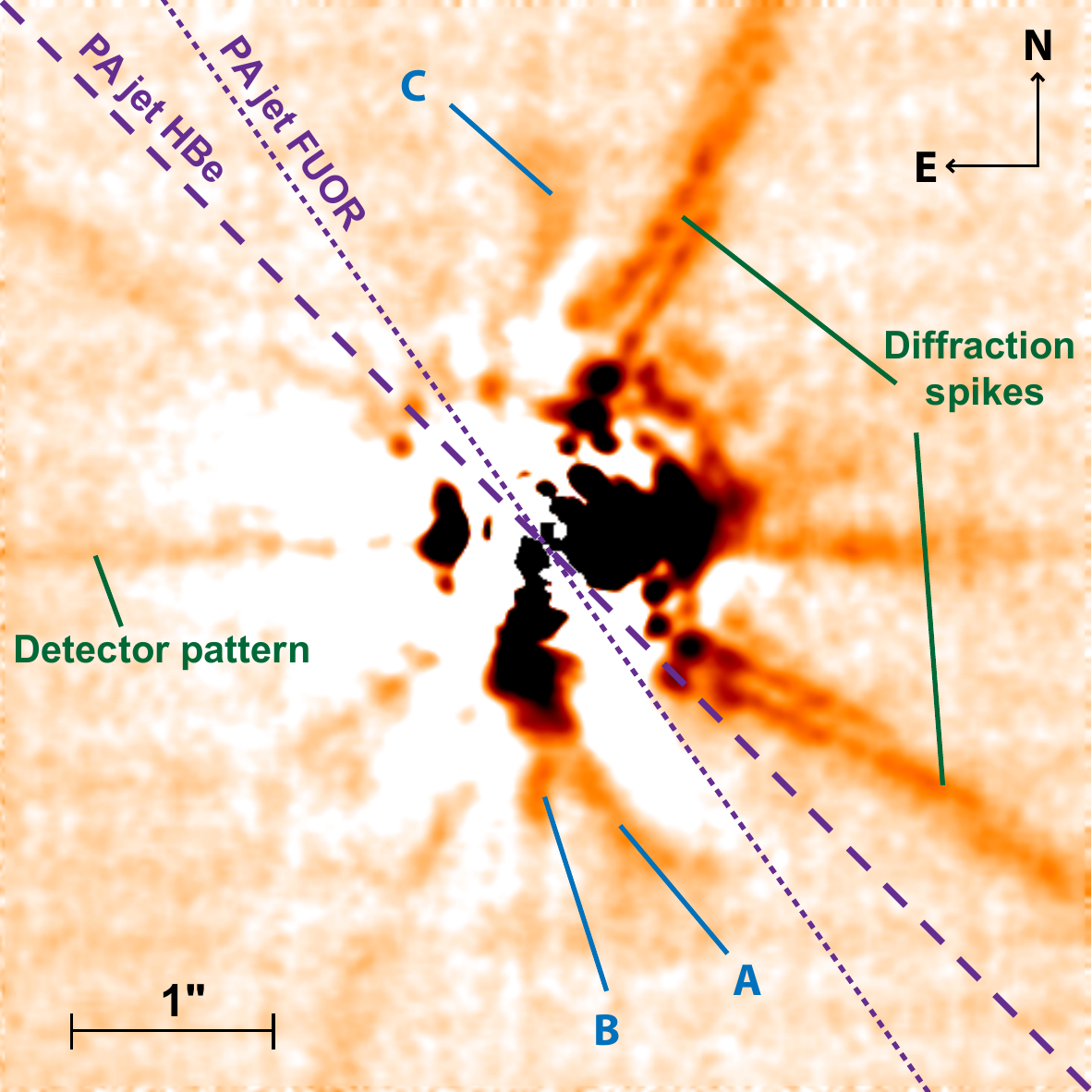}
\caption{VLT/NACO PSF-subtracted $L'$-band image of Z CMa obtained on January 31,
  2009. See Sect. \ref{section:extstruc} for details.}
\label{fig:zcmalp}
\end{figure}

 In our 2009 NaCo
observation in J or H bands, we did not redetect any extended structures
 because the Keck data have comparatively a higher S/N,
as we did not intend to saturate our observations to derive the
photometry of the two components. An extended structure at a projection angle (PA) compatible with that reported by \cite{2002ApJ...580L.167M}, \cite{2015A&A...578L...1C}, and \cite{2016SciA....200875L} is  clearly seen in the  $\mathrm{L\!'}$ band (annoted ``A" in Fig.~\ref{fig:zcmalp}) after applying a radial profile subtraction.  Its PA ranges from 160 to 250$^{\circ}$ and extends up to 2160 au. Two other structures (``B" and ``C" in Fig.~\ref{fig:zcmalp}) are marginally detected in this band at PA$\sim$0 and 180$^{\circ}$. They are not detected at shorter NIR wavelengths or in polarimetry \citep{2015A&A...578L...1C, 2016SciA....200875L}. Their orientation does not coincide with  the shocked emission regions seen in the wide-field images of Z CMa \citep[][Bouy et al. 2011, unpublished]{1989A&A...224L..13P}. They could be part of the cavity  whose NW extension displays a complex structure \citep{1989A&A...224L..13P} or be residuals from the stellar halo subtraction. 
%\cite{2002ApJ...580L.167M} resolved an extended  and curved
%structure in their J-band Keck images at the South-West of Z CMa. They
%suggested that its origin could arose from light scattered off the
%walls of a jet-blown cavity or from a new jet. \cite{2010ApJ...720L.119W} evidenced twin microjets
%arizing the HBe and FUOR components at position angle of about $235^{\circ}$ and $245^{\circ}$ respectively. 

%----------------------------------
\subsection{Astrometry and orbital motion}

\begin{figure}[t]
\centering
\vspace{-0.1cm}
\includegraphics[width = \columnwidth]{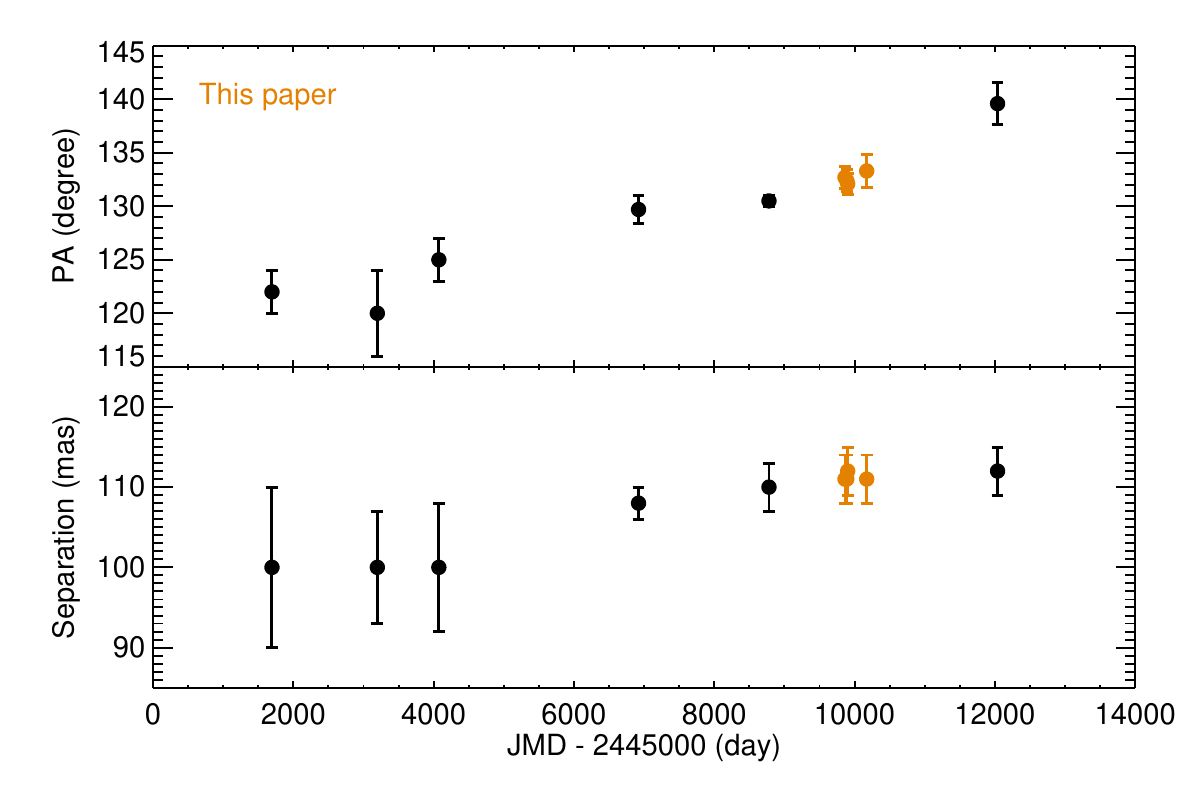}
\caption{Position in separation and position angle of the FUOR component relative to the HBe, based on the literature and our NaCo and Keck observations.}
\label{fig:astro}
\end{figure}

The astrometry of the FUOR relative to the HBe was measured
at each epoch for our NaCo and Keck data. The results are reported in
Table~\ref{tab:relastrophot} and shown in Fig.~\ref{fig:astro}. They complement  older epochs obtained using AO and speckle \citep{2001ApJ...546..358M, 1991AJ....102.2073K, 1993A&A...269..282H, 1994A&A...291..500B, 1995A&A...303..795T} and  more recent measurements obtained with the polarimetric imaging mode of NaCo and SPHERE \citep[][Antoniucci et al. 2016, in prep.]{2015A&A...578L...1C}. While the separation does not seem to vary significantly,
our  astrometric measurements confirm a significant variation of
position angle since 1986. This might be consistent with a coplanar configuration between the binary
and the HBe and FUOR disks \citep{1993A&A...271L...9M, 2010A&A...517L...3B}
as often expected in binary systems, although the jet orientations
differ by at least 10 degrees between the two components \citep{2010ApJ...720L.119W}.

%For visual binaries, the semimajor axis can statistically be estimated using the observed separation \citep{1960JO.....43...41C}. Considering the position of the FUOR relative to the HBe, we find a semi-major axis of 135~au. Assuming masses of 16 and 3~M$_{\odot}$ for the HBe and the FUOR, respectively, we obtain a range of orbital periods between 200-400~yrs, compatible with the orbital motion observed of a few mas/yr.

%------------------------------------------------------------------------
%------------------------------------------------------------------------
\section{Spectro-imagery of the components of Z CMa}
\label{section:IFU}

\begin{figure*}[t]
\centering
\vspace{-0.1cm}
\includegraphics[height=10.5cm]{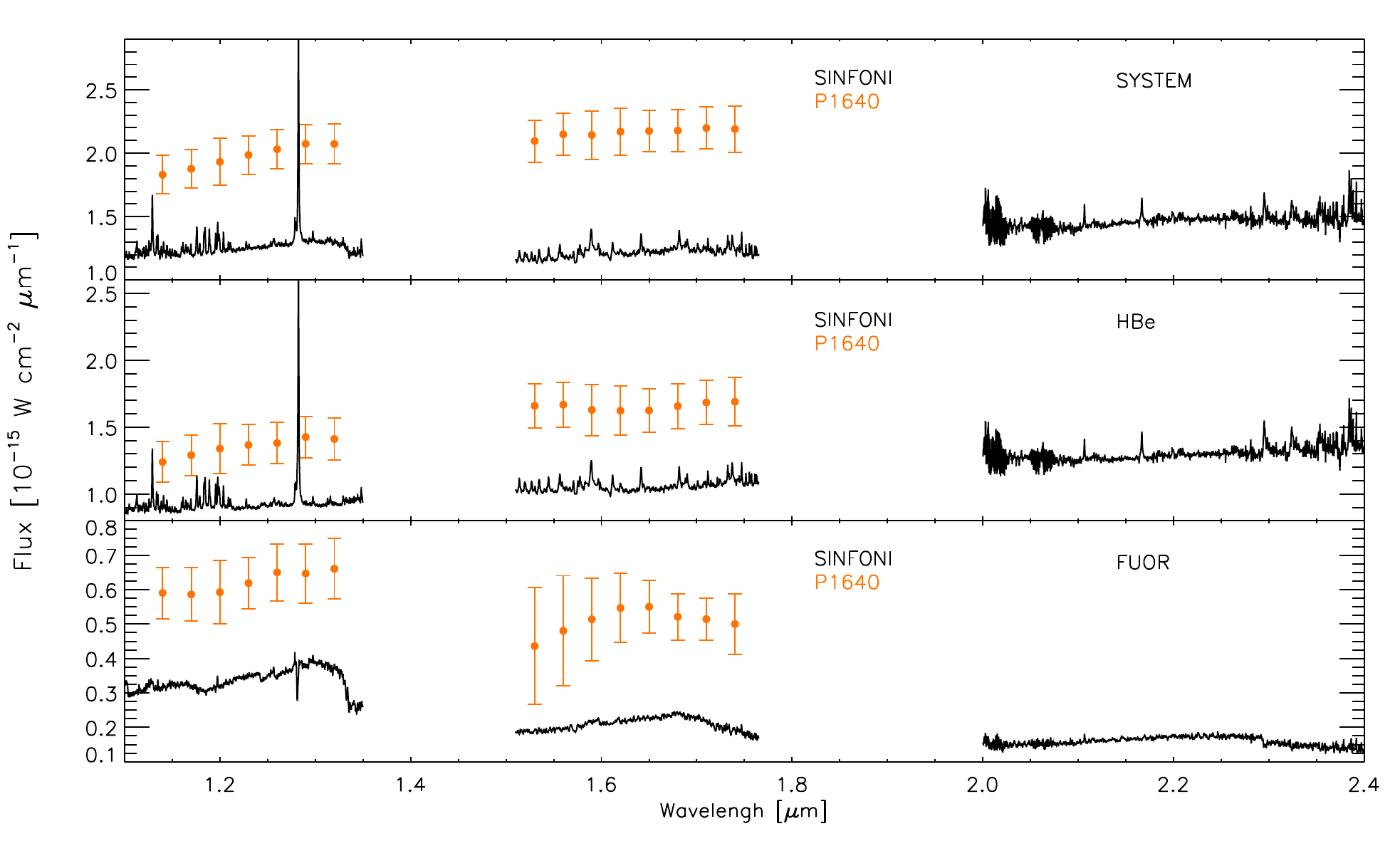}
\caption{Flux-calibrated SINFONI spectra (from February 6, 2009) of the system (top), the HBe component (middle), and of the FUOR (bottom) in the 1.1-2.4 $\mu$m range while the binary was in outburst. The lower resolution (R$\sim$45) P1640 spectrum of the system components also acquired during the outburst phase \citep[March 17, 2009,][]{2013ApJ...763L...9H} is shown in orange.}
\label{fig:zcmaJHK}
\end{figure*}

\begin{figure}[t]
\centering
\vspace{-0.1cm}
\includegraphics[width = \columnwidth]{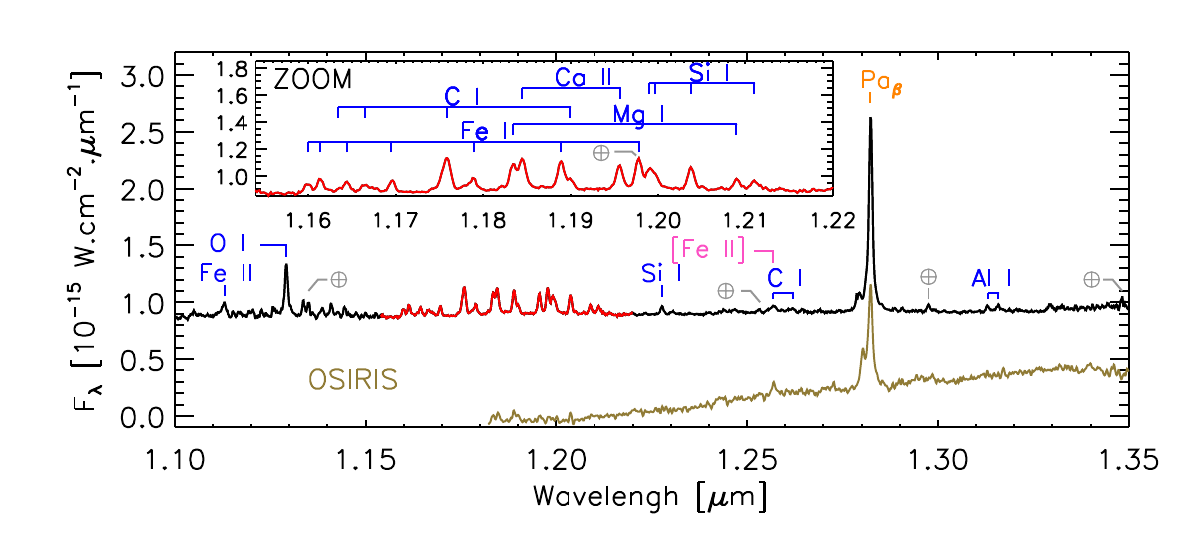}
\includegraphics[width = \columnwidth]{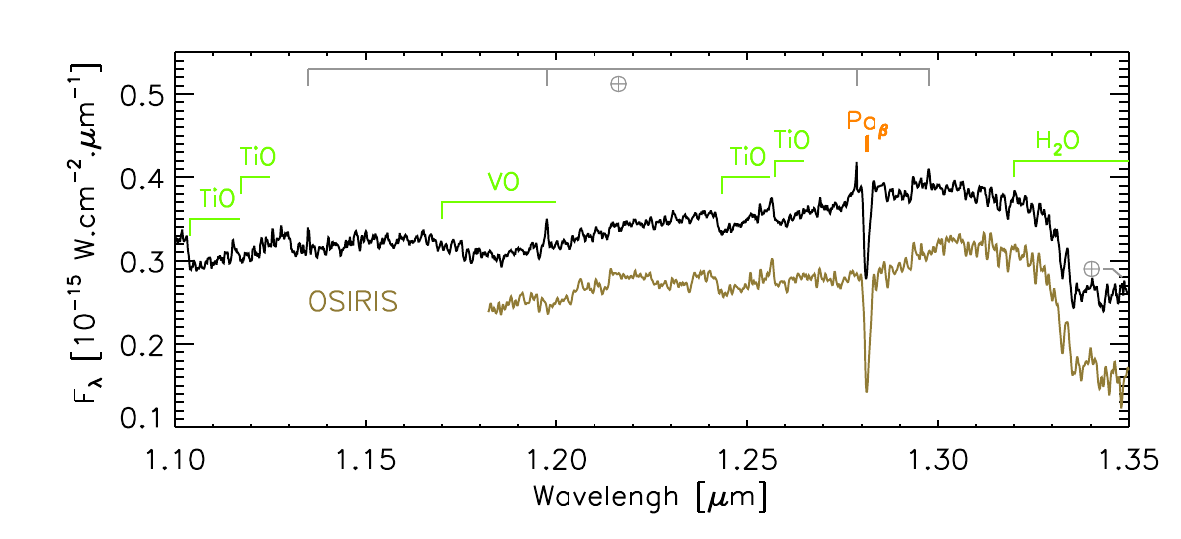}
\caption{J-band spectra of the Z CMa HBe (\textit{top}) and FUOR (\textit{bottom}) components during the outburst phase (black and red line for the zoom; SINFONI spectrum from February 6, 2009) and in the quiescent stage (golden line; OSIRIS spectrum from December 22, 2009). The fluxes of the OSIRIS spectra are normalized to the median flux of the SINFONI spectra between 1.206 and 1.246 $\mu$m and shifted to lower values for clarity.}
\label{fig:spec_Jband}
\end{figure}

\begin{figure}[t]
\centering
\vspace{-0.1cm}
\includegraphics[width = \columnwidth]{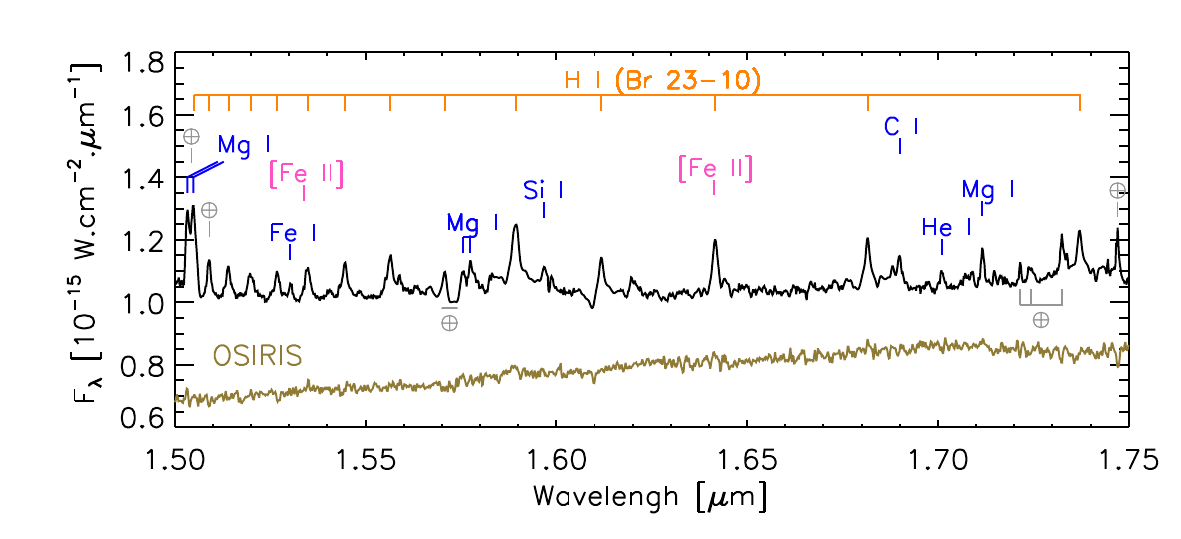}\\
\includegraphics[width = \columnwidth]{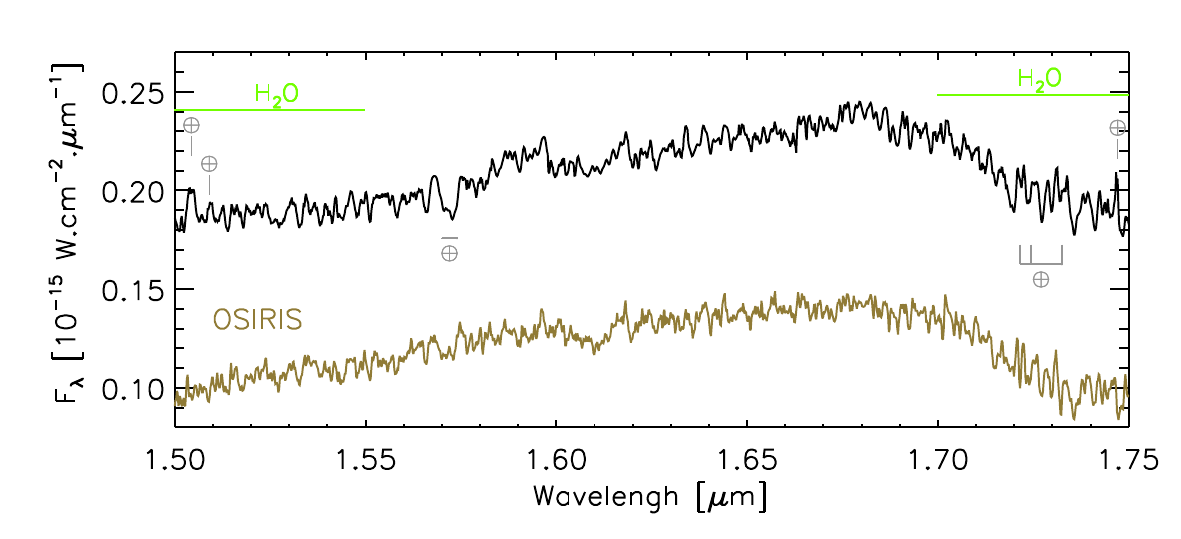}\\
\caption{H-band spectra of the Z CMa HBe (\textit{top}) and FUOR (\textit{bottom}) components during the outburst phase (black line; SINFONI spectrum from February 6, 2009) and in the quiescent stage (golden line; OSIRIS spectrum  from December 22, 2009). The fluxes of the OSIRIS spectra are normalized to the median flux of the SINFONI spectra between 1.66 and 1.70 $\mu$m and shifted to lower values for clarity.}
\label{fig:spec_Hband}
\end{figure}

\begin{figure}[t]
\centering
\vspace{-0.1cm}
\includegraphics[width = \columnwidth]{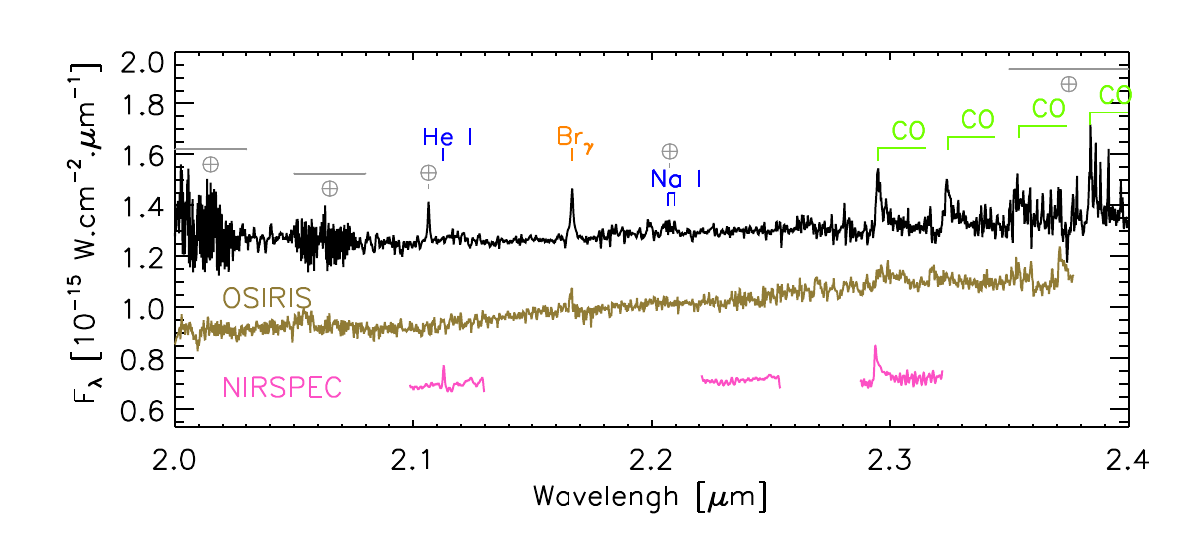}
\includegraphics[width = \columnwidth]{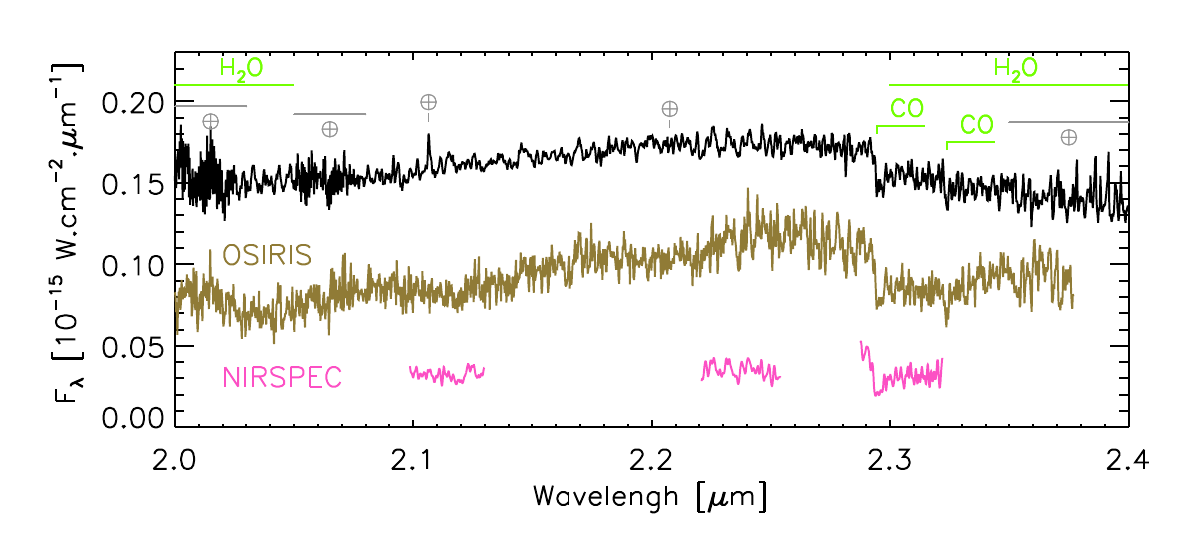}
\caption{Same as Fig. \ref{fig:spec_Hband} but for the K band.  The normalized NIRSPEC spectra (with respect to the continuum of the SINFONI spectra) of the system components obtained while the binary was in a quiescent state (December 17, 2006) are overlaid in pink. The fluxes of the OSIRIS spectra are normalized to the median flux of the SINFONI spectra between 2.18 and 2.22 $\mu$m and shifted to lower values for clarity.} 
\label{fig:spec_Kband}
\end{figure}

The flux-calibrated SINFONI spectra are shown in the 1.1-2.4 $\mu$m range in Fig~\ref{fig:zcmaJHK}. The OSIRIS and SINFONI spectra of the FUOR and of the HBe are compared in each individual band in Figs.~\ref{fig:spec_Jband}, \ref{fig:spec_Hband}, and \ref{fig:spec_Kband}. 

%----------------------------------
\subsection{Near-infrared spectrum of the HBe component}
The 1.1-2.5 $\mu$m spectrum of the HBe component is dominated by  a complex set of emission lines  overlaid on a  continuum with a red slope. These lines are decreased or absent in the OSIRIS spectrum obtained during the quiescent phase. The NIR color-variation taken close in time with the OSIRIS spectra indicates that the continuum became  redder when the system returned to quiescence. This is consistent with the slope variation of the continuum in the J-, H-, and K-band spectra, although there are some uncertainties on the calibration of this slope in the OSIRIS data (see Sect. \ref{subsec:LineIDFUOR}).  The position of the emission lines emerging at more than 5\% of the continuum flux are reported in Table~\ref{tab:specid}. They were retrieved using an interactive least-squares multiple Gaussian fitting tool from the FUSE IDL library\footnote{The tool can be downloaded at \textit{http//fuse.pha.jhu.edu/analysis/fuse\_idl\_tools.html}} with an accuracy of $\sim$1 \AA~(12-28 km.s$\mathrm{^{-1}}$). The velocity of each identified line, their corresponding transition,  the associated non-dereddened fluxes, full-width at half maximum (FWHM), and equivalent widths (EW) for unblended lines complete Table~\ref{tab:specid}. The velocities were corrected for the radial velocity of the system \citep[+30 km.s$\mathrm{^{-1}}$,][]{1989ApJ...338.1001H} and for the barycentric-to-helocientric velocity shift at the time of the observations (-11.2 $km.s^{-1}$). The fluxes were estimated following the estimation and the removal of the continuum beneath the line using a  low-order Legendre polynomial. The associated error bar was estimated from the RMS of the noise inside the continuum estimation zones surrounding the lines. FWHM are taken from the Gaussian fit. Finally, equivalent widths (EW) and their associated error bars are computed following the \cite{1992ApJS...83..147S} method.

\subsubsection{Line identification}

\begin{table*}[t]
\caption{Line identification in the $1.1-2.5~\mu m$ spectrum of the  HBe components}             % title of Table
\label{tab:specid}
\centering
\renewcommand{\footnoterule}{}  % to avoid a line before footnotes
\begin{tabular*}{\textwidth}{llllllll}     % 7 columns
\hline\hline\noalign{\smallskip}
$\lambda_{obs}$	& Element									&		Transition															& $\mathrm{\Delta V}$ 		&	FWHM	&	Flux&		EW  	&	Remarks	 \\
($\mu$m)   		 	&               					    &                                     		  									&  ($\mathrm{km.s^{-1}}$)	&  (\AA)	&  ($\mathrm{10^{-12}erg.s^{-1}cm^{-2}}$)   &		(\AA)  &   \\
\hline
1.11306				&	Fe II								&	$4p^{4}\!F_{3/2}^{\circ}-4s^{2}\:^{4}\!G_{5/2}$			&	42								&	13.6		&	$1.32\pm0.32$		&	$-1.48\pm0.12$		&	CP, Blend with Fe I			\\
1.12920				&	O I								&	$3p^{3}\!P_{2}-3d^{3}\!D^{\circ}_{3}$						&	42								&	10.8		&	$4.60\pm0.39$		&	$-5.12\pm0.16$		&	CP			\\
1.13370				&	C I	?								&	$3p^{1}\!P_{1}-3d^{1}\!D^{\circ}_{2}$						&	71								&	$\dots$&		$\dots$ 	&	$\dots$&	SB, or TR?			\\	
1.13502				&	TR								&	$\dots$																		&		$\dots$						& $\dots$	& $\dots$&	$\dots$&		\\			
1.13861				&	Na I								&	$3p^{2}\!P^{\circ}_{1/2}-4s^{2}\!S_{1/2}$					&	28									&	8.2		&	$0.49\pm0.21$		&	$-0.53\pm0.09$		& IP \\	
1.14100				&	Na I								&	$3p^{2}\!P^{\circ}_{3/2}-4s^{2}\!S_{1/2}$					&	68								&	7.2		&	$0.65\pm0.19$		&	$-0.72\pm0.12$		&	\\	
1.14451				&	Fe I								&	$b^{3}\!P_{1}-z^{3}\!D^{\circ}_{2}$							&	62								&	$\dots$	&	$\dots$&	$\dots$&	SB, IP \\	
1.15996				&	Fe I								&	$a^{5}\!P_{1}-z^{5}\!D^{\circ}_{0}$								&	63								&	8.9		&	$0.63\pm0.07$	&	$-0.73\pm0.04$		&	FB	\\
1.16138				&	Fe I								&	$a^{5}\!P_{2}-z^{5}\!D^{\circ}_{2}$								&	68								&	7.9		&	$1.13\pm0.07$	&	$-1.29\pm0.04$		&	FB	\\		
1.16348				&	C I									&	$3p^{3}\!D_{2}-3d^{3}\!D^{\circ}_{2}$						&	61								&	$\dots$	&	$\dots$&	$\dots$&	SB			\\
1.16442				&	Fe I								&	$a^{5}\!P_{3}-z^{5}\!D^{\circ}_{3}$								&	59								&	9.3		&	$0.99\pm0.08$		&	$-1.12\pm0.04$		&	FB			\\
1.16643				&	C I									&	$3p^{3}\!S_{1}-3d^{3}\!P^{\circ}_{0}$							&	47								&	$\dots$	&	$\dots$&	$\dots$&	SB			\\
1.16959				&	Fe I								&	$a^{5}\!P_{1}-z^{5}\!D^{\circ}_{1}$								&	59								&	7.8		&	$0.78\pm0.07$		&	$-0.88\pm0.04$		&	Blended with 		\\
							&										&																					&										&				&			&			& K I line?	\\	
1.17586				&	C I									&	$3p^{3}\!D_{3}-3d^{3}\!F^{\circ}_{4}$						&	41									&	11.2			&		$3.23\pm0.08$	&	$-3.65\pm0.03$		&	Blend of  \\
							&										&																					&										&				&			&			& C I transitions \\
1.17891				&	Fe I								&	$b^{3}\!P_{2}-z^{3}\!D^{\circ}_{3}$							&	54								&	7.8		&	 $1.37\pm0.08$		&	$-1.54\pm0.03$		&	$\dots$				\\
1.18340				&	Mg I								&	$3p^{1}\!P^{\circ}_{1}-4s^{1}\!S_{0}$							&	54								&	$\dots$	&	$\dots$&	$\dots$&	SB				\\
1.18440				&	Ca II								&	$5s^{2}\!S_{1/2}-5p^{2}\!P^{\circ}_{3/2}$					&	58								&	$\dots$	&	$\dots$&	$\dots$&	SB				\\
1.18893				&	Fe I								&	$a^{5}\!P_{2}-z^{5}\!D^{\circ}_{3}$								&	69								&	$\dots$	&	$\dots$&	$\dots$&	SB with 	\\	
							&										&																					&										&				&			&			& Fe I and C I lines? \\	
1.19001				&	C I									&	$3p^{3}\!D_{3}-3d^{3}\!F^{\circ}_{4}$						&	88								&	$\dots$	&	$\dots$&	$\dots$& SB, Blend of  \\
							&										&																					&										&				&			&			&C I transitions \\
1.19560				&	Ca II								&	$5s^{2}\!S_{1/2}-5p^{2}\!P^{\circ}_{1/2}$					&	63								&	9.9		&	$1.69\pm0.08$		&	$-1.87\pm0.04$		&	FB			\\	
1.19745				&	TR								&	$\dots$																		&	$\dots$							& $\dots$	& $\dots$&	$\dots$&	$\dots$		\\			
1.19780				&	Fe I								&	$b^{3}\!P_{3}-z^{3}\!D^{\circ}_{4}$							&	30								&	8.5		&	$\dots$&	$\dots$&	IP, blended \\
							&										&																					&										&				&			&			&with TR \\
1.19897				&	Si I								&	$4s^{3}\!P^{\circ}_{1}-4p^{3}\!D_{2}$							&	44								& $\dots$	&	$\dots$&	$\dots$&	SB, IP				\\
1.19972				&	Si I								&	$4s^{3}\!P^{\circ}_{0}-4p^{3}\!D_{1}$							&	46								& $\dots$	&	$\dots$&	$\dots$&	SB, IP				\\	
1.20377				&	Si I								&	$4s^{3}\!P^{\circ}_{2}-4p^{3}\!D_{3}$							&	61								&	9			&	$1.46\pm0.12$		&	$-1.62\pm0.05$		&	$\dots$			\\
1.20902				&	Mg I								&	$3d^{1}\!D_{2}-4f^{1}\!F^{\circ}_{3}$							&	69								&	9			&	$0.75\pm0.12$		&	$-0.84\pm0.05$		&	$\dots$			\\
1.21105				&	Si I								&	$4s^{3}\!P^{\circ}_{1}-4p^{3}\!D_{1}$							&	80								&	12.2		&		$0.78\pm0.13$	&	$-0.87\pm0.05$		&	$\dots$			\\
1.22768				&	Si I								&	$4s^{3}\!P^{\circ}_{2}-4p^{3}\!D_{2}$							&	56								&	10.4		&	$0.71\pm0.12$		&	$-0.79\pm0.05$		&	CP, blended ?		 \\
1.24399				&	UF									&	$\dots$																		&	$\dots$								&	$\dots$			&	$\dots$&	$\dots$&	SB 	\\					
1.24665				&	UF									&	$\dots$																		&	$\dots$							&	$\dots$		&	$\dots$&	$\dots$&	SB	\\
1.25681				&	C I									&	$3p^{3}\!P_{1}-3d^{3}\!P^{\circ}_{0}$							&	50								&	$\dots$	&	$\dots$&	$\dots$&	SB				\\
1.25681				&	[Fe II]								&	$a^{6}\!D_{9/2}-a^{4}\!D_{7/2}$								&	-62								&	$\dots$	&	$\dots$&	$\dots$&	CP, SB		\\
1.26200				&	C I									&	$3p^{3}\!P_{2}-3d^{3}\!P^{\circ}_{2}$							&	47								&	$\dots$	&	$\dots$&	$\dots$&	IP				\\
1.27870				&	TR								&	$\dots$																	&		$\dots$						& $\dots$	& $\dots$&	$\dots$&	Weak TR		\\			
1.28239				&	H I (Pa$\mathrm{\beta}$)				&	$3-5$													&	42								&	8.7		&	$28.17\pm0.68$		&	$-30.53\pm0.14$&	Component 			\\
							&										&																					&										&				&			&			& at -649	km.s$^{-1}$ \\
1.29754				&	TR								&	$\dots$																		&	$\dots$							& $\dots$	&  $\dots$&	$\dots$&	$\dots$\\	
1.31305				&	Al I								&	$^{2}\!S_{1/2}-^{2}\!P^{\circ}_{3/2}$							&	67								&	8.2		&	$0.45\pm0.13$		&	$-0.49\pm0.05$		&	$\dots$	\\
1.31583				&	Al I								&	$^{2}\!S_{1/2}-^{2}\!P^{\circ}_{3/2}$							&	78								&	6.6		&	$0.53\pm0.08$		&	$-0.58\pm0.04$		&	$\dots$	\\
1.32941				&	Fe I ?								&	$ b^{3}\!G-z^{3}\!F^{\odot}$										&	47							&	10.8		& $0.66\pm0.26$&$-0.71\pm0.10$	&	$\dots$	\\	
1.34830				&	TR								&	$\dots$																		&	$\dots$							&	$\dots$	&	$\dots$&	$\dots$&	$\dots$	\\	
\hline
1.48863				&	Mg I								&	$3d^{3}\!D_{3}-4f^{3}\!F^{\circ}_{4}$							&	85								&	$\dots$	&	$\dots$&	$\dots$&	IB, IP		\\
1.50327				&	Mg I								&	$4s^{3}\!S_{1}-4p^{3}\!P^{\circ}_{2}$							&	60								&	$\dots$	&	$\dots$&	$\dots$&	SB			\\	
1.50484				&	Mg I								&	$4s^{3}\!S_{1}-4p^{3}\!P^{\circ}_{1}$							&	69								&	$\dots$	&	$\dots$&	$\dots$&	SB			\\	
1.50484				&	H I (Br 23)						&	$4-23$																		&	$\dots$							&	$\dots$	&	$\dots$&	$\dots$&	ND \\
1.50436				&	TR								&	$\dots$																		&	$\dots$							&	$\dots$	&	$\dots$&	$\dots$&	$\dots$	\\	
1.50898				&	H I (Br 22)						&	$4-22$																		&	47								&	9.9		&	$1.48\pm0.09$		&	$-1.45\pm0.04$		&	Overlaping TR \\
1.51405				&	H I (Br 21)						&	$4-21$																		&	51								&	14.6		&	$1.39\pm0.13$		&	$-1.36\pm0.04$		&	$\dots$			\\
1.51994				&	H I (Br 20)						&	$4-20$																		&	55								&	19.4		& $1.71\pm0.44$			&	$-1.42\pm0.09$		&		$\dots$		\\
1.52684				&	H I (Br 19)						&	$4-19$																		&	62								&	17.5		&	$1.13\pm0.14$		&	$-1.11\pm0.07$		&	FB			\\
1.53034				&	Fe I								&	$e^{7}\!D_{5}-n^{7}\!D^{\circ}_{5}$									&	80								&	$\dots$	&	$\dots$&	$\dots$&	SB			 \\
1.53495				&	[Fe II]								&	$a^{4}\!F_{9/2}-a^{4}\!D_{5/2}$										&	$\dots$							&	$\dots$	&	$\dots$&	$\dots$&	ND \\
\noalign{\smallskip}\hline
\end{tabular*}
\end{table*}

\begin{table*}[t]
\caption{Continuation of Table~\ref{tab:specid}.}             % title of Table
\label{tab:specidHBe2}
\centering
\renewcommand{\footnoterule}{}  % to avoid a line before footnotes
\begin{tabular*}{\textwidth}{@{\excs}llllllll}     % 7 columns
\hline\hline\noalign{\smallskip}
$\lambda_{obs}$	& Element									&		Transition															& $\mathrm{\Delta V}$ 		&	FWHM	&	Flux&		EW  	&	Remarks	 \\
($\mu$m)   		 	&               					    &                                     		  									&  ($\mathrm{km.s^{-1}}$)	&  (\AA)	&  ($\mathrm{10^{-12}erg.s^{-1}.cm^{-2}}$)   &		(\AA)  &   \\
\hline \noalign{\smallskip}		
1.53495				&	H I (Br 18)						&	$4-18$																		&	57								&	16.7		&		$1.91\pm0.20$	&	$-1.89\pm0.05$		&	$\dots$			 \\
1.54470				&	H I (Br 17)						&	$4-17$																		&	64								&	14		&	$1.85\pm0.19$		&	$-1.81\pm0.06$		&	$\dots$			 \\					
1.55640				&	H I (Br 16)						&	$4-16$																		&	52								&	11.5		&	$3.23\pm0.38$		&	$-3.15\pm0.10$		&	$\dots$			 \\
1.55825				&	${}^{12}\!\mathrm{CO}$?				&	3--0 of $X {}^1\!\Sigma^{+}-X {}^1\!\Sigma^{+}$		&	$\dots$							&	$\dots$ & $\dots$	&	$\dots$ &	SB	\\	1.57083				&	H I (Br 15)						&	$4-15$																		&	52								&	13.6		&	$\dots$&	$\dots$&	Overlaping TR \\
1.570-1.574		&	TR								&	$\dots$																		&	$\dots$							&	$\dots$	&	$\dots$&	$\dots$&	$\dots$	\\			
1.57558				&	Mg I								&	$4p^{3}\!P^{\circ}_{0}-4d^{3}\!D_{1}$								&	19								&	$\dots$	&	$\dots$&	$\dots$&	SB, IP		\\
1.57746				&	Mg I								&	$4p^{3}\!P^{\circ}_{2}-4d^{3}\!D_{3}$								&	73								&	$\dots$	&	$\dots$&	$\dots$&	SB, IP		\\			
1.57746				&	${}^{12}\!\mathrm{CO}$?				&	4--1 of $X {}^1\!\Sigma^{+}-X {}^1\!\Sigma^{+}$		&	$\dots$							&	$\dots$ & $\dots$	&	$\dots$ &	SB	\\			
1.58927				&	H I (Br 14)						&	$4-14$																		&	136								&	$\dots$	&	$\dots$&	$\dots$&	Blended with \\
							&										&																					&										&				&			&			& Si I line \&  TR \\
1.59677				&	Si I								&	$4p^{3}\!D_{3}-5s^{3}\!P^{\circ}_{2}$								&	51								&	$\dots$		&	$\dots$		&	$\dots$		&	SB				\\
1.59871				&	${}^{12}\!\mathrm{CO}$?				&	5--2 of $X {}^1\!\Sigma^{+}-X {}^1\!\Sigma^{+}$		&	$\dots$							&	$\dots$ & ??	&	?? &	$\dots$	\\			
1.61173				&	H I (Br 13)						&	$4-13$																		&	55								&	18.8		&	$1.79\pm0.10$		&	$-1.73\pm0.05$		&	Close TR?	\\
1.61962				&	${}^{12}\!\mathrm{CO}$?				&	6--3 of $X {}^1\!\Sigma^{+}-X {}^1\!\Sigma^{+}$		&	$\dots$							&	$\dots$ 	& $0.94\pm0.27$	&	$-0.91\pm0.09$&	$\dots$	\\				
1.64163				&	H I (Br 12)						&	$4-12$																		&	74								&	18.8		&	$\dots$&	$\dots$&	CP, SB		\\
1.64117				&	[Fe II]								&	$a^{4}\!F_{9/2}-a^{4}\!D_{7/2}$										&	$\dots$							&	$\dots$	&	$\dots$&	$\dots$&	ND\\	
1.68159				&	H I (Br 11)						&	$4-11$																		&	75								&	17.8		&		$2.62\pm0.44 $	&	 $-2.48\pm0.10$		&	$\dots$				\\
1.68813				&	UF									&	$\dots$																		&	$\dots$							&	$\dots$	&	$\dots$&	$\dots$&	SB	\\				
1.68990				&	C I									&	$3p^{1}\!D_{2}-3d^{1}\!F^{\circ}_{3}$								&	60								&	$\dots$		&	 $\dots$		&	$\dots$		&	SB				\\
1.70105				&	He I								&	$3p^{3}\!P^{\circ}-4d^{3}\!D$										&	48								&	9.1		&	$\dots$&	$\dots$&	Overlaping TR \\	
1.71171				&	Mg I								&	$4s^{1}\!S_{0}-4p^{1}\!P^{\circ}_{1}$								&	54								&	8.8		&	$0.95\pm0.70$		&	$-0.89\pm0.17$		&	$\dots$				\\
1.721-1.732				&	TR								&	$\dots$																		&	$\dots$							&	$\dots$	&	$\dots$&	$\dots$&	$\dots$	\\				
1.73713				&	H I (Br 10)						&	$4-10$																		&	66								&	15.6		&	$\dots$&	$\dots$&	Close TR \\
1.747-1.800			&	TR								&	$\dots$																		&	$\dots$							&	$\dots$	&	$\dots$&	$\dots$&	$\dots$	\\				
\hline
2.00-2.03				&	TR								&	$\dots$																	&		$\dots$						& $\dots$	& $\dots$&	$\dots$&	$\dots$\\			
2.05-2.08				&	TR								&	$\dots$																	&		$\dots$						& $\dots$	& $\dots$&	$\dots$&	$\dots$\\			
2.10645				&	TR								&	$\dots$																	&		$\dots$						& $\dots$	& $\dots$&	$\dots$&	$\dots$\\			
2.16657				&	H I (Br$\gamma$)		&	$4-7$																		&	51								&	17.4		&	$2.87\pm0.83$		&	$-2.33\pm0.09$		&	$\dots$			\\
2.20690				&	Na I								&	$4s^{2}\!S_{1/2}-4p^{2}\!P^{\circ}_{3/2}$						&	78								&	$\dots$	&	$\dots$&	$\dots$&	IB, IP		\\
2.20770				&	TR								&	$\dots$																		&	$\dots$							&	$\dots$	&	$\dots$&	$\dots$&	$\dots$	\\				
2.20951				&	Na I								&	$4s^{2}\!S_{1/2}-4p^{2}\!P^{\circ}_{1/2}$						&	63								&	$\dots$	&	$\dots$&	$\dots$&	IB, IP		\\	
2.29512				&	${}^{12}\!\mathrm{CO}$				&	2--0 of $X {}^1\!\Sigma^{+}-X {}^1\!\Sigma^{+}$		&	$\dots$							&	$\dots$	&		$13.97\pm4.40$	&	$-10.82\pm0.54$		&	$\dots$			\\
2.32417				&	${}^{12}\!\mathrm{CO}$				&	3--1 of $X {}^1\!\Sigma^{+}-X {}^1\!\Sigma^{+}$		&	$\dots$							&	$\dots$	&	$8.91\pm4.37$		&	$-6.78\pm0.56$		&	$\dots$			\\
2.35391				&	${}^{12}\!\mathrm{CO}$				&	4--2 of $X {}^1\!\Sigma^{+}-X {}^1\!\Sigma^{+}$		&	$\dots$							&	$\dots$	&	$8.66\pm4.09$		&	$-6.55\pm0.63$		&	$\dots$			\\
2.35-2.40				&	TR								&	$\dots$																	&		$\dots$						& $\dots$	& $\dots$&	$\dots$&	$\dots$\\			
2.38411				&	${}^{12}\!\mathrm{CO}$				&	5--3 of $X {}^1\!\Sigma^{+}-X {}^1\!\Sigma^{+}$		&	$\dots$							&	$\dots$	&	$\dots$&		$\dots$ & Overlaping TR				\\
\noalign{\smallskip}\hline
\end{tabular*}
\flushleft \footnotesize
Notes: \noindent $TR$: Telluric residuals. $FB$: Line foot blended with other lines. $SB$: Line strongly blended. $CP$: Complex profile. $IB$: Identified only in emission in the spectrum not corrected for telluric absorptions. $IP$: Poor accuracy on the estimation of the line position. $UF$: Unidentified feature. $ND$ Not seen directly.
\end{table*}

 We propose an identification of most of  the lines, for which we rely on analogies between our spectrum and those of several young embedded targets found in the literature \citep{1994ApJ...425..231K, 1996AJ....112.2184G, 2005A&A...429..543N, 2006ApJ...641..383G, 2008A&A...479..503A, 2010AJ....140.1214C, 2011AJ....141...40C, 2011ApJ...736...72K, 2013A&A...554A..66C, 2013MNRAS.430.1125C}. We  completed and double-checked the identification using three databases for atomic and ionic transitions\footnote{\textit{http://physics.nist.gov/PhysRefData/ASD/lines\_form.html}}$^{,}$\footnote{\textit{http://www.pa.uky.edu/$\sim$peter/atomic/}}$^{,}$\footnote{\textit{http://www.cfa.harvard.edu/amp/ampdata/kurucz23/sekur.html}}. We also carefully flagged spurious lines that were introduced  while dividing the spectrum by the telluric standards. \\

%\caption{Comparison of the near-infrared spectrum of the HBe component of Z CMa to spectra of variable young stellar objects PV Cephei \citep{2013A&A...554A..66C}, V1647 ORI \citep{2006ApJ...641..383G}, PTF 10nvg \citep{2011AJ....141...40C, 2013AJ....145...59H}, and EX Lupi \citep{2011ApJ...736...72K}. All spectra were smoothed to the lowest commun resolution. Spectra of  PV Cephei,  V1647 ORI and PTF 10nvg have been dereddened by $A_{V}=8$, 9,  and 6 mag, respectively to reproduce the Z CMa's HBe pseudo-continuum slope. The H-band spectrum of EX Lupi is dereddened by $A_{V}=1$ while the K-band spectrum is redenned by  $A_{V}=1$.}
%\label{fig:PVCephei}

\begin{figure}[t]
\centering
\vspace{-0.1cm}
\includegraphics[width = \columnwidth]{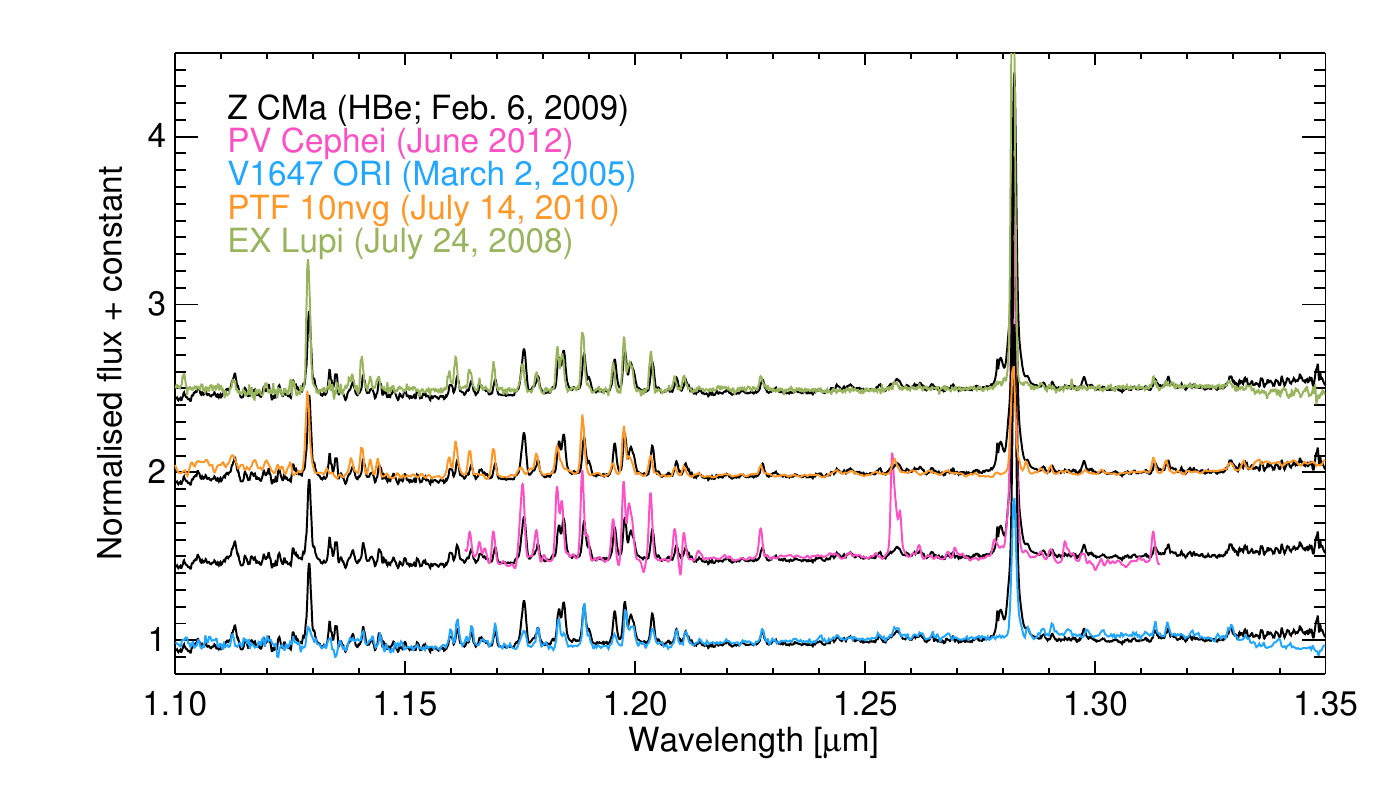}
\includegraphics[width = \columnwidth]{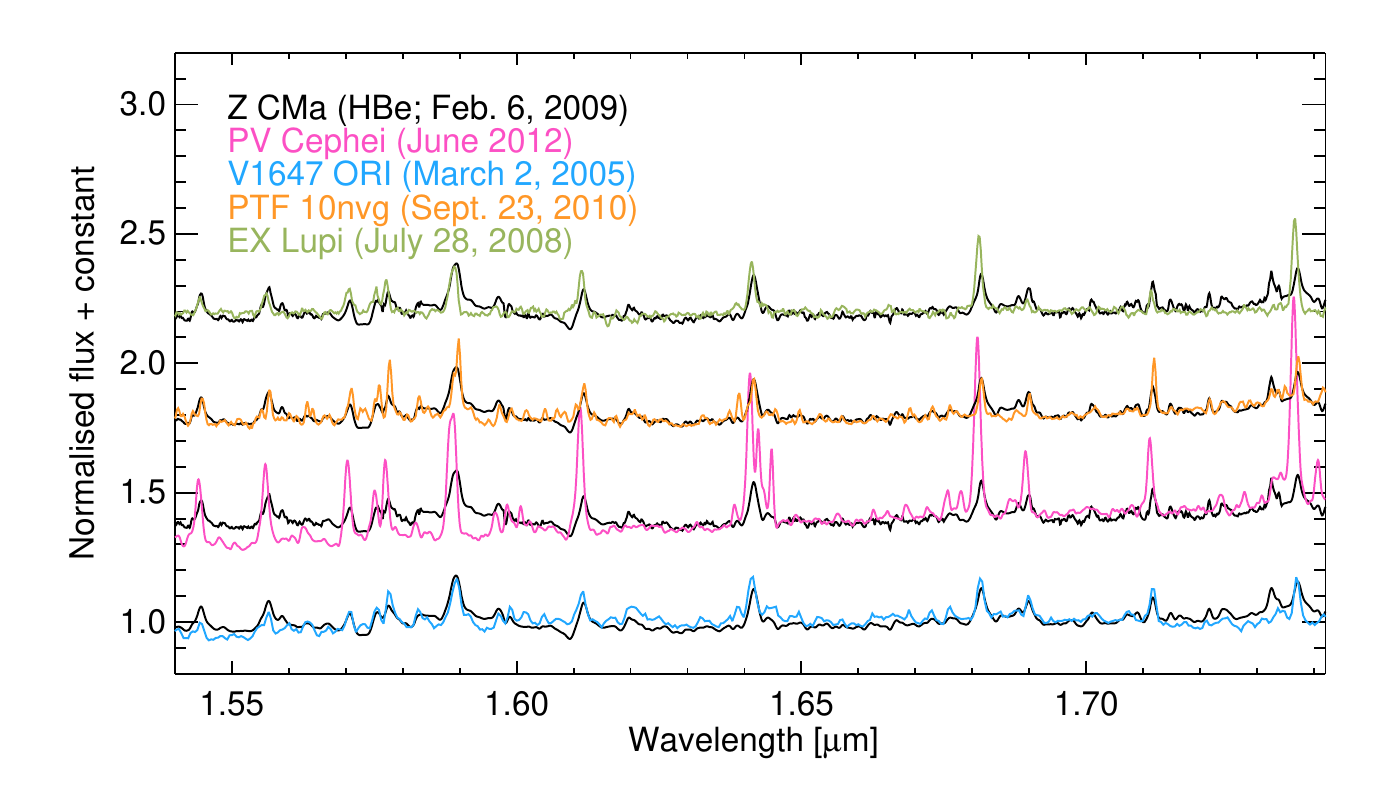}
\includegraphics[width = \columnwidth]{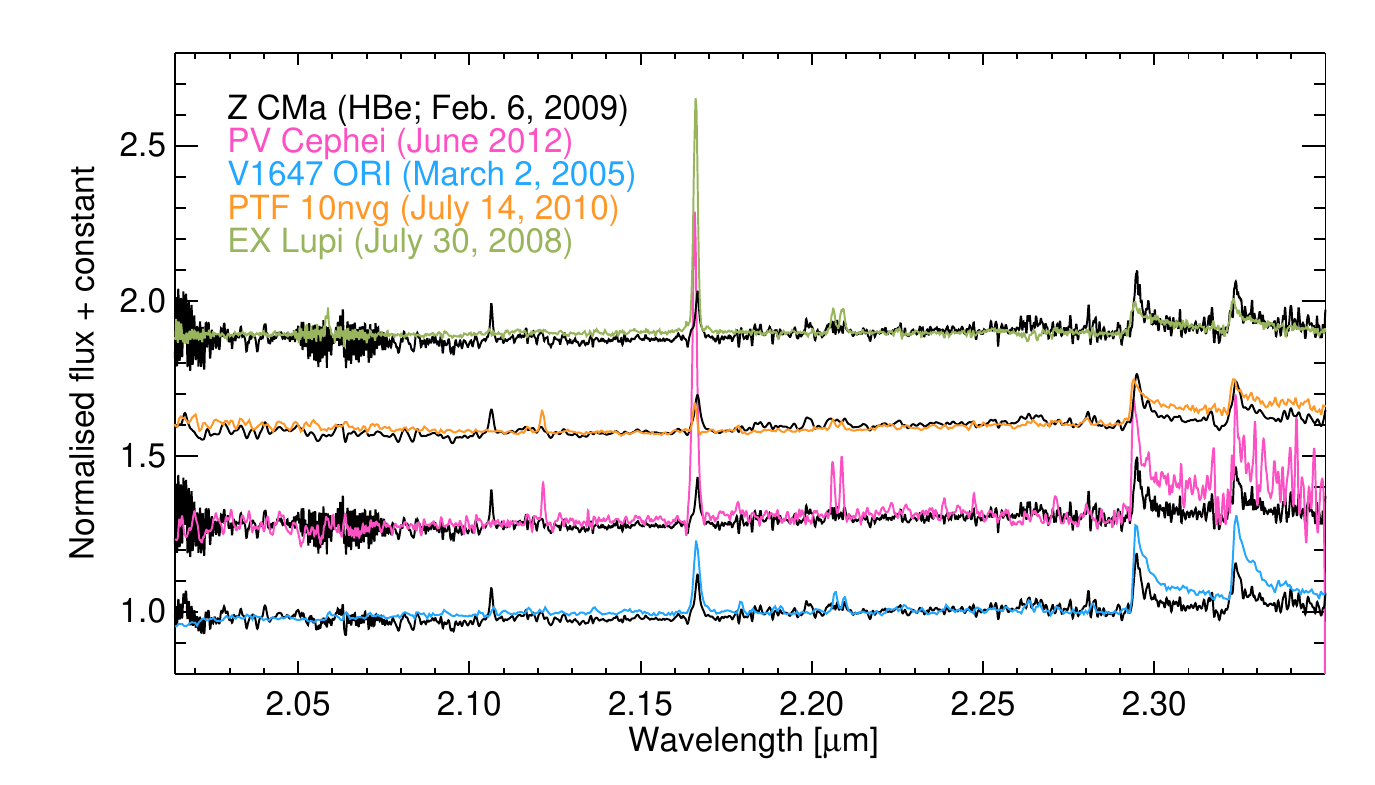}
\caption{Comparison of the NIR spectrum of the HBe component of Z CMa to spectra of variable young stellar objects PV Cephei \citep{2013A&A...554A..66C}, V1647 ORI \citep{2006ApJ...641..383G}, PTF 10nvg \citep{2011AJ....141...40C, 2013AJ....145...59H}, and EX Lupi \citep{2011ApJ...736...72K}. All spectra were smoothed to the lowest common resolution. Spectra of  PV Cephei,  V1647 ORI, and PTF 10nvg were dereddened by $A_{V}=8$, 9,  and 6 mag, respectively to reproduce the Z CMa's HBe pseudo-continuum slope. The H-band spectrum of EX Lupi is dereddened by $A_{V}=1$, while the K-band spectrum is redenned by  $A_{V}=1$.}
\label{fig:PVCephei}
\end{figure}

We retrieved the same emission lines in the EX Or prototype EX Lup captured during an outburst \citep{2011ApJ...736...72K}. The spectrum also resembles those of the outbursting protostars V1647 Ori, PTF 10nvg, and PV Cephei \citep[][see Fig. \ref{fig:PVCephei}]{2006ApJ...641..383G, 2011AJ....141...40C,  2013AJ....145...59H, 2013A&A...554A..66C}. We also find a good match with those of class I stars IRAS 03220+3035N, IRAS 16289-4449, and IRAS 20453+6746 \citep{2010AJ....140.1214C}, and of the embedded R CrA objects IRS2 and HH100 IR \citep[see Figs 2, 3, and 4 of][]{2005A&A...429..543N}.  

%PV Cephei 
%Av=6-12 mag. 8 mag de deredenning additionel pour Z CMa. 
%PTF 10nvg
%Av=9.5.  6 mag de dérougissement additionel

The spectrum is dominated by hydrogen emission lines. The Paschen $\beta$ represents the strongest line of the spectrum.  In the H band, the more remarkable emission features are the series of Brackett lines (order 10 to 22). The Brackett 23 line is expected to be blended with a Mg I emission feature at 1.504 $\mu$m. The K band  also has a moderate Br $\gamma$ emission line \citep[also present in the AMBER spectra of the HBe,][]{2010A&A...517L...3B}. 

The J-band spectrum is also dominated by  an O I line  at 1.1292 \AA. The O I line is commonly attributed to fluorescence excitation by a UV continuum and/or to resonant absorption of Ly $\beta$ photons \citep{1994ApJ...425..231K, 2005A&A...429..543N}.  The non-detection of  the O I line at 1.316 $\mu$m favors the second hypothesis \citep{2011ApJ...736...72K}.   

 A forest of lines is also present from 1.15 to  1.22 $\mu$m. This forest most likely corresponds to the broad feature seen in the spectra of HH100 IR and R Cra IRS2. It is composed of permitted lines of the neutral species of Fe, C,  Si, Mg, and of inonized Ca. We also identify other lines of Fe I, Si I , C I, Al I, Na I,  and Mg I at longer wavelengths in the J, H, and K  bands.

 Only  one ionized species can be firmly identified in the spectrum (Ca II). The first ionization potential of Ca (6.11 eV) is lower than those of Si (8.15 eV)  but remains comparable or higher than those of Mg (7.64 eV), Al (5.98 eV) and Na I (5.14 eV). The absence of Al II, Na II, and Mg II emission can be explained by the energies of the upper transition levels (12.1-37.2 eV) in the 1.1-2.5 $\mu$m range. These energies are significantly higher than those involved for the Ca II (7.5 eV). As noted by \cite{2011ApJ...736...72K}, the ionization potentials of the metallic species are lower (5.14-8.15 eV) than the one of carbon (11.26 eV), oxygen (13.62 eV), and hydrogen (13.6 eV).  This suggests that metals lies in a medium where hydrogen is mostly neutral. 
  
 We report a weak [Fe II] emissions around 1.257 $\mu$m, 1.534, and 1.644 $\mu$m with several velocity components, associated with a micro jet studied in \cite{2010ApJ...720L.119W}.   The 1.644 $\mu$m line is partially blended with the Br$_{12}$ line. We analyze structures associated with the 1.534 $\mu$m line in Sect. \ref{subsec:spatstruct}. 

The spectrum has strong $\mathrm{^{12}\!CO}$  overtones of the $X {}^1\!\Sigma^{+}-X {}^1\!\Sigma^{+}$ system seen in emission in the K band. These overtones have previously been reported by    \cite{2013ApJ...763L...9H}, while the system was in the quiescence phase before the 2008 outburst. This is retrieved in many young stellar objects \citep[e.g.,][]{2010AJ....140.1214C, 2013MNRAS.430.1125C}. \cite{2010AJ....140.1214C} showed that Br$\mathrm{\gamma}$  coincides with the emergence of the CO band-head seen in emission.  Since Br$\mathrm{{\gamma}}$ is a well-known accretion tracer \citep{1998AJ....116.2965M}, the authors proposed that the CO emission might arise from disk surfaces when the systems are quite veiled and accretion rates are high. We follow this hypothesis and model them in Sect. \ref{subsec:COlines}. Alternatively, CO overtones might be produced inside magnetospheric accretion funnels \citep{1997ApJ...478L..33M}.  This line is not detected in the spectrum that \cite{1997ApJ...478..381L} obtained while the system was in quiescence.

 To conclude, the comparison of the HBe spectrum to the one of PTF 10nvg during its brighter stages suggests that we may also be seeing water-band emission in  the spectrum of the HBe at 1.33-1.35 $\mu$m, 1.49-1.55  $\mu$m, and 1.70-1.75 $\mu$m.

\subsubsection{Line profiles}
\label{subsubsection:lineprof}
\begin{figure}[t]
\centering
\vspace{-0.1cm}
\begin{tabular}{cc}
\includegraphics[width = 4.2cm]{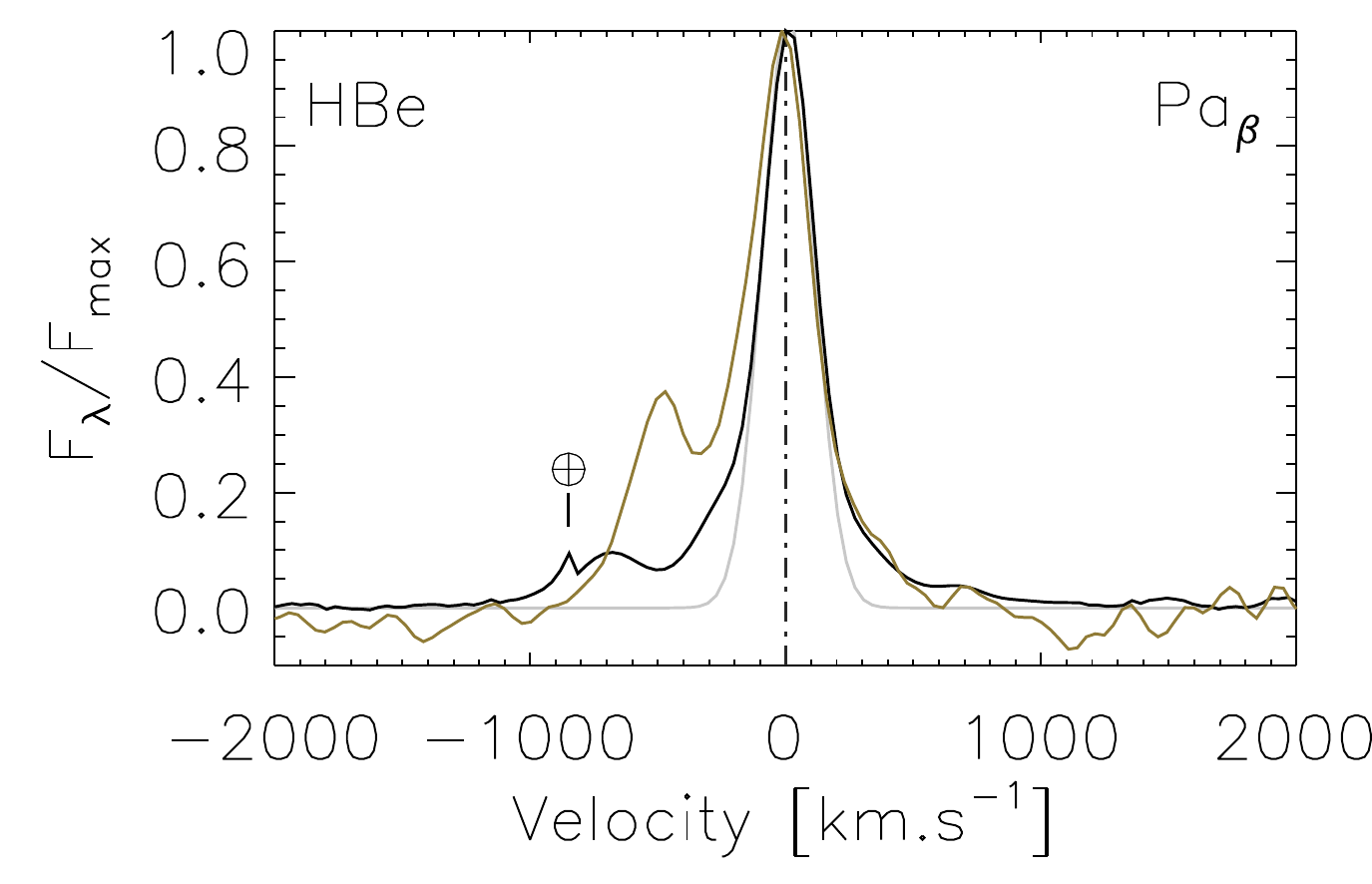} &
\includegraphics[width = 4.2cm]{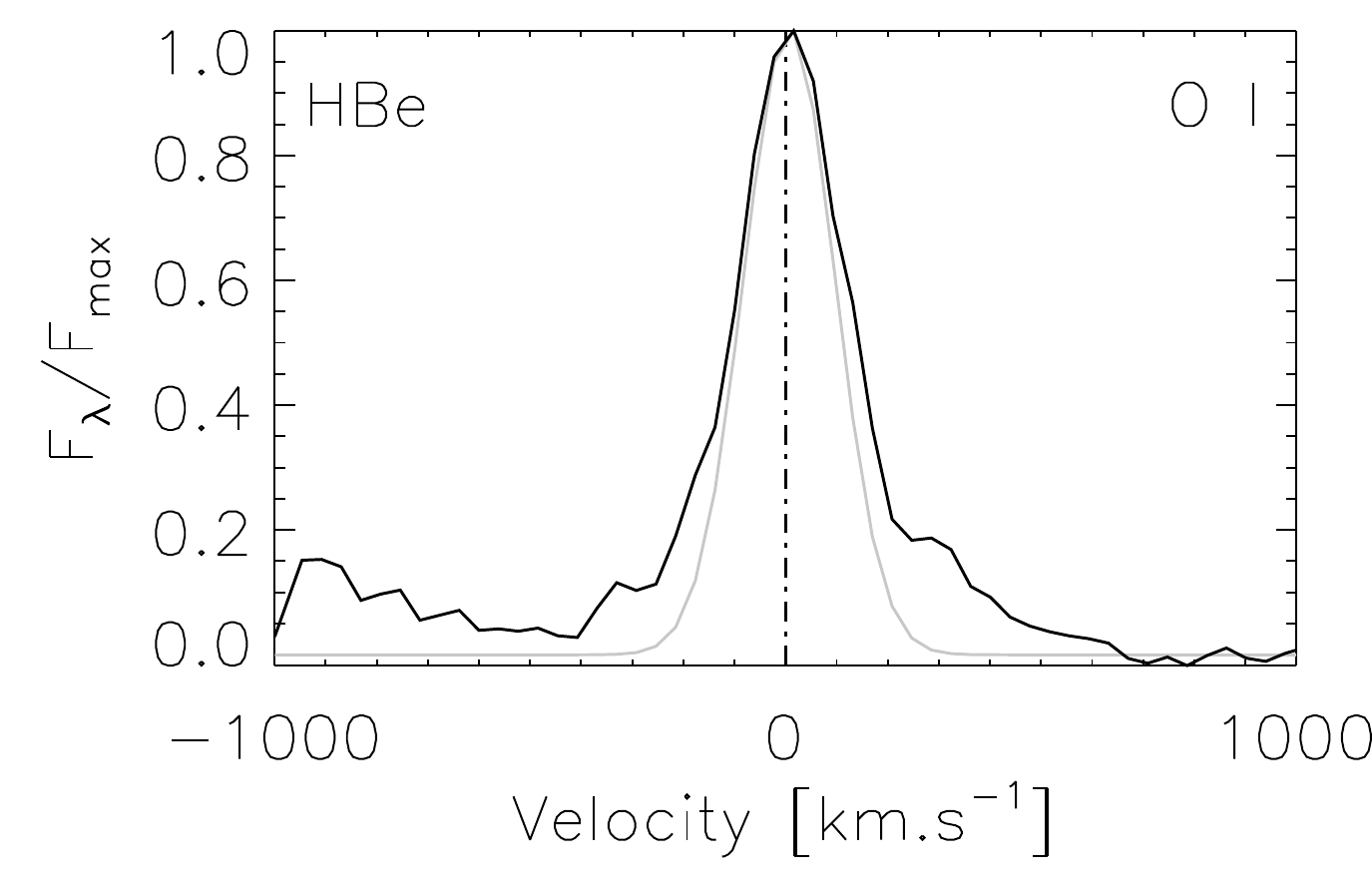}\\
\includegraphics[width = 4.2cm]{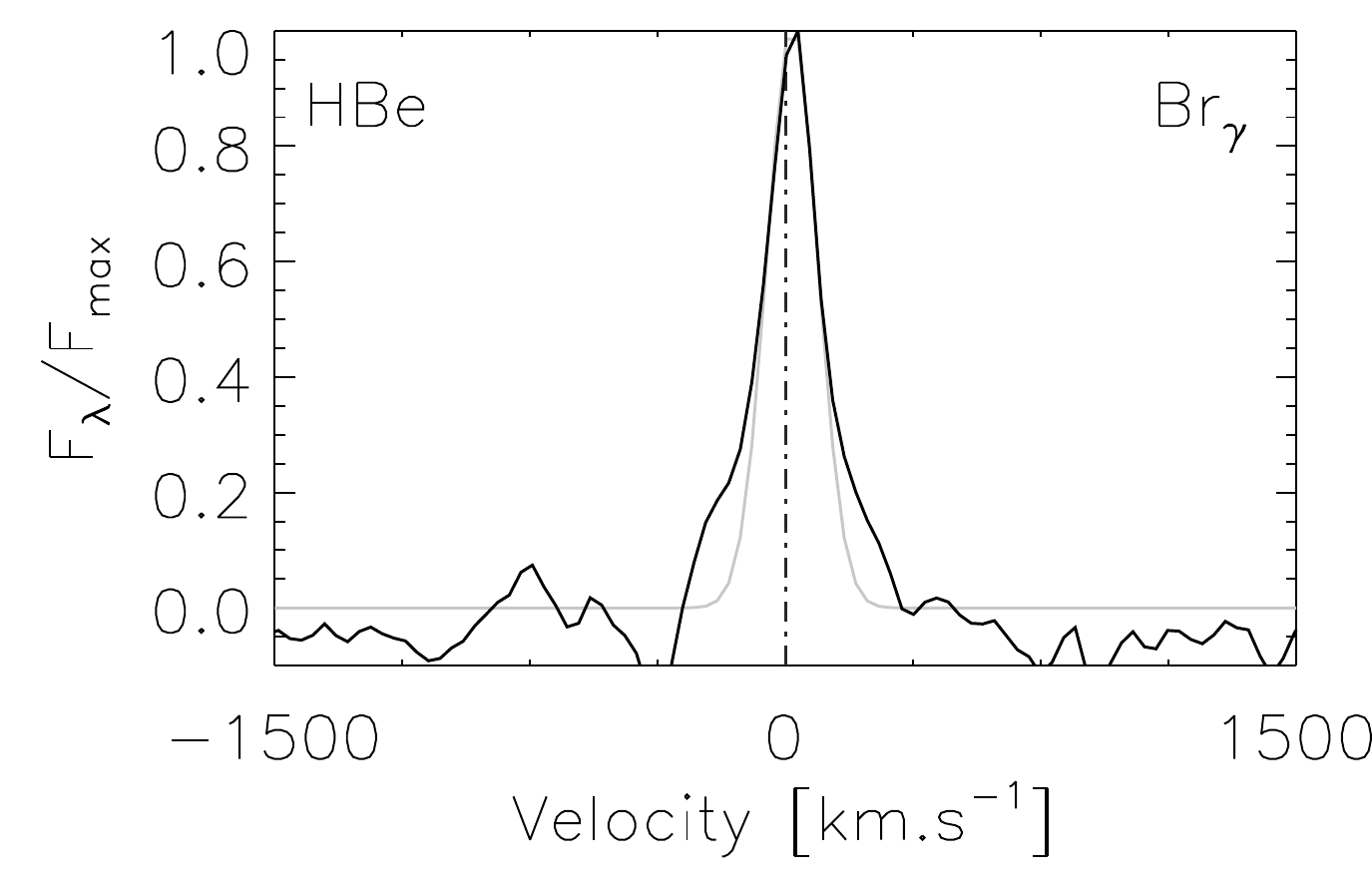} &
\includegraphics[width = 4.2cm]{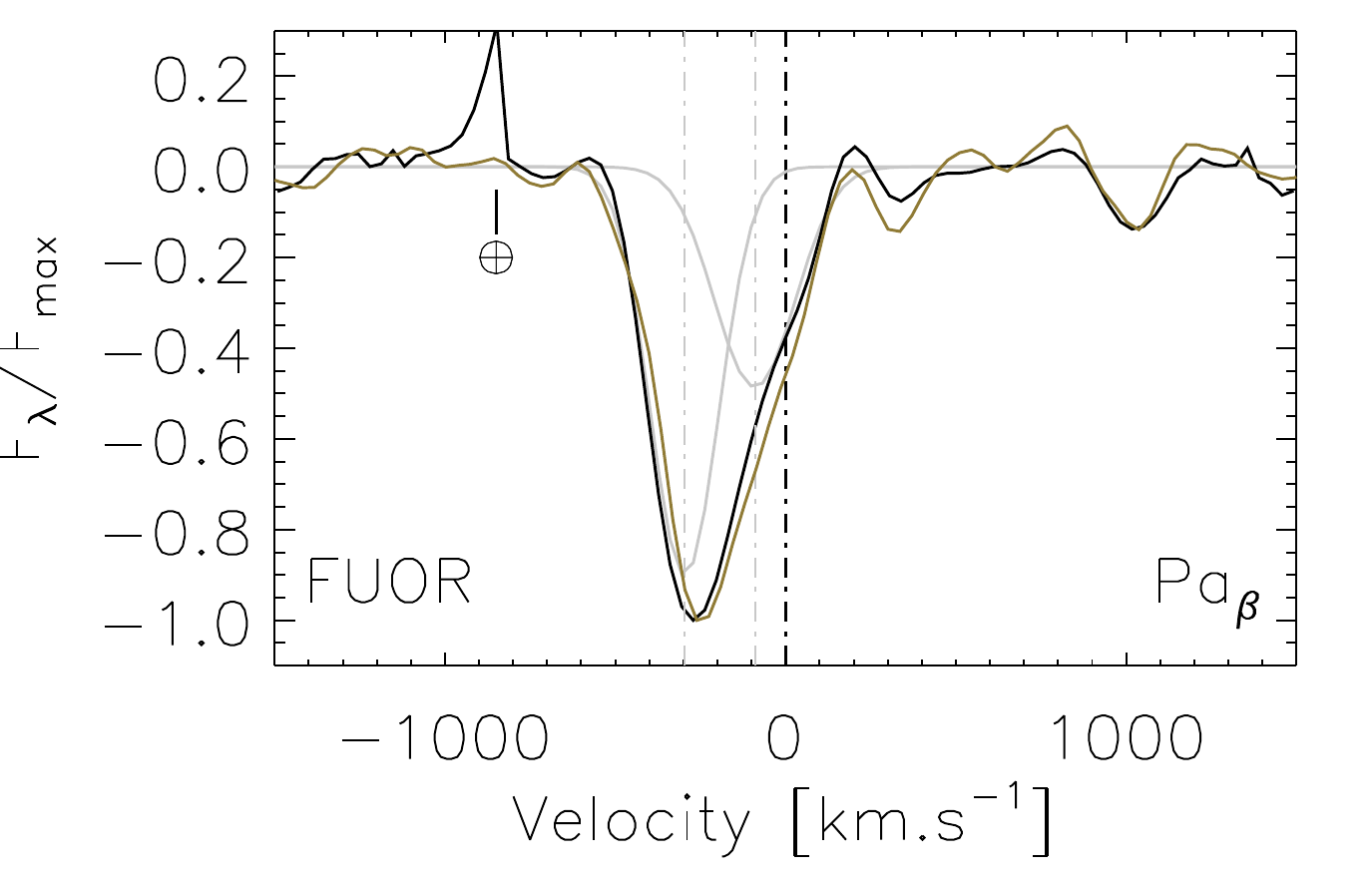}\\
\end{tabular}
\includegraphics[width =\linewidth]{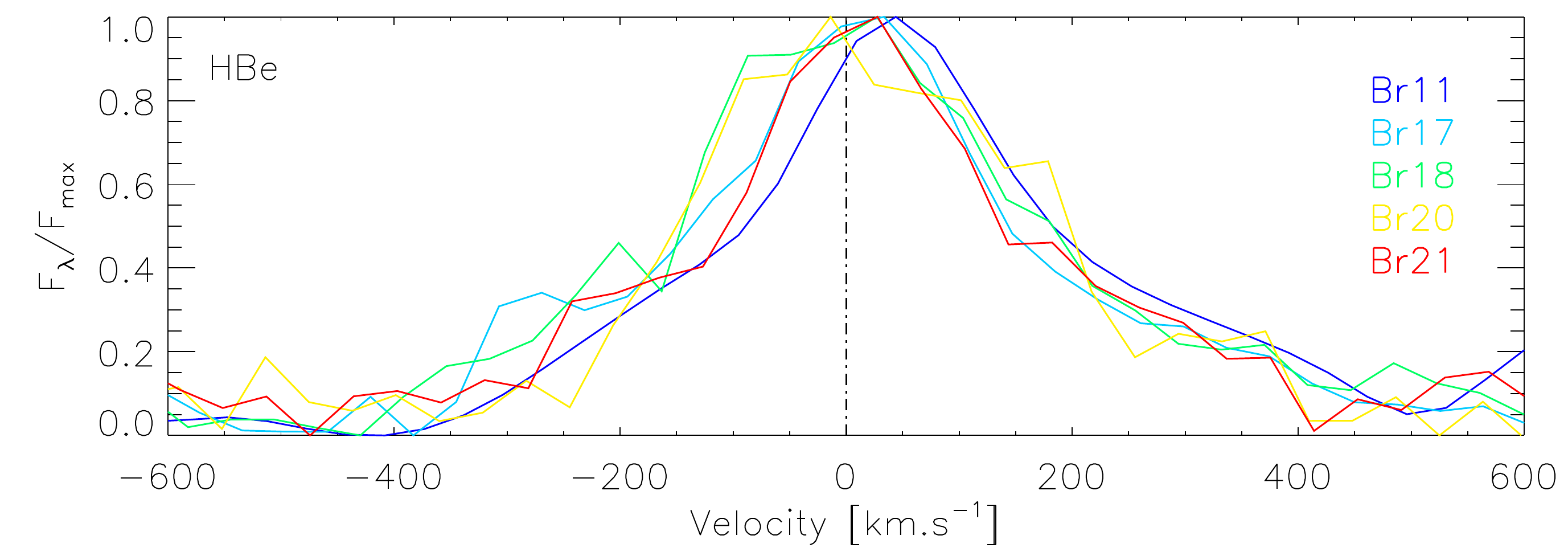}
\caption{Profile of strong and weakly blended lines for the HBe and FUOR components. The black solid lines correspond to the SINFONI (outburst) data, while the golden lines correspond to the OSIRIS spectra (post-outburst).}
\label{fig:lineprofile}
\end{figure}

Most of the lines have a FWHM greater than the instrumental line-width (5.11 \AA~in the J band, 5.00 \AA~in the H band, 3.68\AA~in the K band; measured on thorium-argon calibration lines).  The profiles of  the $\lambda$1.1292 O I, Pa $\beta$, and of several other Brackett lines that are unaffected by a strong blend (Br $\gamma$, Br 11, 17, 18, 20,  and 21) are reported in Fig.~\ref{fig:lineprofile}.  They were subtracted from their continuum and normalized to their highest value. The profiles of the Pa $\beta$ lines extracted from  the SINFONI (black line) and OSIRIS (golden line) spectra are compared in the upper left panel. A Gaussian is fitted on the line core and is overlaid (gray line). 

The oxygen line core has a non-significant velocity.  The line has two broad asymmetrical wings that extend up to $-400$ and $+600$km.s$\mathrm{^{-1}}$. An additional unrelated feature (possibly an Al I line) is present at $-900$km.s$\mathrm{^{-1}}$. 

The resolved profiles of  isolated H I emission lines (Pa$\beta$, Br 11, Br 17, Br 18, Br 20, and Br 21) are shown in Fig. \ref{fig:lineprofile}. The Pa $\beta$ profile is divided into a main component characterized by asymmetrical wings and into a blueshifted lobe at $-650$km.s$^{-1}$. The lobe appears at lower velocities ($-480$km.s$^{-1}$) in the OSIRIS spectrum and contributes more to the total line flux. We retrieved these two components in the spectrum of EX Lup \citep{2011ApJ...736...72K},  PV Cephei \citep{2013A&A...554A..66C}, and PTF 10nvg \citep{2011AJ....141...40C, 2013AJ....145...59H}.  The velocity is consistent with an emission coming from the basis of the HBe jet  \citep{2010ApJ...720L.119W}.  The line peak is slightly redshifted (42km.s$^{-1}$), as is the one of PV Cephei \citep{2013A&A...554A..66C}.  The profile also has a red tail that extends up to $\sim$1000 km.s$^{-1}$  with a bump at $+700$km.s$^{-1}$.  We retrieved this long tail in the H$\alpha$ and H$\beta$ lines seen in the 1996 and 2000 spectra of Z CMa \citep{VDA04}. VDA04 proposed that the tail originates from the extended atmosphere of the HBe. The bump is very consistent with the one seen in the spectrum of EX Lup during outburst \citep{2011ApJ...736...72K}. In summary, the complex profile of Paschen $\beta$ (and related time variations) indicates that the line forms in (at least) two distinct regions: one at the basis of the ouflow, and another closer to the star.

Brackett lines in the H band are slightly redshifted at the peak ($+25$km.s$^{-1}$) and have a symmetrical profile extending to $\pm 400$km.s$^{-1}$.  The Br $\gamma$ profile is slightly asymmetric and redshifted. This profile is coherent with the double-peaked asymmetric profile (peaks at $\sim -40$ and 120km.s$^{-1}$) studied at R$\sim$12 000 while the system was in outburst by \cite{2010A&A...517L...3B}.   We showed that the Br$\mathrm{\gamma}$ line originates from a  bipolar wind at the au scale. The same is expected for the other lines of the series. 

\subsubsection{H I line ratio} 
\label{lineratio}
\begin{figure}
\includegraphics[width=\linewidth]{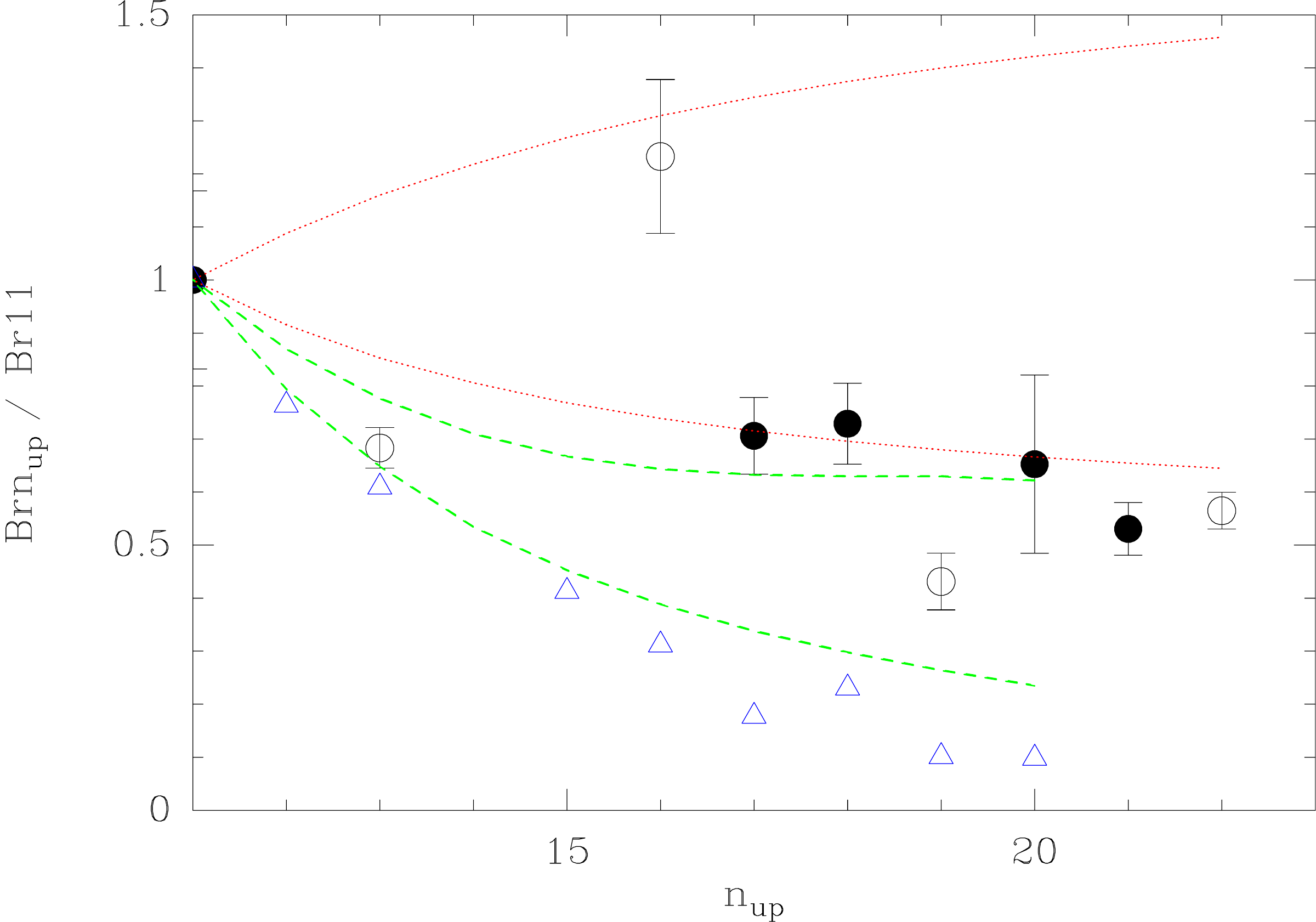}
\caption{H band HI Brackett  excitation diagram for the HBe of ZCMa (circles). All line fluxes are normalized to the HI Brackett 11 line. 
Open circles show the lines that are probably contaminated. Blue triangles show the line ratios detected in EX Lupi \citep{2011ApJ...736...72K}. The green curves are predictions from case B recombination from \cite{1987MNRAS.224..801H} for T=$10^4$ K and n=$10^8$ cm$^{-3}$ (lower curve) and for T=500 K and n=$10^7$ cm$^{-3}$ (upper curve).   The red dotted lines show expected line ratios from blackbody emission at T=1000 K (lower curve) and T=10$^{4}$ K (upper curve).}
\label{fig:HIratios}
\end{figure} 

We show in Fig. \ref{fig:HIratios} the HI line ratios observed during the photometric outburst phase for the HBe component. We only include the HI Brackett  lines detected in the H band to overcome the effect of poorly constrained reddening correction. The HI excitation diagram is significantly flatter than observed in  EX Lupi \citep{2011ApJ...736...72K} and  in other UX Ori outbursting stars
\citep{2009ApJ...693.1056L}.  Both optically thin case-B recombination and optically thick blackbody emission can reproduce the observed line ratios but would require very low gas temperatures
(500-1000 K) and moderate densities of $10^7$-$10^8$ cm$^{-3}$, suggesting that the lines do not originate in the immediate vicinity of the star. Such temperatures and densities could be
found in the surface layers of the accretion disk, which are strongly irradiated by the central star and accretion luminosity, or at the base of the outflow.  We derive a lower limit to the Pa$\beta$/Br$\gamma$ ratio of 7.7 without reddening correction. For the Pa$\beta$ line, we integrated the emission in the same velocity range as observed in Br$\gamma$, that is between -300 and +300 km s$^{-1}$. This very high ratio
is marginally compatible with case-B recombination and would exclude optically thick blackbody emission. The limit for optically thick blackbody emission is Pa$\beta$/Br$\gamma$ $\leq$ 4  \citep{2011A&A...534A..32A}. However, it is not clear that the Pa$\beta$ and Br$\gamma$ lines originate from the same regions as they show both distinct emission profiles (Sect. \ref{subsubsection:lineprof}), and different behaviors with time  (Sect. \ref{subsubsection:variability}).  It is  therefore likely that simple models assuming constant density and temperature are not an adequate representation of  the HI-emitting medium. Models of spherically expanding wind in LTE  predict both high Pa$\beta$/Br$\gamma$ ratios and flat HI excitation diagrams \citep{2009ApJ...693.1056L, 2011A&A...534A..32A}. The observed Pa$\beta$ flux and Pa$\beta$/Br$\gamma$ ratio would require mass-loss rates $\ge$ 10$^{-6}$ M$_{\odot}$ yr$^{-1}$ and an envelope thickness $> 100 R_{\odot}$. Full radiative transfer modeling in either a disk or outflow model  would be required, but this is beyond the scope of this paper.

 		\subsubsection{Modeling of the CO emission lines}
 		\label{subsec:COlines}
\begin{figure}[t]
\centering
\vspace{-0.1cm}
\includegraphics[width=\linewidth]{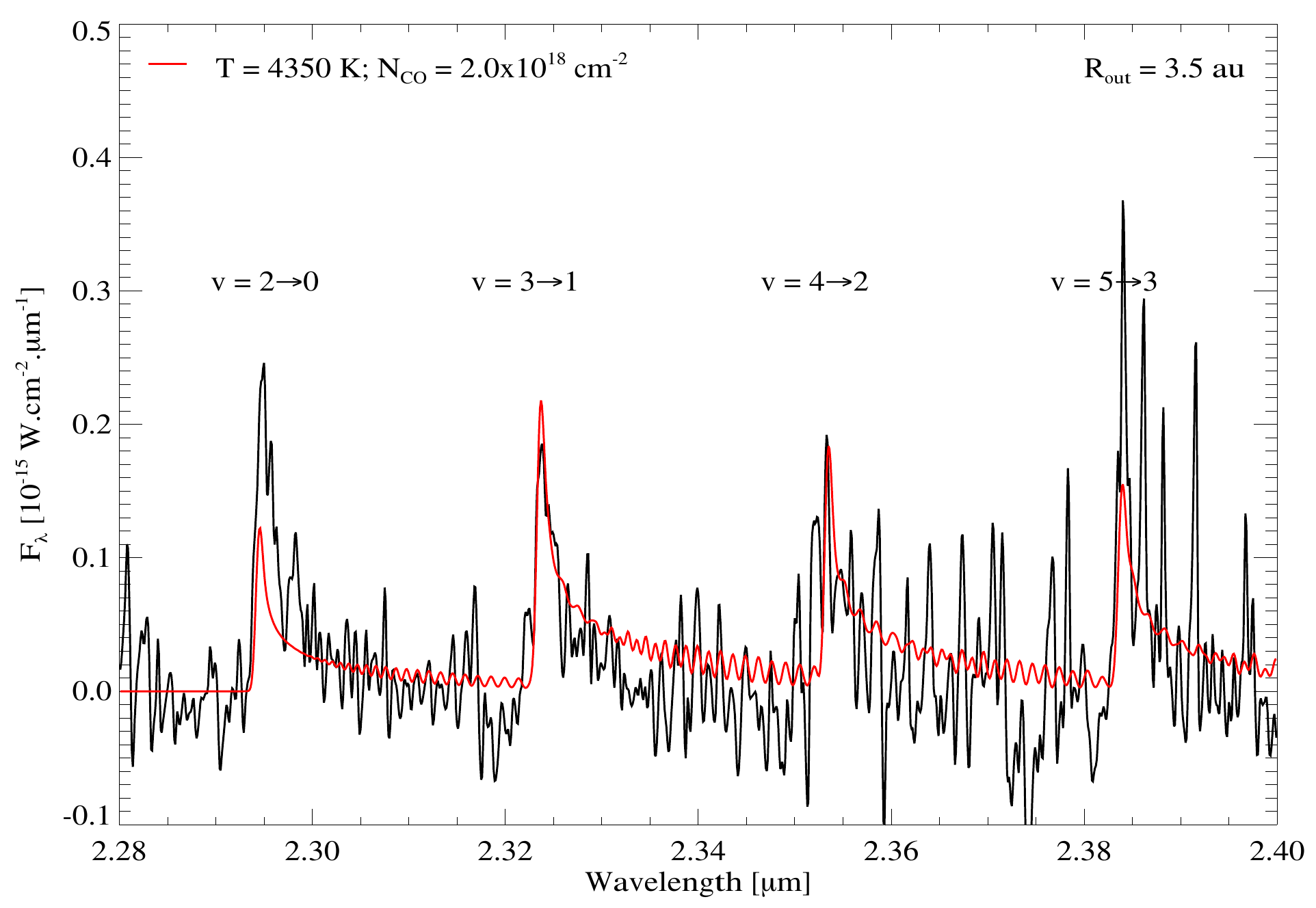}
\caption{CO bandhead seen in the SINFONI spectrum of the HBe (black) compared to our best -fit model for an outer radius of the line-emitting region of 3.5 au (red).}
\label{fig:modCO}
\end{figure}

		The CO overtones seen in emission in the HBe component are also a characteristic feature of other EX Or (see Fig. \ref{fig:PVCephei}). \cite{2014MNRAS.443.1916E} have used interferometry to resolve the CO line emitting areas around Herbig stars. They found that the regions of the disks emitting in the CO bandhead lines are located between  0.05 and 2 au from the central star. We  therefore modeled the CO lines of the HBe spectrum acquired in outburst following a similar approach as \cite{2011ApJ...736...72K} for EX Lupi. The  CO disk emission model and related fitting procedure are described in Appendix \ref{appA:detailsCO}. We conclude that the CO gas in Z CMa has an excitation temperature of about $4300 \pm 100$ K, and the CO column density is between $\mathrm{2\times10^{18} cm^{-2}}$ and $\mathrm{2\times10^{19} cm^{-2}}$, depending on the disk size.  Combinations of the outer radius of the emitting disk region $R_{out}$ and of the CO column density $N_{\mathrm{CO}}$ can give similar fit. We show in Fig. \ref{fig:modCO} the best-fitting model for  $R_{out}$=3.5 au.

We therefore conclude that the CO emission can be reproduced to  first order by the emission from a disk model.  The excitation temperature is almost 2000 K hotter than the one derived for EX Lup by \cite{2011ApJ...736...72K} with the same disk modeling procedure.  

%If we assume a $R_{out}$ = 2 au (see Section \ref{subsection:spectroastrom}), we find best fitting $N_{CO}=\mathrm{6\times10^{18} cm^{2}}$. We also get a consistent $T_{ext}=4300$ K, and  $N_{CO}=\mathrm{1\times10^{19} cm^{2}}$  and $N_{CO}=\mathrm{4\times10^{18} cm^{2}}$ when the $\nu=2\rightarrow0$ and $\nu=3\rightarrow1$ are fitted separatly (respectively). 

%\begin{figure}[t]
%\centering
%\vspace{-0.1cm}
%\includegraphics[width=\linewidth]{zcma_co.pdf}
%\caption{CO evertones of the HBe component compared to a synthetic spectra infered from disk models.}
%\label{fig:CO}
%\end{figure}

\subsubsection{Spectro-photometric variability}
\label{subsubsection:variability}
We report in Fig. \ref{fig:lightcurve} the  dates of the spectroscopic observations of the HBe on the light curve of the system.

The low-resolution spectrum (R$\sim$250) of the system gathered during ourburst by \cite{2009ATel.2024....1A} also displays the O I, the Paschen $\beta$ emission line, and the Brackett lines  seen in the SINFONI H-band spectrum of the HBe (Anto+09 in Fig. \ref{fig:lightcurve}). The comparison of the SINFONI (this work) and P1640 \citep{2013ApJ...763L...9H} spectra  obtained during outburst, the OSIRIS spectrum (while the system was returning to quiescence; this work), and the pre-outburst (quiescent state) NIRSPEC spectrum   from 2006 \citep{2013ApJ...763L...9H} enable us to pinpoint the evolution of the features related to the HBe during outburst. This collection of data is complemented by a low-resolution (R$\sim$690) K-band spectrum of each component from November 1992 \citep{1997ApJ...478..381L} while the system was in quiescence, but still at a magnitude lower than the one of 2006 and 2009. The evolutions of the equivalent widths of the Pa$\beta$, Br$\gamma$, and $\nu=2\longrightarrow0\:^{12}$CO  are reported in Table \ref{tab:EWvar}. 

The evolution of the Pa $\beta$  line (Fig. \ref{fig:lineprofile}) of the HBe shares some behavior with the one of PV Cephei. The blueshifted lobe that is seen in absorption in the spectrum of PV Cephei (P Cygni profile) only appears in the ourburst phase. The equivalent widths of the lines for the HBe and this object decreased by $\sim$50\% in 10 and 9 months, respectively, after the outburst.  The decline of the equivalent width is consistent with the one of the $\nu=2\longrightarrow0\:^{12}$CO overtone.  

\begin{figure*}[t]
\centering
\vspace{-0.1cm}
\begin{tabular}{cc}
\includegraphics[height = 8cm]{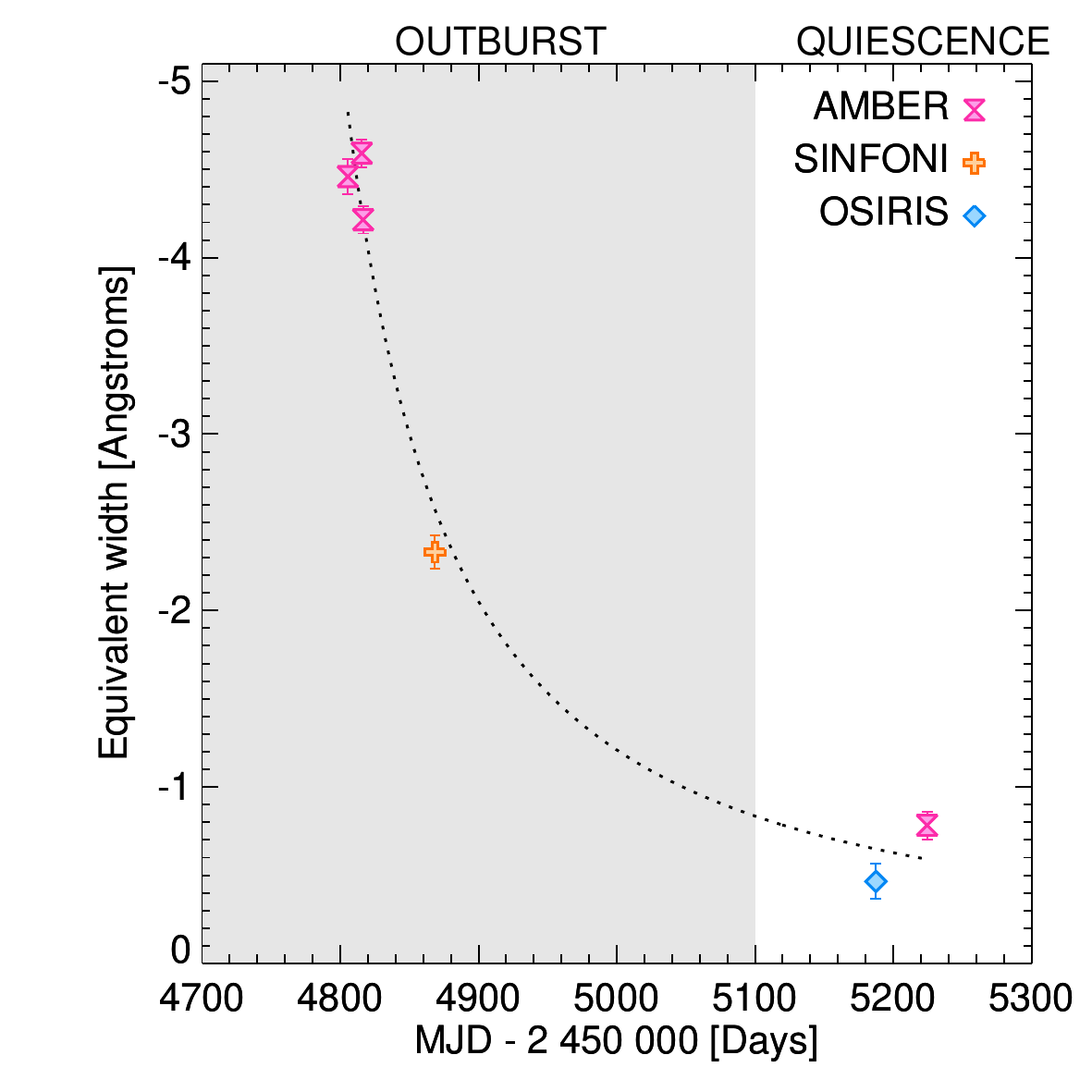} &
\includegraphics[height = 8cm]{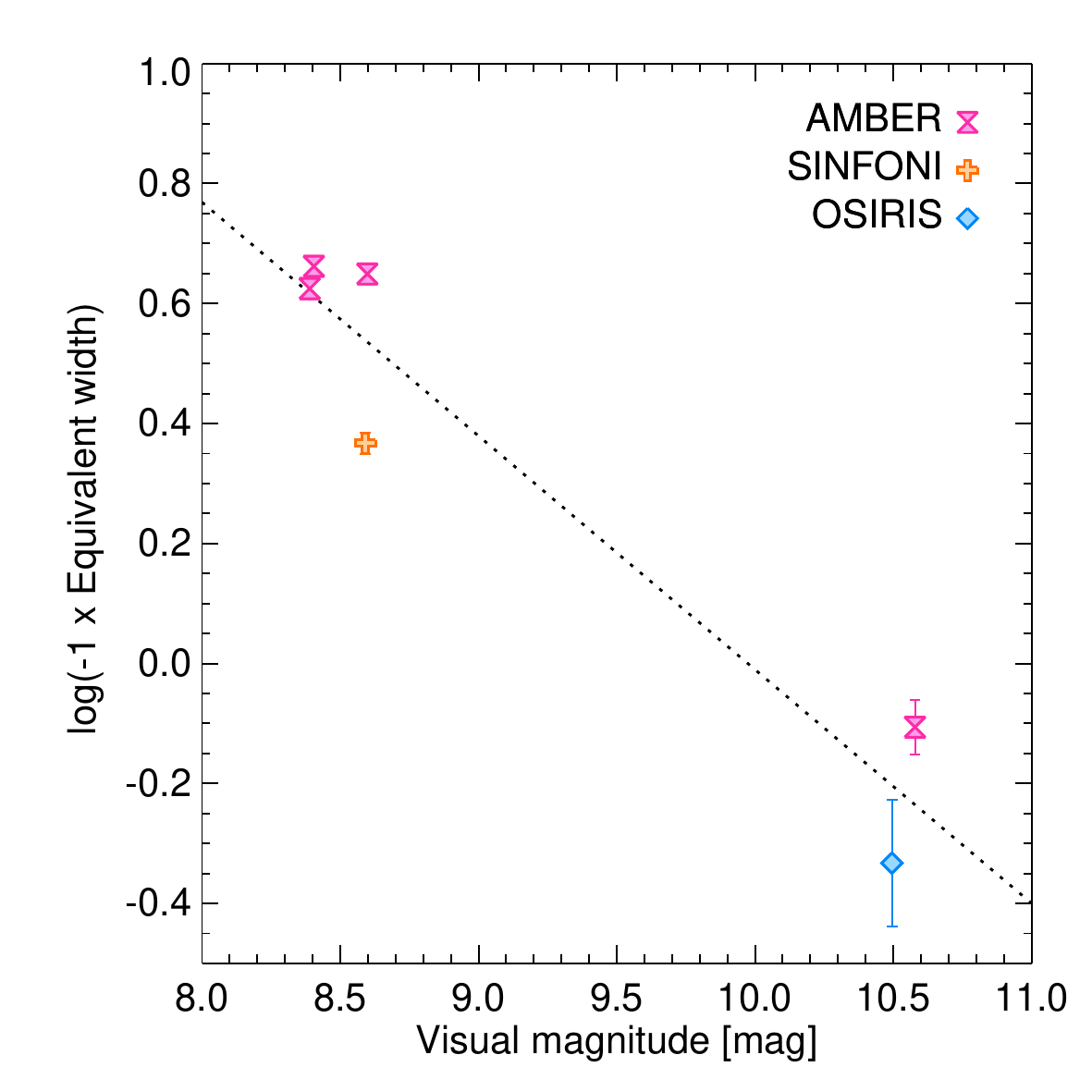} \\
\end{tabular}
\caption{Decline of the Br $\gamma$ equivalent width with time (left) and with the system V-band magnitude (right).}
\label{fig:varbrg}
\end{figure*}

In  Fig. \ref{fig:varbrg},  we show that the equivalent width of the Br $\gamma$ line decreases after the photometric outburst.  This decline can be reproduced by the function $EW(t)=A/(t-B)^{\beta}$, where $A$ and $B$ are constant values modulating the amplitude and epoch of maximum outburst, and $\beta$ the decline factor.  We used a Levenberg-Marquardt fitting tool  to estimate  $A=-1516\pm1828$, $B=4708\pm25$, and  $\beta=1.25\pm0.20$.  The $B$ parameter corresponds to an epoch of maximum outburst on  August 30, 2008. Unfortunately, this falls at a time when the V band was not monitored, but the system was considered to be in outburst at that time. We caution of course that the  analysis is limited by the small number of epochs. The  equivalent widths of the line decline with that of the V-band magnitude of the system (correlation factor=0.95) taken from the AAVSO International database\footnote{\textit{http://www.aavso.org}} (Kafka, S., 2016). This line is not detected in the November 1992 low-resolution spectrum of the HBe, which confirms that it is related to the system activity.

The NIRSPEC spectrum obtained while the system was in a quiescent state shows a He I line (2.1125 $\mu$m, 3p$^{3}$P$^{\circ}$-4s$^{3}$S). This line is not retrieved in the SINFONI and OSIRIS spectra.  It is seen in some YSO spectra and is commonly associated with the stronger 2.0508$\mu$m He I line  (2s$^{1}$S-2p$^{1}$P$^{\circ}$) \citep[e.g.,][]{2013MNRAS.430.1125C, 2013MNRAS.436..511M} which has been detected by \cite{VDA04} in the IRTF  spectrum of the system taken when it was returning to quiescence (post-2000 outburst). This second line is located outside of the NIRSPEC coverage, but it is still missing from our SINFONI and OSIRIS spectra of the HBe.  \cite{2013MNRAS.430.1125C} proposed that these lines form by collision-induced excitation in a wind. The disppearance of this line system in the outburst (SINFONI) and post-outburst (OSIRIS) state might be due to the evolving contrast with the continuum emission, or to the variable extinction in the line of sight (see below).

%Braket Gamma: Declin de lq rai qpres lòutburst. Declin de l'accretion uniquement ? 
%Raie de Fe II a 1.257 plus visible dans le spectre OSIRIS. 
%Raies de He I observee uniquement dans le spectre de 2006. Correspond a  Raies de He I a 2.1125 microns (3p3P°-4s3S). Retrouve dans certains spectres de YSO de ISHII et al. 2001 et Cooper 2013 et associee a autre raie de He I plus forte a 2.0508 microns (2s1S-2p1P°) vue dans 15% des YSO de Cooper et al. 2013, mais située en dehors de l`'interval spectral de NIRSPEC.   Cooper et al. 2013 propose que les raies de He I se forment par excitation par collision, et provient d'un vent (voir aussi Edwards et al. 2007, 2009). 
%Voir le papier de Murakawa et al. 2013

%Variability - CO
\begin{table*}[t]
\caption{Variability of the equivalent widths of the Paschen $\beta$ and Brackett $\gamma$ lines, and of the $\nu=2\longrightarrow0$ $^{12}$CO overtone for the HBe component}             % title of Table
\label{tab:EWvar}
\centering
\renewcommand{\footnoterule}{}  % to avoid a line before footnotes
\begin{tabular}{lllll}     % 3 columns
\hline\hline\noalign{\smallskip}
Date 	&  Instrument   &  EW Pa$\mathrm{\beta}$ &  EW Br$\mathrm{\gamma}$ & EW $^{12}$CO 	\\
(DD/MM/YYYY)  &   		 	&       (\AA) &       (\AA) &       (\AA) \\
\hline\noalign{\smallskip}
17/12/2006   &  NIRSPEC & \dots & \dots & $-3.14\pm0.17$ \\
05/12/2008   &  AMBER   & \dots &$-4.46\pm0.10$ &  \dots \\
15/12/2008   &  AMBER   & \dots  &$-4.59\pm0.08$ &  \dots  \\
16/12/2008   &  AMBER   & \dots &$-4.21\pm0.08$ &   \dots \\
06/02/2009  &  SINFONI & $-30.53\pm0.14$ & $-2.33\pm0.09$& $-5.51\pm0.11$ \\
22/12/2009  &  OSIRIS &  $-14.67\pm0.22$ &$-0.46\pm0.10$ & $-2.29\pm0.10$  \\
10/01/2010   &  AMBER   & \dots &$-0.78\pm0.08$ &   \dots \\
\noalign{\smallskip}\hline
\end{tabular}
\end{table*}

The NIR broadband photometry is mostly sensitive to the continuum emission. Therefore, the  decrease of the line-equivalent width coupled to the stagnation (K band) or decrease (V band, J and H bands) of the continuum flux indicates that the flux produced by the excited regions does decline as well or that the absorption increases. The bluing of the slope of the HBe in the outburst phase supports  the hypothesis of a decrease in extinction as well as an increase of the internal  heating  of the inner portions of the disk due to an increase in accretion rate.  We note that the continuum slope of the SINFONI spectrum is consistent with the one obtained with P1640. Both the SINFONI and OSIRIS K-band spectral continua look redder than the one obtained in November 1992  \citep[Fig. 1 of][]{1997ApJ...478..381L} while the system was in quiescence.

\begin{figure*}[t]
\centering
\begin{tabular}{ccc}
\includegraphics[width=6cm]{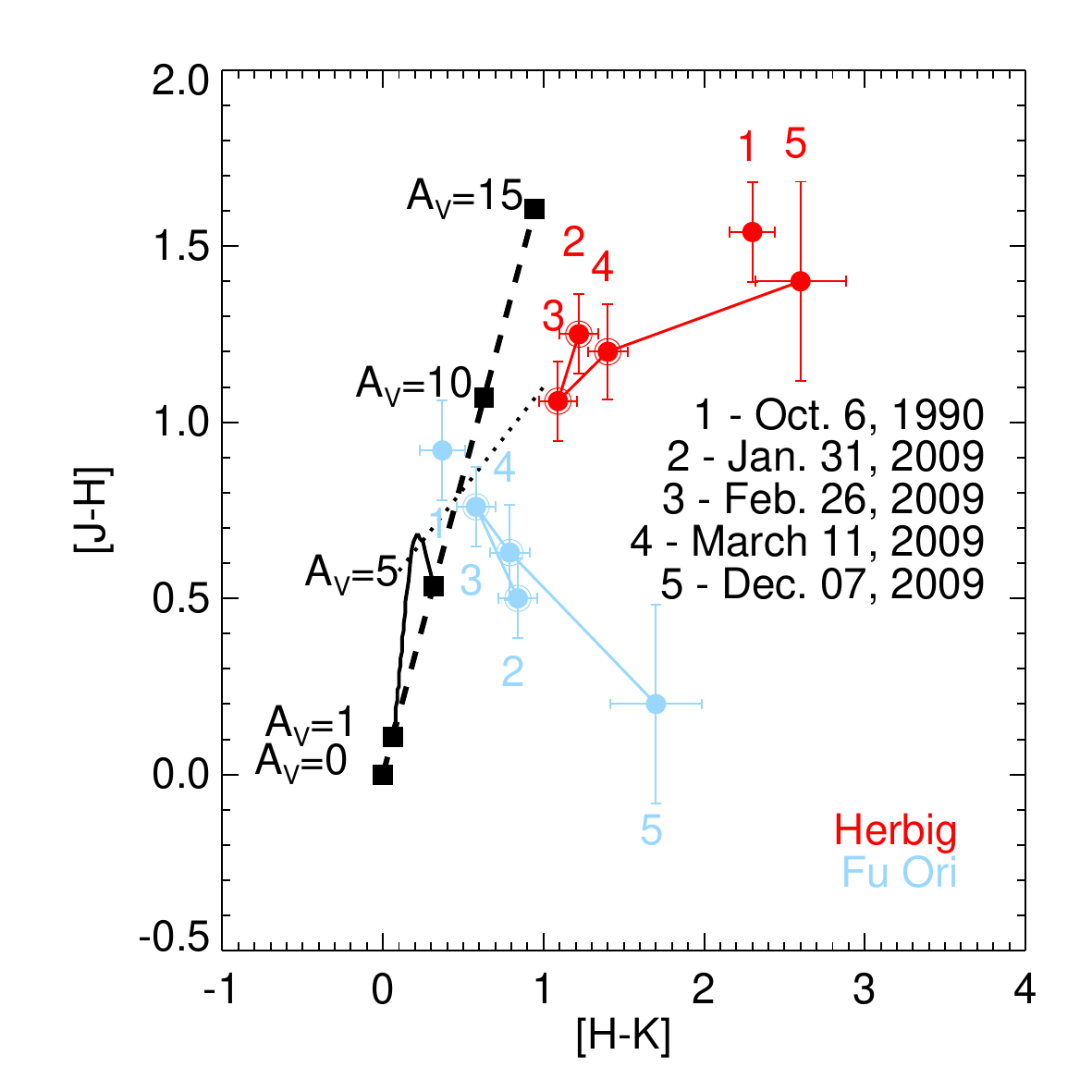}
\includegraphics[width=6cm]{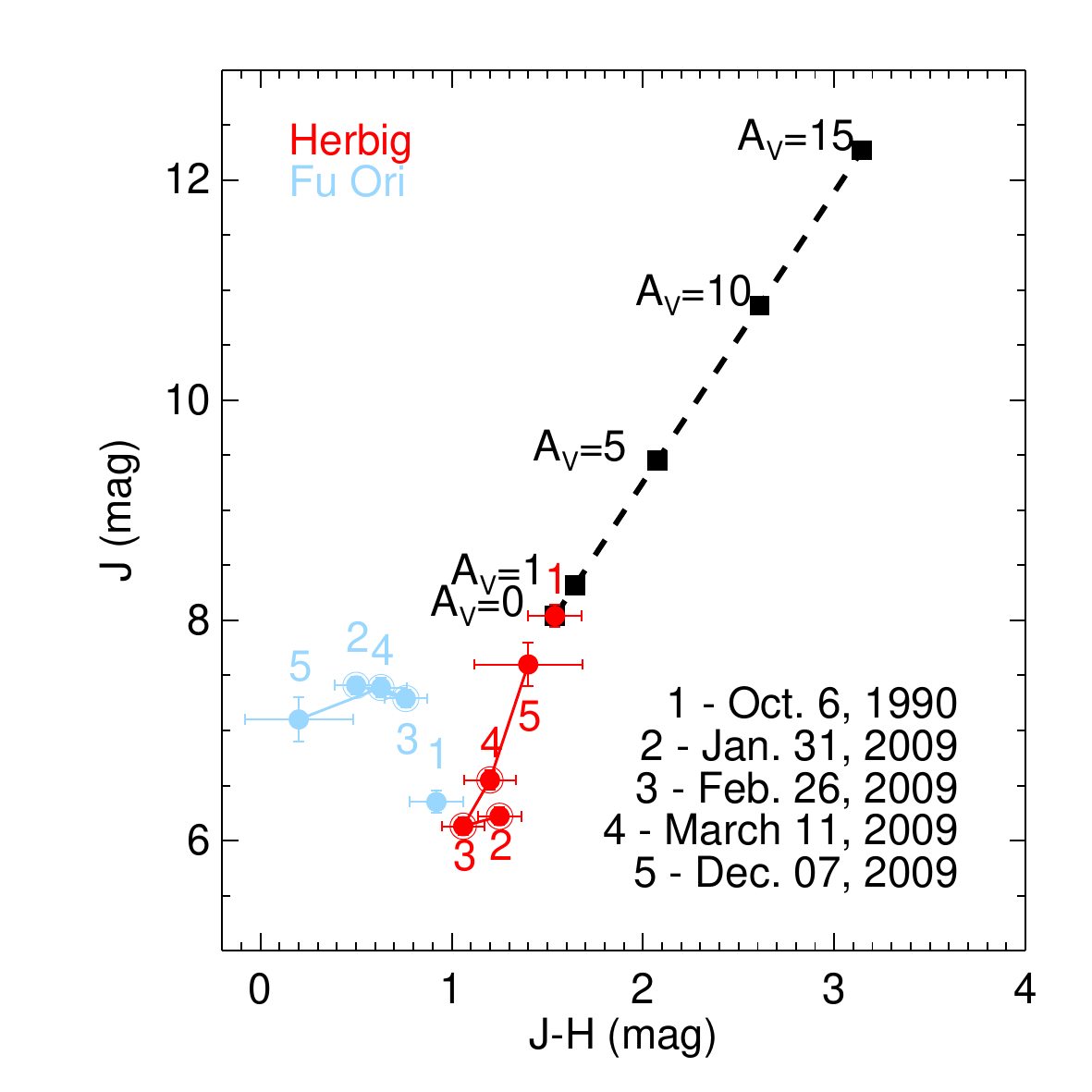} &
\includegraphics[width=6cm]{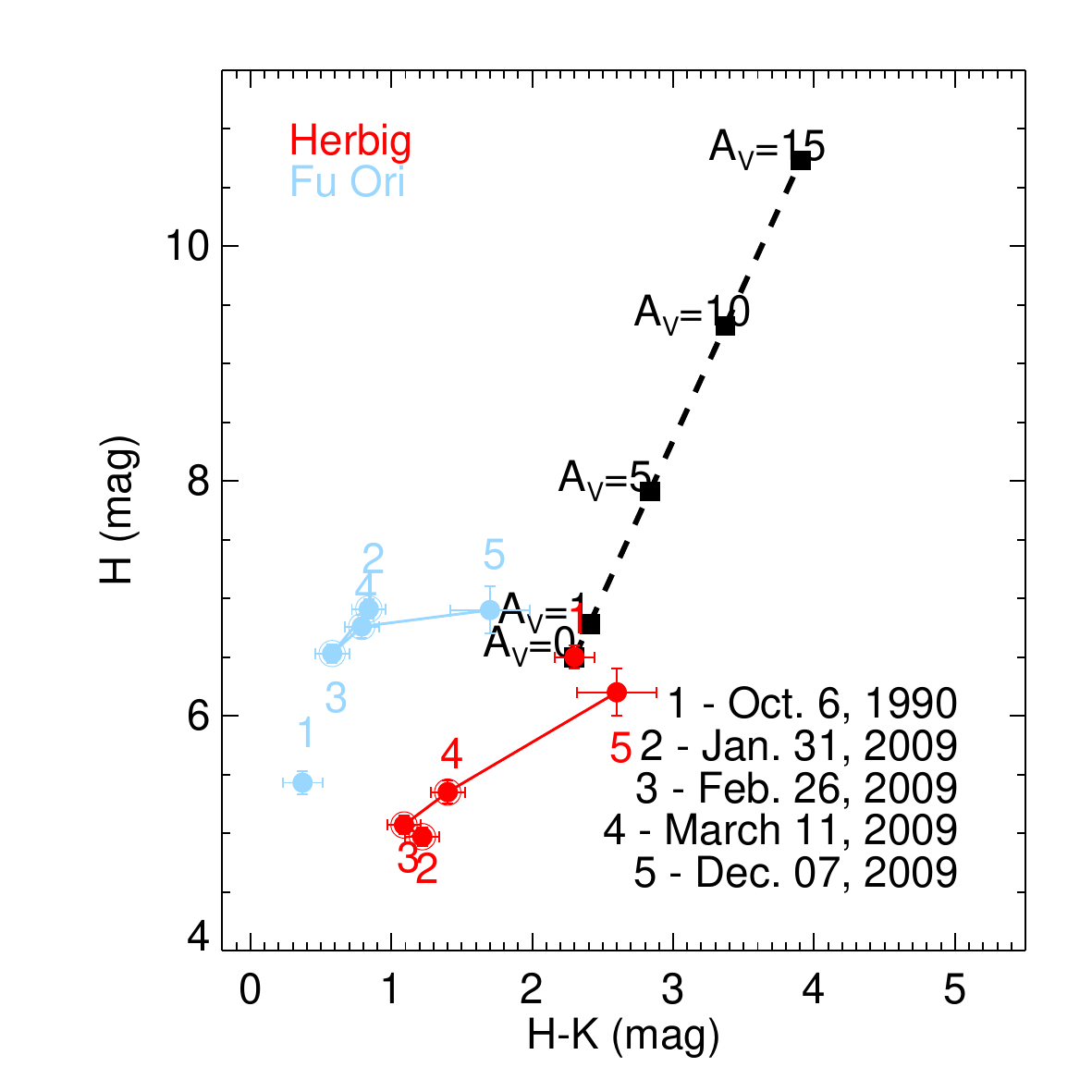}  \\
\end{tabular}
\caption{Evolution of the FUOR and HBe NIR colors and photometry when Z CMa was in a quiescent state (1; October 1990 observations), in outburst  (2-3-4; encircled disks; corresponding to NaCo observations on January 31, February 26, and March 11, 2009 respectively) and returning to a quiescent state in December 2009 (5). The reddening vector for visual extinctions of 0, 1,5,10, and 15 mag as well as the colors of pre-main-sequence stars \citep[solid line,][]{2013ApJS..208....9P} and the locus of CTTS \citep[dashed line,][]{1997AJ....114..288M} are overlaid for comparison.}
\label{fig:zcmacm}
\end{figure*}

The evolution of the HBe in color-color and color-magnitude diagrams (Fig. \ref{fig:zcmacm}) indicates that the  photometry of the star has one component that follows the interstellar reddening vector. This suggests that the outburst is at least partly due to a reduced extinction in the line of sight. \cite{VDA04} reported a spectral type B0IIIe for the HBe. Assuming that these stars have NIR photometric colors close to 0, the location of the HBe in  the Fig. \ref{fig:zcmacm}  indicates that the star was extincted by A$_{v}=$12 mag during the 2008 outburst.  This is  consistent with A$_{v}=$10 mag measured by \cite{2013ApJ...763L...9H} from the SED analysis of the star obtained during the outburst. This also agrees with the lower limit on A$_{v}$ derived from line ratio in Sect. \ref{lineratio}. The reduced A$_{v}$ found by \cite{2013ApJ...763L...9H}  is also consistent with the higher flux level of their spectra compared to SINFONI's and shown in Fig. \ref{fig:zcmaJHK}, although we cannot exclude that the different flux levels may come from differences in the flux-calibration methods used the two sets of spectra.

The variation in H-K  color of the HBe  component shows an additional contribution perpendicular to the reddening vector.  This deviation is reproduced by several EX Ors \citep{2006A&A...453..579L, 2011A&A...527A.133K, 2012ApJ...749..188L}, and in particular EX Lup. EX Lup represents the NIR spectrum of the HBe well (Fig. \ref{fig:PVCephei}). \cite{2012ApJ...749..188L} showed that during photometric outbursts of EX Ors, an additional blackbody component appears with temperatures between 1000 and 4500 K and a flux corresponding to blackbody emission from a uniform disk of radius between 0.01 and 0.1 au. \cite{2013ApJ...763L...9H} fit the SED of the HBe with two blackbodies at $\mathrm{T_{eff}}$=8500 and 1100 K behind $A_{V} = 10$. But their analysis relies on photometric points obtained while the system was at different stages. As shown in Fig. 3 of \cite{2016A&A...588A..20G}, an increase in temperature of the blackbody component in the SED of the HBe during the outburst phase might well explain the variation in colors perpendicular to the interstellar reddening vector.

%----------------------------------
\subsection{FUOR component}

%CO en abso typique des Fu Ori (Mould et al. 1978, Carr et al. 1989). 

	\label{subsec:LineIDFUOR}
The NIR spectrum of the FUOR component of Z CMa is characteristic of
M8-M9 giants of the IRTF spectral library \citep{2009ApJS..185..289R},  as expected for FU Orionis objects \citep[see][]{1996ARA&A..34..207H} with
a luminosity in NIR totally dominated by the accretion disk flux and
with the cooler regions of the disk producing deep molecular
absorptions. Strong broad absorption bands of $\mathrm{H_{2}O}$, TiO, VO, and CO
lines are therefore detected. The CO lines in absorption of the FUOR
were previously marginally detected in the unresolved spectrum of Z
CMa by \cite{1997AJ....114.2700R}, through the strong contribution of
the HBe component the K-band flux. These absorptions are typical of FU Orionis objects (Mould et al. 1978, Carr et al. 1989). The individual spectrum of the component obtained by \cite{1997ApJ...478..381L} and \cite{2013ApJ...763L...9H} showed that the CO in absorption can be associated  with this component. The SINFONI (outburst) and OSIRIS (returning to quiescence) spectra shown in Figs. \ref{fig:spec_Jband}, \ref{fig:spec_Hband}, and \ref{fig:spec_Kband} unambiguously confirm this result.

The OSIRIS and SINFONI spectra display a strong Pa $\beta$ absorption. This line is retrieved in the spectra of  FU Orionis objects HBC687 (=IRAS 19266+0932) and V1735 Cyg (=IRAS 21454+4718) of the \cite{2010AJ....140.1214C} sample. The overall spectrum of HBC687 better reproduces the slope of the FUOR component of Z CMa. 

The profile of the Paschen $\beta$ absorption can be fit by two Gaussian functions with FWHM of 8 and 11.5 \AA~(broadly consistent with the spectral resolution), intensity ratio beneath the pseudo-continuum of 1.5, and velocities of $-265$ km.s$^{-1}$ and $-102$ km.s$^{-1}$, respectively. This double-peaked profile is characteristic of FU Orionis stars in the optical \citep[e.g.,][]{1985ApJ...299..462H}. It  is reminiscent of the $H_{\gamma}$ and $H_{\delta}$ line profiles extracted from the optical spectrum of Z CMa acquired in February 1983, while the system was not in outburst \citep{1984AJ.....89.1868C}.  The spectra of HBC687 and  V1735 Cyg seem to exhibit an asymetric Pa $\beta$ line profile similar to the one of Z CMa, although the wavelength sampling of these comparison spectra is lower than SINFONI's. The line profile of the FUOR does not change significantly during the one-year lag corresponding to the OSIRIS and SINFONI spectra. The line properties are clearly not compatible with  the hypothesis of an unresolved binary.

\begin{table}[t]
\caption{Line identification in the $1.1-2.5~\mu m$ spectrum of the FUOR components}             % title of Table
\label{tab:specidFUOR}
\centering
\renewcommand{\footnoterule}{}  % to avoid a line before footnotes
\begin{tabular}{lll}     % 3 columns
\hline\hline\noalign{\smallskip}
$\lambda_{obs}$	& Element							&		Transition																	 		\\
($\mu$m)   		 	&               					    & 												          										\\
\hline
1.10399				&	TiO								&	0--0	 band of 	$\Phi (b {}^1\!\Pi -d  {}^1\!\Sigma)$					\\
1.11752				&	TiO								&	1--1	band of 	$\Phi (b {}^1\!\Pi -d  {}^1\!\Sigma)$ 					\\	
1.13502				&	TR								&	$\dots$																				\\			
1.19745				&	TR								&	$\dots$																				\\			
1.17-1.20			&	VO								&	0--1 band of	 $A  {}^4\!\Pi -X{}^4\!\Sigma-$							\\
1.24339				&	TiO								&	0--1	band of 	$\Phi (b {}^1\!\Pi -d  {}^1\!\Sigma)$					\\
1.25341				&	TR								&	$\dots$																				\\			
1.25643				&	[Fe II] ?							&	$a^{6}\!D_{9/2}-a^{4}\!D_{7/2}$							  				\\			
1.25739				&	TiO								&	1--2	band of 	$\Phi (b {}^1\!\Pi -d  {}^1\!\Sigma)$					\\					
1.27870				&	TR								&	$\dots$																				\\			
1.28104				&H I (Pa$\mathrm{\beta}$)&$3_{*}-5_{*}$																	\\	
1.34830				&	TR								&	$\dots$																				\\	
1.3-1.55				&$\mathrm{H_{2}O}$			&		2$\nu_{3}$, $\nu_{1}$ + $\nu_{2}$, 2$\nu_{1}$,								\\
							&										&	  2$\nu_{2}$ + $\nu_{3}$, $\nu_{1}$ + 2$\nu_{2}$									\\	
\hline
1.50436				&	TR								&	$\dots$																				\\	
1.50898				&	TR								&	$\dots$																				\\	
1.570-1.574		&	TR								&	$\dots$																				\\			
1.597-1.618			&	$\mathrm{{}^{12}\!CO}$ ?	& 5--2 band of $X {}^1\!\Sigma^{+}-X {}^1\!\Sigma^{+}$	\\
1.619-1.634			&	$\mathrm{{}^{12}\!CO}$ ?	& 6--3 band of $X {}^1\!\Sigma^{+}-X {}^1\!\Sigma^{+}$	\\
1.721-1.732				&	TR								&	$\dots$																		\\				
1.70-2.05			&$\mathrm{H_{2}O}$	&		$\nu_{2}$+$\nu_{3}$, $\nu_{1}$ + $\nu_{2}$,  3$\nu_{2}$\\
1.747-1.800			&	TR								&	$\dots$																			\\				
\hline
2.00-2.03				&	TR								&	$\dots$																			\\
2.05-2.08				&	TR								&	$\dots$																			\\
2.10645				&	TR								&	$\dots$																				\\
2.20770				&	TR								&	$\dots$																				\\
2.29452 				& 	$\mathrm{{}^{12}\!CO}$	&2--0 band of $X {}^1\!\Sigma^{+}-X {}^1\!\Sigma^{+}$			\\
2.32382				&	$\mathrm{{}^{12}\!CO}$	&3--1 band of $X {}^1\!\Sigma^{+}-X {}^1\!\Sigma^{+}$			\\
2.35-2.40				&	TR								&	$\dots$																			\\
2.3-2.4				&	$\mathrm{H_{2}O}$		&		$\nu_{1}$,$\nu_{3}$, 2$\nu_{2}$									\\
\noalign{\smallskip}\hline
\end{tabular}
\end{table}

%Decallage vitesse Fe II FuOri: -190 km.s-1
%Pa beta FuOri: Vrad=-309 km.s-1, EW=$-1.53\pm0.10$, Flux=$3.99\pm0.08$
%CO 2.3 microns FuOri : EW=$-1.07\pm0.74$, Flux=$6.60\pm0.67$

The colors of the FUOR recorded in 1990 and during the 2008 outburst (Fig. \ref{fig:zcmacm}) are redder than those of pre-main-sequence stars earlier than K4 \citep[J-H=0.60, H-K=0.17,][]{2013ApJS..208....9P} and typical of classical FU Orionis stars \citep[see Fig. 4 of][]{2008AJ....135.1421G}.  The variations in position of the source in color-color and color-magnitude diagrams during and after the 2008 outburst is almost perpendicular to the interstellar extinction vector. The colors while the system was returning to quiescence fall outside of the locus of FU Ors and classical T Tauri stars of \cite{1997AJ....114..288M}. It is difficult to relate this behavior to those of other FU Ors since there has not been an extensive follow-up of the NIR photometry of these objects so far. Our observations  are consistent with the variation in optical colors, in the period which  show a progressive bluing of Z CMa colors consistent with the behavior of the HBe in the NIR, but also another dependence that could be induced by the variability of the FUOR component \citep{2009IBVS.5905....1G}. 

The intrinsic variation of the FUOR colors could translate into NIR slope variation of our continuum. In that case, the slope of the spectral continuum of the HBe  in the H-band OSIRIS data could be uncertain because it has been based on the hypothesis of a non-variation of the spectrum of the FUOR (see Sect. \ref{subsec:IFS}). This  does not change our conclusions on the variation of the HBe continuum slope, which mostly relies on the broadband NIR colors reported in Table \ref{tab:phot}.  
 
 \subsection{Resolved structures}
\label{subsec:spatstruct}
The spatial sampling of the  SINFONI cubes enable us to look for extended structures down to $\sim$0.15''. We detected  in \cite{2010ApJ...720L.119W} the FUOR  and HBe jets at a position angle of $\sim235^{\circ}$ and $245^{\circ}$, respectively into 1.257 $\mu$m and 1.644 $\mu$m $\mathrm{[Fe II]}$ lines.  In a recent analysis of  $H_{\alpha}$ and $[OI]$ (655.6 nm and 629.5 nm) AO-imaging data (Antoniucci et al. 2016, submitted to A\&A), we report the detection of the wiggling of the FUOR jet. We chose here to re-investigate our data to look for structures in other emission lines in the context of the results of Antoniucci et al. 2016.

\begin{figure*}[t]
\centering
\vspace{-0.1cm}
\begin{tabular}{cccc}
\includegraphics[width =4.3cm]{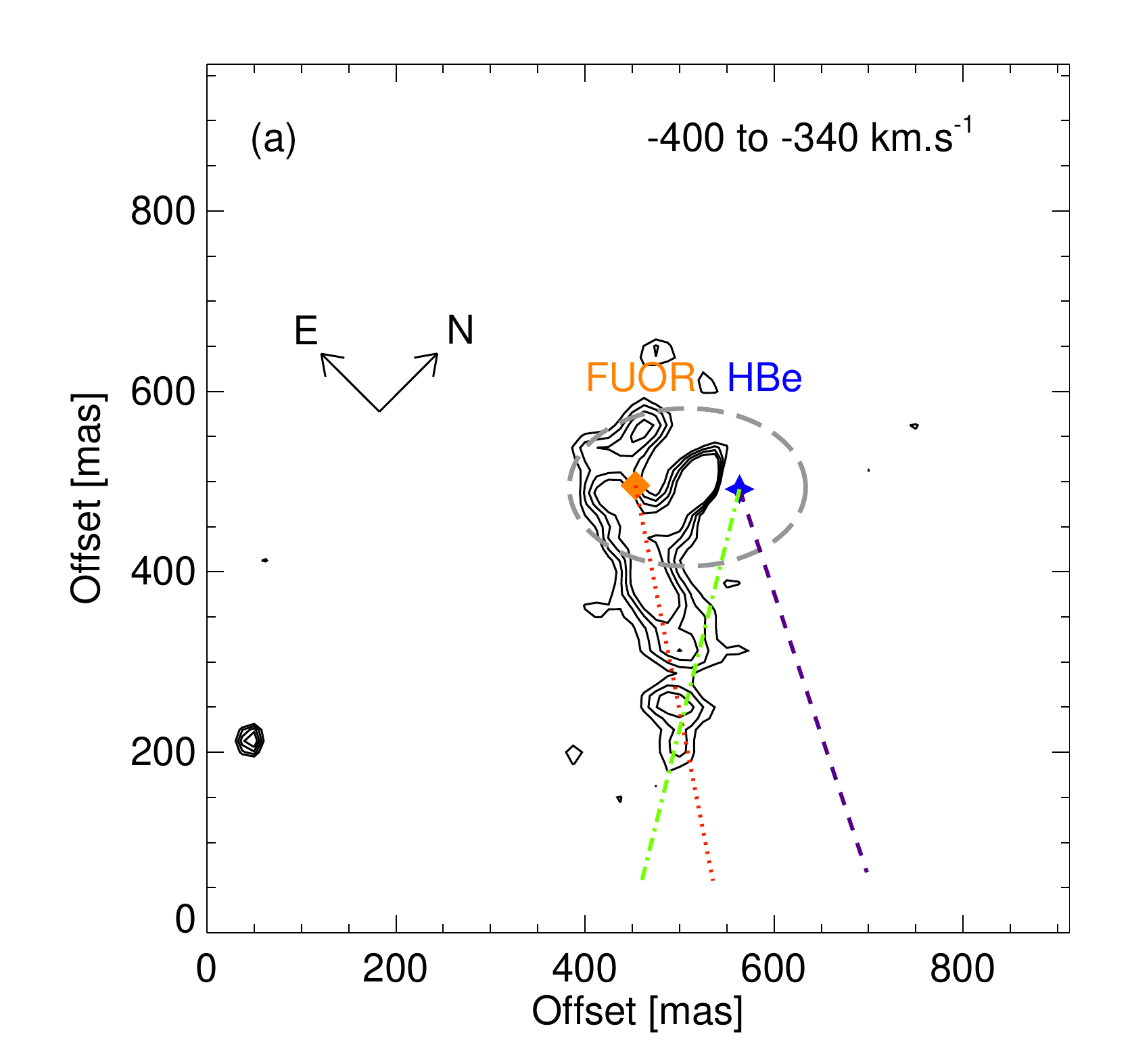} & 
\includegraphics[width =4.3cm]{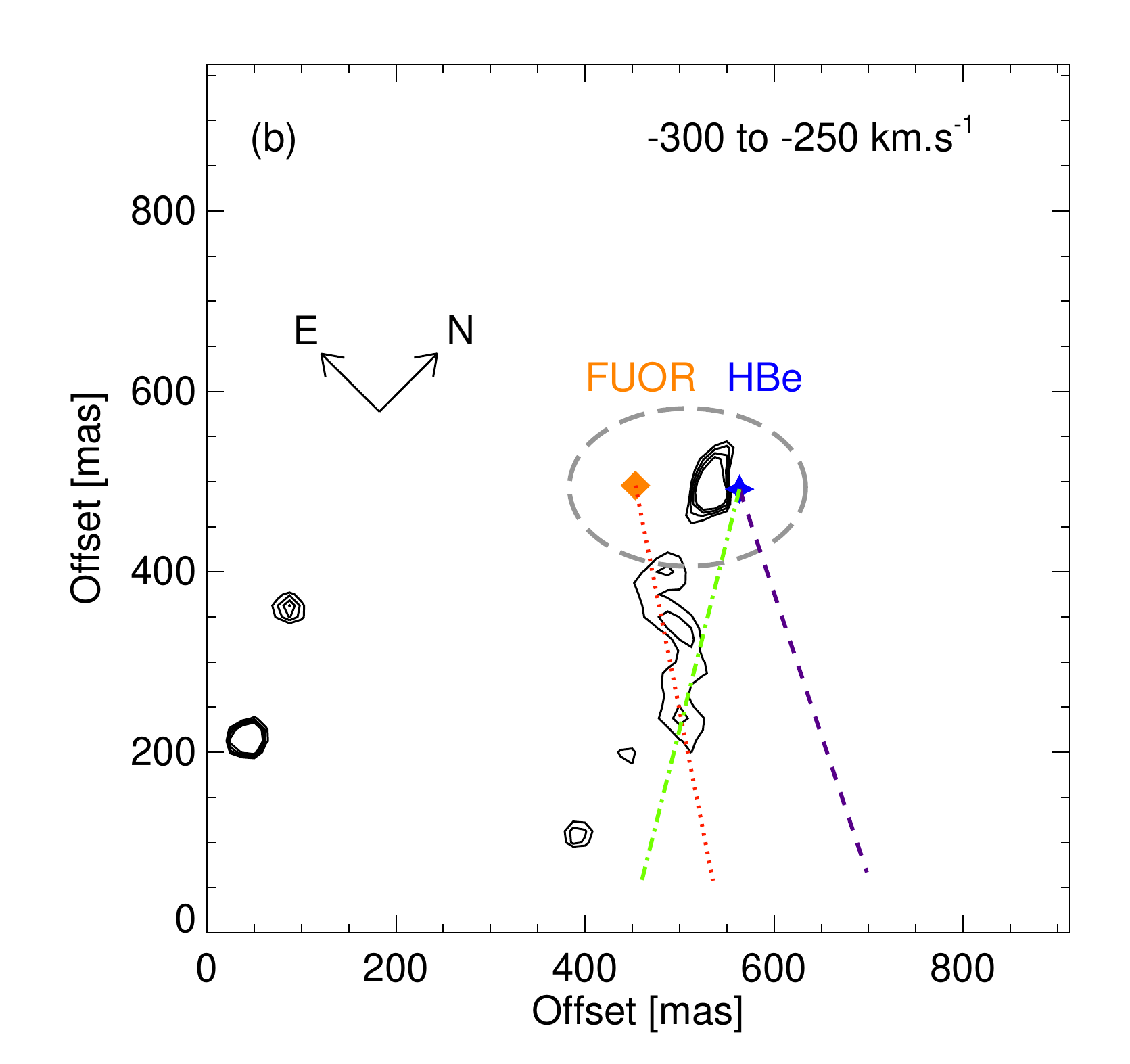} 
\includegraphics[width =4.3cm]{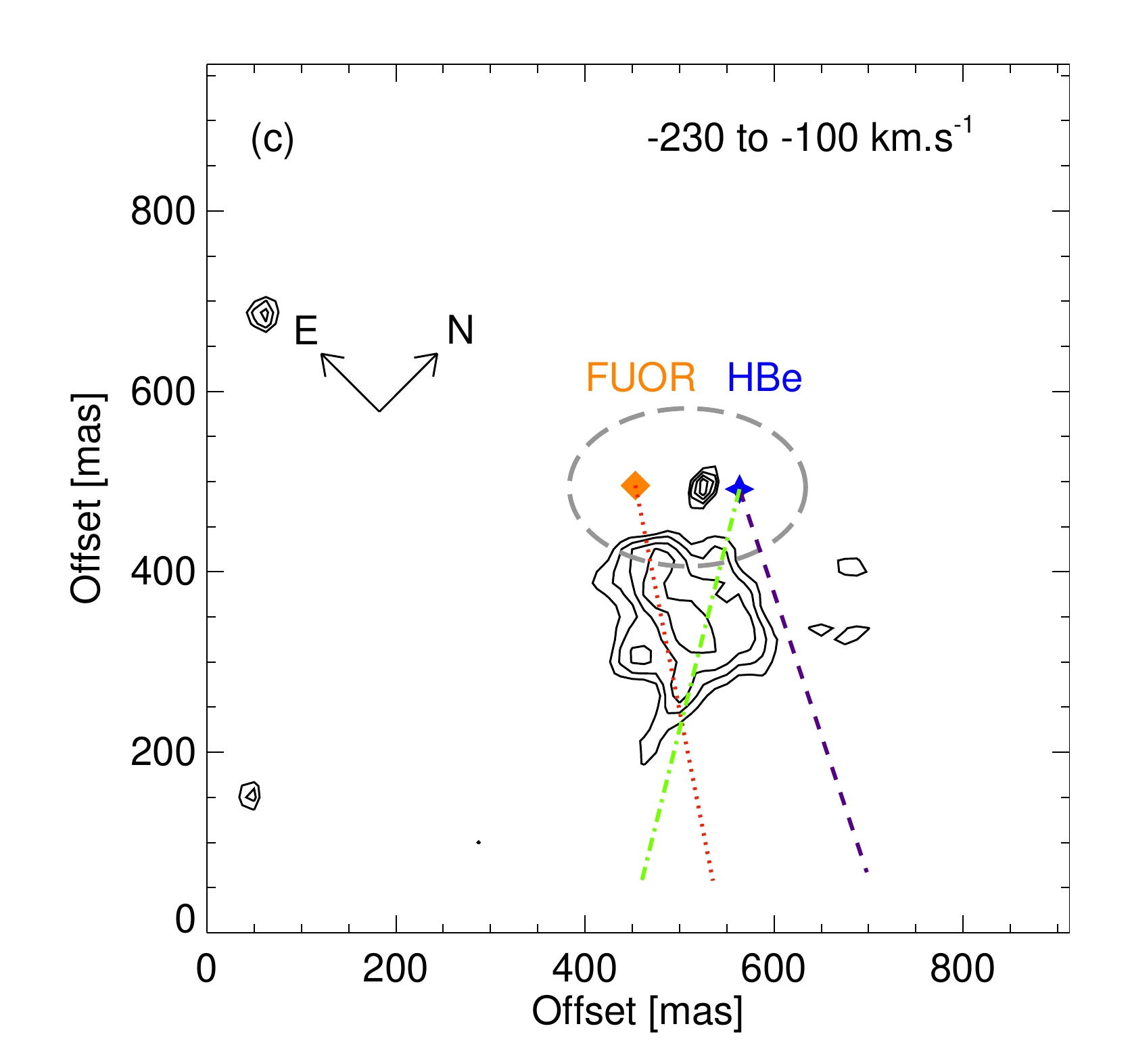} & 
\includegraphics[width =4.3cm]{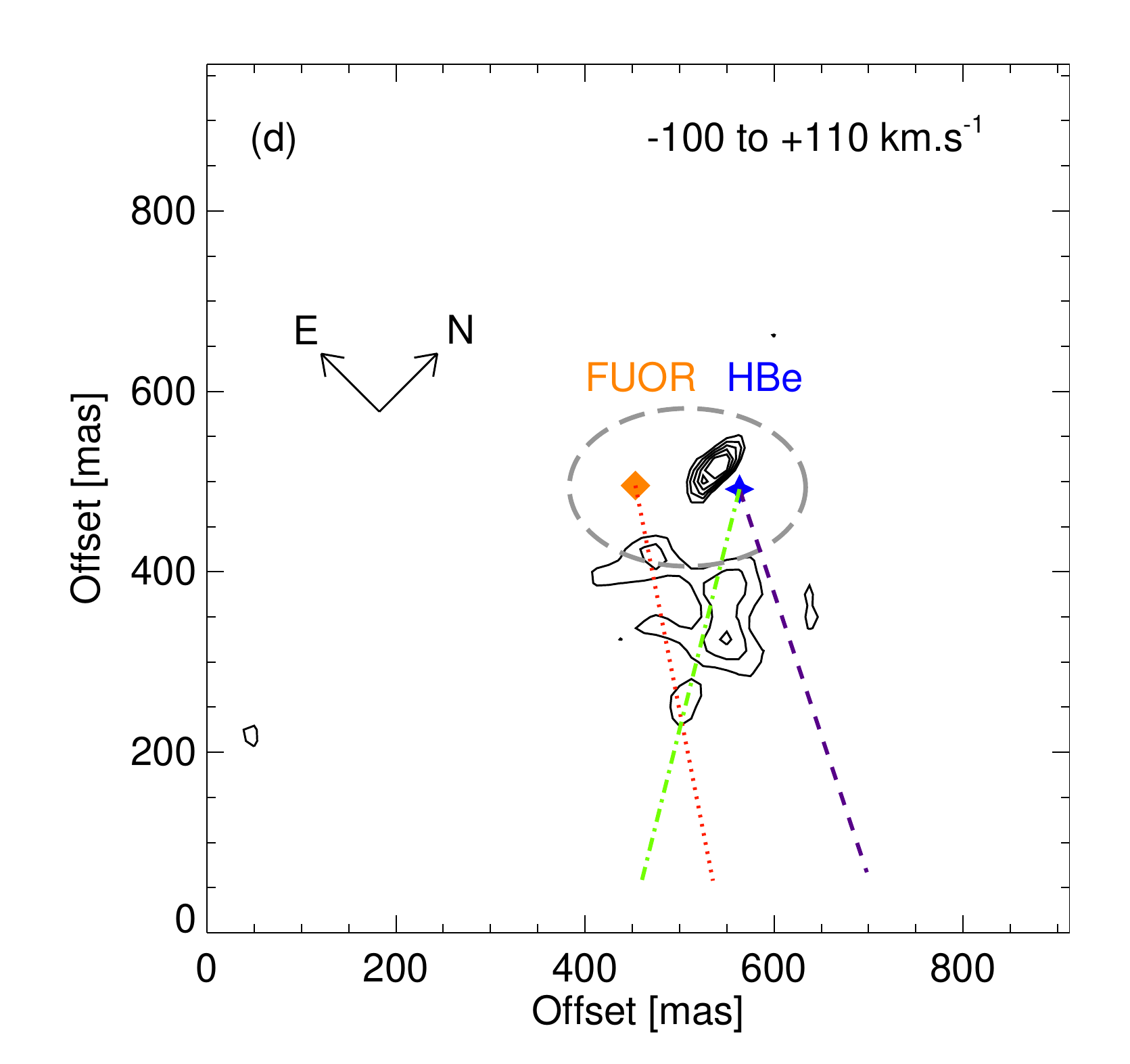} \\ 
\end{tabular}
\caption{Extended structures observed in the $[Fe II]$ line at 1.53 $\mu$m}
\label{fig:FeII1p533}
\end{figure*}

\begin{figure*}[t]
\centering
\vspace{-0.1cm}
\begin{tabular}{ccc}
\includegraphics[width =5.5cm]{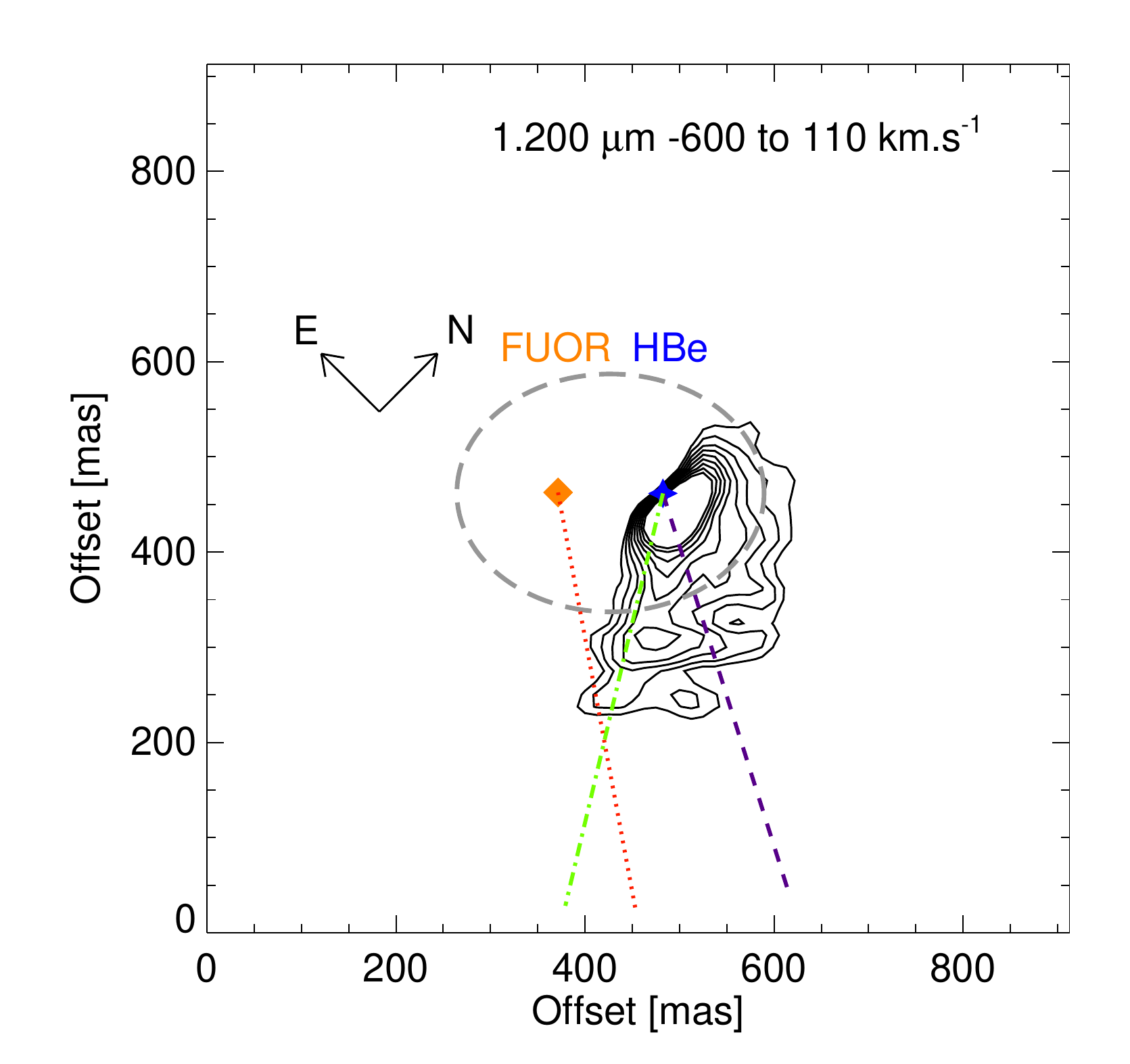} & 
\includegraphics[width =5.5cm]{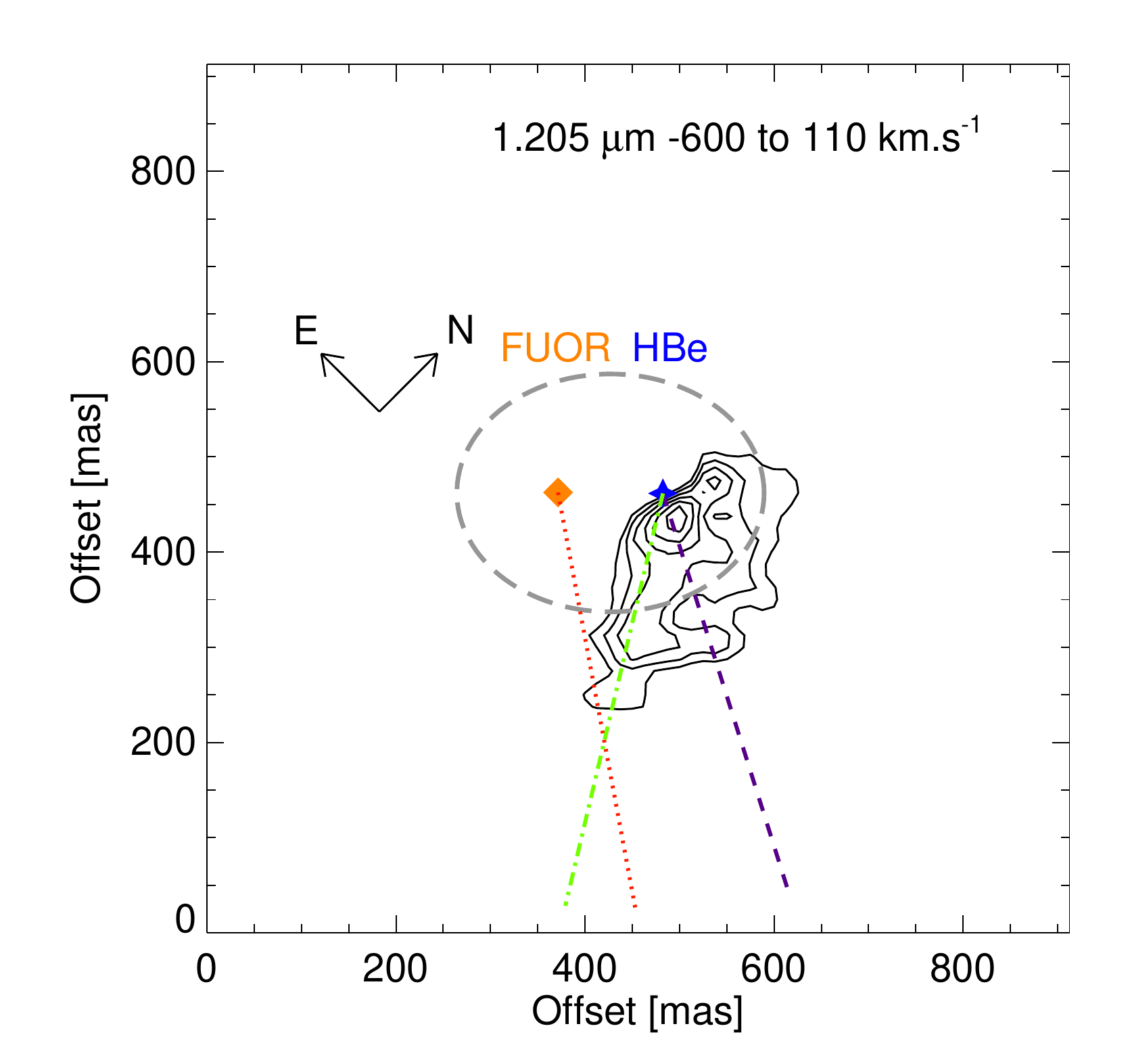} &
\includegraphics[width =5.5cm]{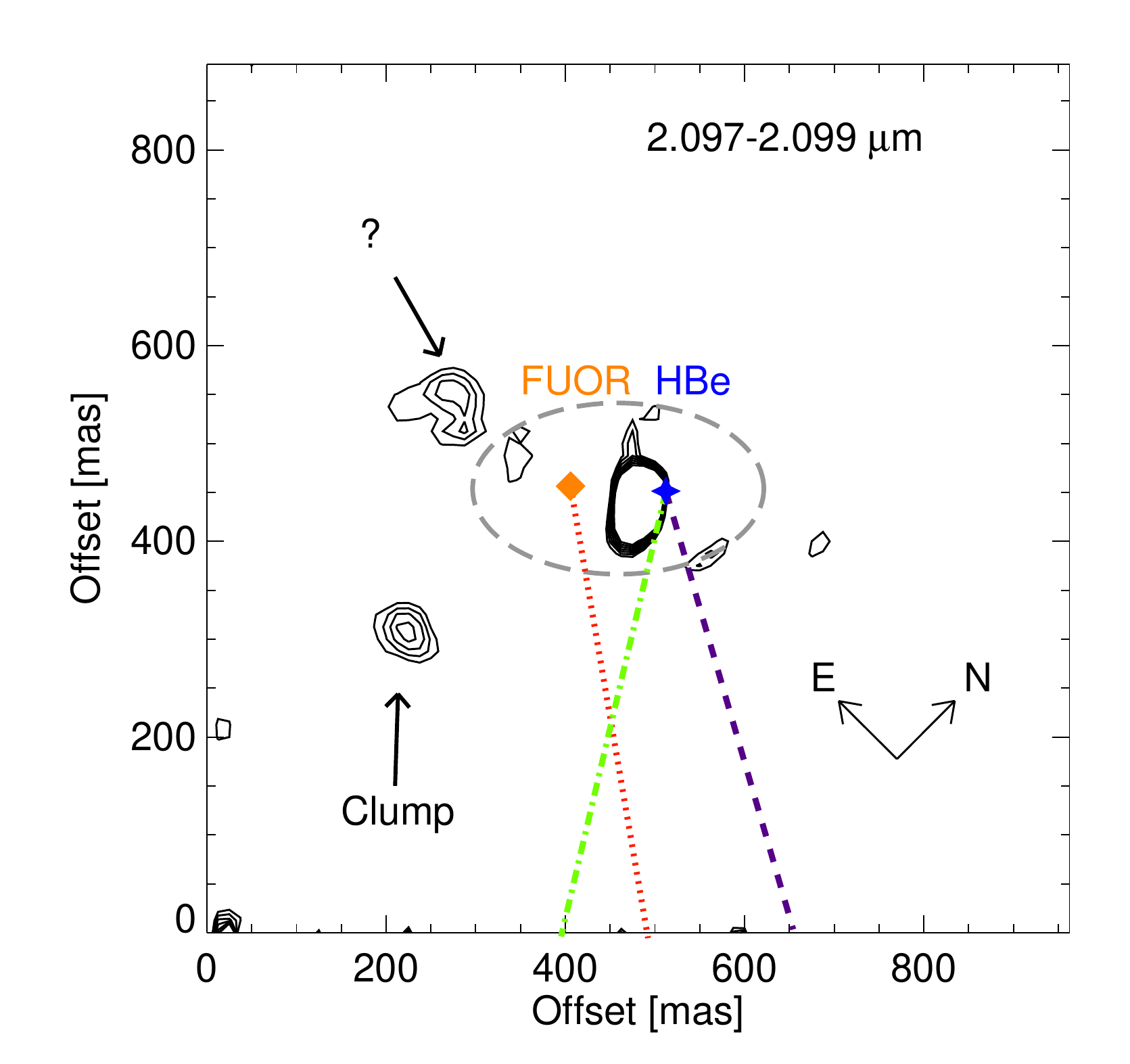} \\ 
\end{tabular}
\caption{Extended structures observed at 1.200, 1.205, and 2.098 $\mu$m}
\label{fig:1p205}
\end{figure*}

We report extended emission at 1.200, 1.205, and 1.533, and 2.098 $\mu$m following the subtraction of the continuum emission (see Appendix \ref{appB:contemi} for the details). We classify these emissions into three categories: 

\begin{itemize}
\item The 1.533 $\mu$m emission corresponds to the $[Fe II]$ line seen in our spectrum of the system (Table \ref{tab:specid} and Fig. \ref{fig:spec_Hband}). The emission map at more than 3 $\sigma$ the noise level  is shown in Fig. \ref{fig:FeII1p533}.  The dashed ellipse corresponds to the era within which the continuum substraction leaves strong residuals. The emission is associated with the FUOR micro jet (orange dashed line in the figures). It is elongated along a position angle of $\sim$238$^{\circ}$\footnote{This value is consistent with the interval given in \cite{1989A&A...224L..13P} for the large-scale jet and by \cite{2010ApJ...720L.119W} and Antoniucci et al. (2016) for the FUOR micro jet.}. The emission is detected in the same velocity range at 1.257 and 1.644 $\mu$m.  The emission has a sinusoid shape with a $\sim$75 mas semi-period from $-400$ to $-100$ km.s$^{-1}$ that is consistent with the wiggling found by Antoniucci et al. 2016.

\item  The 1.200 and 1.205 $\mu$m line emissions may be associated with $[Fe II]$ lines at 1.2000278$\mu$m and 1.2054490$\mu$m, respectively. We assume that we correctly identified the lines to map the emission in the velocity space in Fig.  \ref{fig:1p205}. These lines trace an elongated emission at a position angle of $\sim$214$^{\circ}$ (green dashed lines in the Figs. \ref{fig:FeII1p533} to \ref{fig:1p205}). The emission is seen from -600 to 110 km.s$^{-1}$.  Its elongation is in the same direction as the $K1$ knot reported by \cite{2010ApJ...720L.119W}. The structure is not seen in the polarized band imaging data of \cite{2015A&A...578L...1C} and \cite{2016SciA....200875L},  but the position angle corresponds to the one of the elongated stream (see Sect. \ref{section:extstruc}). The stream  does not extend down to 0.15" however, and does not point toward the HBe  \citep{2015A&A...578L...1C}. Additional observations carried out at these wavelengths with high-order adaptive optic instruments such as VLT/SPHERE (\texttt{N\_CntJ} filter) may help to clarify the nature of this structure.
 
 \item A clump is detected at 2.098 $\mu$m (Fig. \ref{fig:1p205}) at a PA of $164\pm3^{\circ}$ and separation of $313\pm13$ mas (291-360au). This clump is at the same position as the polarized clump or arm detected in the H (1.66 $\mu$m) and Ks (2.18 $\mu$m) band  by \cite{2015A&A...578L...1C} and \cite{2016SciA....200875L}.  The He I and H$_{2}$ emission lines seen in jets and embedded objects in this wavelength range translate into an absolute velocity $\gg$ 1000 km.s$^{-1}$. Plausible explanations are that the line is incorrectly identified, or that the structure is unrelated to the system. We detect another clump at a position angle of $\sim95^{\circ}$ (labeled as $?$ in the Fig. \ref{fig:1p205}). This additional clump is also found at 2.11$\mu$m, and could correspond to an emission in the He I line (2.112583 $\mu$m) blueshifted by $-400$ to $-200$ km.s$^{-1}$.  Nonetheless, it falls close to the PSF Airy ring of the FUOR which moves with wavelength and might produce some residual emission at the continuum subtraction step.    
\end{itemize}
 
 We were unable to detect any extended structures apart from the probable diffraction spikes associated with the HBe and FUOR point sources in  images derived from the cubes collapsed in wavelengths and with the stellar halo of each component removed with a Gaussian smoothing (with a FWHM=5, 10, and 15 pixels). 
 
 %Figure FeII line at 1.2054241 microns
%Figure FeII line at 1.2000278 microns

\section{Origins of the outburst:  extinction or accretion?}
\label{section:discussion}
%The behavior of the HBe during the 2008 ouburst can be summarized as follow: 
%\begin{itemize}
%\item a bluering of the spectral continuum;
%\item an increase in the equivalent widths of the Hydrogen lines, among others; 
%\item an increase of the V band and JHK band flux;
%\item a bluering of the colors consistent with the one found on the spectra, and that appears to only partially follow the interstellar extinction vector. The rest could be due to a black-body component whose temperature varies between the quiescence and outburst stages; 
%\item a shift of the blue component of the line profile of the Paschen $\beta$ line toward higher velocities (in absolute value). 
%\end{itemize}
%

The architecture of the Z CMa system is reminiscent of those of other EX Ors objects with  binary companions such as V1118 Ori \citep{2007AJ....133.1000R}, VY Tau \citep{1993A&A...278..129L}, and XZ Tau \citep{2003ApJ...583..334H}, altough Z CMa may be the most massive system of this class of outbursting binaries, and the only one with a FU Or companion. These three systems have projected separations of 72 au, 95 au, and 43 au respectively. It is possible that the outbursts of  Z CMa are related to instabilities in the disk surrounding the HBe at the corotation radius (e.g., much closer to the star) where gas accumulates until it is accreted during outbursts \citep{2012MNRAS.420..416D}. This scenario may explain the behavior of EX Lup \citep{2012MNRAS.420..416D, 2015ApJ...798L..16B}. In this case, are our spectrophotometric data  of the HBe of Z CMa more compatible  with the scenario of an accretion+ejection event, or a variation of the extinction  in the line of sight?  %In that respect, Z CMa shares a similar architecture, but scaled-up in mass, to  the HD 100453 system. This 1.7 M$_{\odot}$ Herbig Ae star has a companion at a projected separation of 120 au which forms a pair of objects with an identical mass ratio as Z CMa. The disk surrounding the Herbig is structured \citep{2015ApJ...813L...2W} with spiral arms seein in scattered light (dust) extending to 42 au. \cite{2016ApJ...816L..12D} has shown that the spirals could be induced by the companion. 

The fact that the jet of the HBe wiggles \citep{2010ApJ...720L.119W}, the existence of knots along the direction of the HBe outflow, and the change in velocity of the blueside wing of the Paschen $\beta$ line associated with the HBe outflow suggest that the 2008 outburst and spectrophotometric properties of the most massive star of the system are related to an accretion+ejection event. The spectrum of the HBe, quite similar to the one of the prototype EX Lup, strengthens this conclusion. Nonetheless, because (i) the Br $\gamma$ line emission may originate at least partly in a disk wind \citep{2010A&A...517L...3B} and (ii) the empirical relation between the Br $\gamma$ flux and the accretion rate seems to break down in the Be regime \citep[][]{2011AJ....141...46D},  the accretion rate as well as the extinctions from line ratio (Pa$\beta$/Br$\gamma$)  cannot be derived reliably from our NIR spectra.

\cite{2010A&A...509L...7S} claimed that the 2008  outburst is related to a change in extinction from the dust cocoon surrounding the HBe based on the level of optical polarization in both continuum and spectral lines along a position angle roughly perpendicular to the outflow launched by the HBe. The characteristics of the HBe component during the 2008 outburst are not fully typical of EX Or objects and outbursting stars. The variation in  the NIR continuum slope (colors) of the HBe with respect to the absolute NIR flux is  \textit{anticorrelated} with the one of V1647 Ori, PV Cep, or V1118 Ori \citep{2006ApJ...641..383G, 2009ApJ...693.1056L, 2016ApJ...819L...5G}. The emission lines of these four reference objects are \textit{correlated} with their overall NIR brightness (outburst or quiescence) like Z CMa, based on the current stage of observations (temporal baseline, frequency of the measurements) of all those systems.  \cite{2009A&A...499..529S} noted that V1118 Ori  had a different X-ray luminosity and temperature variation during an outburst with respect to the behavior of Z CMa during the 2008 outburst, thus pointing to a different outburst mechanism for these systems.   Z CMa may resemble the case of  PTF 10nvg where the spectrophotometry of the system seems to be driven by the rotating circumstellar disk material located at short separation  ($<$ 1 au) which causes semi-periodic dimming, and by the emission of excited zones in that same disk \citep{2013A&A...551A..62K}. PTF 10nvg reproduces the behavior of the absolute flux, NIR spectral slope, and NIR emission lines of Z CMa. 

The 2008 outburst can be placed in context now that the system has been monitored for several years before and after the outburst (Fig. \ref{fig:lightcurve}).  VDA04 noted that photometric variations of ZCMa between 1987 February and 1991 closely follow a behavior of redder optical colors when the system becomes fainter.  It is therefore quite possible that the 2008 and 1987 outbursts of Z CMa have close origins. At least, both outburst were caused by the HBe. We also note that the 2008 outburst was followed closely by another one in 2011 with similar V-band brightness.  We can speculate that the 2008 and 2011 outbursts might have been caused by  one single accretion+ejection event modulated by absorption in the line of sight during the quiescent stage of 2010 (OSIRIS spectra). This would explain why the CO bandhead is seen in emission during the 2006 quiescent stage (the last one before the 2008 outburst) and then could not be seen at all in the OSIRIS spectrum taken during the 2009 quiescent period (veiling). 

%\subsection{On the system architecture}
%\label{section:discussion}

%Companion evidenced around other EX Ors ?
%Knots in polarized intensity ==>instability 
%Downscaled version of HD 100453? with primary =1.7 Msun, companion =0.3-0.5 MSun at 110 au, then same sep and mass ratio? Dong et al.  talk about 

%------------------------------------------------------------------------
\section{Conclusion}
\label{section:summary}
We have obtained multi-epoch NIR astrometric, photometric, and spectroscopic data of the young system Z CMa using adaptive-optics-fed instruments. Our data enabled us to obtain resolved 1-4 $\mu$m photometry as well as medium-resolution (R$\sim$2000-4000) spectra of each component. 

Our photometry confirms that the Herbig component of the system is the source driving the ourburst. Its spectrum in the outburst stage is characteristic of EX Ors such as EX Lupi and of embedded young stellar objects. The emission lines, NIR luminosity,  spectral continuum (colors) of the Herbig component of ZCMa, and their evolution with time, suggest that the properties of the object are driven both by accretion+ejection changes and variable extinction in the line of sight. We find in particular a correlation between the strength of the Br $\gamma$ emission and the brightness of the system.  Nevertheless, we cannot have a direct access to the reddenning and accretion rate because the emission lines that would enable for an estimation originate in different locations around the Herbig objet (disk, outflow). This prevents us from firmly linking the spectrophotometric properties of the system to their origins. The FU Or companion also experienced color variations at a constant luminosity during the outburst that do not seem to be related to variable extinction. 

We detected  the extended emission the FUOR jet at 1.53 $\mu$m ([Fe II] line). We also resolved an extended emission at 1.200 and 1.205 $\mu$m that extends along a position angle of 214$^{\circ}$, and whose origin is unclear. To conclude, we identified a point source at 2.098 $\mu$m that is concomitant with a polarized clump seen in  multiple NIR imaging data. 

The system is undergoing a new outburst in 2016. Similar observations as conducted in our study, but with a better temporal sampling, coupled to observations at longer wavelengths with the mid-infrared instrument VISIR \citep{2004Msngr.117...12L} and ALMA may dramatically improve our understanding of the architecture of the system, and how it relates to the accretion+ejection diagnostics used in the NIR.

%------------------------------------------------------------------------

\begin{appendix}
\section{Details on modeling the CO emission}
\label{appA:detailsCO}
We first considered a disk defined by inner and outer radii ($R_{in}$, $R_{out}$) and an inclination ($i$). The radial velocity was computed for each point of the disk assuming that the disk material is in Keplerian rotation around a star with a mass $M_{\star}$. As for the CO emission, we  used Eq. 10  from \cite{2000A&A...362..158K} to calculate the absorption coefficient $\kappa$.  A simple Gaussian was assumed for the spectral function. This  accounts for the thermal and turbulent broadening and for a medium-resolution spectrum. $\kappa$ was multipled with the column density of CO to obtain the optical depth $\tau$. We assumed a simple slab, so that the column density and excitation temperature of CO ($T_{ex}$) were identical at every location in the disk. We calculated the intensity $I$ following: 		
		\begin{equation}
		I = BB \times (1-e^{(-\tau / \cos(i))})
		\end{equation}
where BB is the Planck function corresponding to the excitation temperature.
		 
The computed spectrum  was redshifted (or blueshifted) by the radial velocity appropriate for each point of the disk. We finally integrated over the whole disk in each spectral channel and smoothed the resulting spectrum to account for the finite instrumental spectral resolution of SINFONI.
		
We varied  $T_{ext}$ between 1500 K and 4500 K, in steps of 150 K, and a column density $N_{\mathrm{CO}}$ between $\mathrm{10^{17.6} cm^{-2}}$ and  $\mathrm{10^{19.8} cm^{-2}}$  in logarithmic scale, in steps of 0.1 in the exponent.  We took $R_{in}=0$ au, $i$=20$^{\circ}$, and  $\mathrm{M_{\star}=16 M_{\odot}}$. We considered $R_{out}$  from 0.5 to 3.5 au, with 0.5 au increments. 

We compared the first two overtones of the HBe (following the removal of the continuum in the SINFONI spectrum) to the grid of 3381 model spectra using a $\chi_{2}$ minimization. 

\section{Removal of the continuum emission}
\label{appB:contemi}
The removal of the spectral continuum in the SINFONI datacubes is required to look for faint structures close to the binary. This is usually done by fitting spaxel to spaxel the spectral continuum around the object line. Here, the intrinsically different continua from the FUOR and HBe made this removal more complicated. We found that using neighboring cube slices as reference PSF provided a good solution to remove most of the continuum flux of each component. The technique works better than the subtraction in the spectral range (fitting a low-order polynomial) or the spatial subtraction of a PSF model of each component using the extracted cubes derived from the \texttt{CLEAN 3D} algorithm (see Sect. \ref{subsec:IFS}).  

\end{appendix}

\bibliographystyle{aa}

\begin{acknowledgements}
 We would like to thank particularly the staff of ESO-VLT for their support at the telescope as well as  Sasha Hinkley and Mario Van Den Acker, who kindly provided their  NIR photometry and spectra of the Z CMa system. We are very grateful to Mike Connelley, Kevin Covey, Lynne Hillenbrand, Erika Gibb, and Alessio Caratti o Garatti for providing their spectra of outbursting and embedded objects.  We acknowledge with thanks the variable star observations from the \textit{AAVSO International Database} contributed by observers worldwide (Eric Blown in particular) and used in this research. This work was supported by the Momentum grant of the MTA CSFK Lend\"{u}let Disk Research Group.  We  acknowledge partial financial support from the {\sl Programmes Nationaux de Plan\'etologie et de Physique Stellaire} (PNP \&  PNPS) and the {\sl Agence Nationale de la Recherche}, in France.
\end{acknowledgements}

\bibliography{ZCMa_bonnefoy}

\end{document}